\documentclass[11pt,twoside,openright,DIV10,BCOR5mm,pagesize]{scrbook}

\pdfoutput=1

\typearea[current]{last}

\ifx  \Comment \Undefined
      \long \def \Comment #1\EndComment {}
\else
      \ifx  \EndComment \Undefined
            \def \EndComment
                {\message {Error: \noexpand \EndComment encountered
                                without preceding \noexpand \Comment}}
      \else %%% both defined; let's hope they're the right definitions!
      \fi
\fi

\Comment
This file, "letterspacing.tex", created by Philip Taylor of RHBNC
(P.Taylor@Mail.Rhbnc.Ac.Uk) for Kaveh Bazargan of Focal Image
(Kaveh@Focal.Demon.Co.Uk) may be freely distributed provided that
no changes whatsoever are made to its contents
\EndComment

\Comment
As far as is possible, `inaccessible' control sequences (i.e.~control sequences
containing one or more commercial-at$\,$s) are used to minimise the risk of
accidental collision with user-defined macros; however, the control sequences
used to access the natural dimensions of the text are required to be user
accessible, and therefore must contain only letters.
\EndComment

\newdimen \naturalwidth
\newdimen \naturaldepth
\newdimen \naturalheight

\Comment
All control sequences defined hereafter are inaccessible to the casual user,
including the control sequence used to store the category code of commercial-at;
this catcode must be saved in order to be able to re-instate it at end of
module, as we have no idea in what context the module will be |\input|. Having
saved the catcode of commercial-at, we then change it to |11| (i.e.~that of a
letter) in order to allow it to be used in the cs-names which follow.
\EndComment

\expandafter \chardef
        \csname \string @code\endcsname =
                \the \catcode `\@
\catcode `\@ = 11

\Comment
We will need a a box register in order to measure the natural dimensions of
the text, and a token-list register in which to save the tokens to be
letter-spaced.
\EndComment

\newbox \l@tterspacebox
\newtoks \l@tterspacetoks

\Comment
We will need to test whether a particular macro expands to a space, so we will
need another macro which does so expand with which to compare it; we will
also need a `more infinite' variant of |\hss|.
\EndComment

\def \sp@ce { }
\def \hsss
    {\hskip 0 pt plus 1 fill minus 1 fill\relax}

\Comment
Many of the macros will used delimited parameter structures, typically
terminated by the reserved control sequence |\@nd|; we make this a synonym
for the empty macro (which expands to nothing), so that should it accidentally
get expanded there will be no side effects.  We also define a brief synonym
for |\expandafter|, just to keep the individual lines of code reasonably short.
\EndComment

\let \@nd = \empty
\let \@x = \expandafter

\Comment
We will also need to compare a token which has been peeked at by |\futurelet|
with a space token; because of the difficulty of accessing such a space
token (which would usually be absorbed by a preceding control word), we
establish |\sp@cetoken| as a synonym.  The code to achieve this is messy,
because of the very difficulty just outlined, and so we `waste' a control
sequence |\t@mp|; we then return this to the pool of undefined tokens.
\EndComment

\let \sp@cetoken = \relax
\edef \t@mp {\let \sp@cetoken = \sp@ce}
\t@mp \let \t@mp = \undefined

\Comment
The user-level macro |\letterspace| has exactly the same syntax as that for
|\hbox| and |\vbox|, as explained in the introduction; the delimited parameter
structure for this macro ensures that everything up to the open brace which
delimits the beginning of the text to be letter-spaced is absorbed as parameter
to the macro, and the brace-delimited text which follows is then assigned to the
token-list register |\l@tterspacetoks|; |\afterassignment| is used to regain
control once the assignment is complete.
\EndComment

\def \letterspace #1#%
    {\def \hb@xmodifier {#1}%
     \afterassignment \l@tterspace
     \l@tterspacetoks =
    }

\Comment
Control then passes to |\l@tterspace|, which starts by setting an |\hbox|
containing the text to be typeset; the dimensions of this box (and therefore
of the text) are then saved in the open control sequences previously declared,
and the |\hbox| becomes of no further interest.

A new |\hbox| is now created, in which the same text, but this time
letter-spaced, will be set; the box starts and ends with |\hss| glue so that if
only a single character is to be letter-spaced, it will be centered in the box.
If two or more characters are to be letter-spaced, they will be separated by
|\hsss| glue, previously declared, which by virtue of its greater degree of
infinity will completely override the |\hss| glue at the beginning and end;
thus the first and last characters will be set flush with the edges of the box.

The actual mechanism by which letter-spacing takes place is not yet apparent,
but it will be seen that it is crucially dependent on the definition of
|\l@tt@rspace|, which is expanded as the box is being set; the |\@x|
(|\expandafter|) causes the actual tokens stored in |\l@tterspacetoks| to be
made available as parameter to |\l@tt@rspace| rather than the token-list
register itself, as it is these tokens on which |\l@tt@rspace| will operate.
The |\@nd| terminates the parameter list, and the |{}| which immediately
precedes it ensures that that there is always a null element at the end of the
list: without this, the braces which are needed to protect an accent/character
pair would be lost if such a pair formed the final element of the list, at the
point where they are passed as the second (delimited) parameter to |\p@rtiti@n|;
by definition, <TeX> removes the outermost pair of braces from both simple and
delimited parameters during parameter substitution if such braces form the first
and last tokens of the parameter, and thus if a brace-delimited group ever
becomes the second element of a two-element list, the braces will be irrevocably
lost.  The |{}| ensure that such a situation can never occur.
\EndComment

\def \l@tterspace
    {\setbox \l@tterspacebox = \hbox
                {\the \l@tterspacetoks}%
     \naturalwidth =  \wd \l@tterspacebox
     \naturaldepth =  \dp \l@tterspacebox
     \naturalheight = \ht \l@tterspacebox
     \hbox \hb@xmodifier
     \bgroup
           \hss
           \@x \l@tt@rspace
           \the \l@tterspacetoks {}\@nd
           \hss
     \egroup
    }

\Comment
The next macro is |\l@tt@rspace|, which forms the crux of the entire operation.
The text to be letter-spaced is passed as parameter to this macro, and the
first step is to check whether there is, in fact, any such text; this is
necessary both to cope with pathological usage (e.g.~|\letterspace {}|), and to
provide an exit route, as the macro uses tail-recursion to apply itself
iteratively to the `tail' of the text to be letter-spaced; when no elements
remain, the macro must exit.

Once the presence of text has been ensured, the token-list representing
this text is partitioned into a head (the first element), and the tail
(the remainder);  at each iteration, only the head is considered, although
if the tail is empty (i.e.~the end of the list has been reached), special
action is taken.

If the first element is a space, it is treated specially by surrounding it
with two globs of |\hsss| glue, to provide extra stretchability when compared
to the single glob of |\hsss| glue which will separate consecutive non-space
tokens; otherwise, the element itself is yielded, followed by a single glob
of |\hsss| glue.  This glue is suppressed if the element is the last of the
list, to ensure that the last token aligns with the edge of the box.

When the first element has been dealt with, the macro uses tail
recursion to apply itself to the remaining elements (i.e.~to the tail);
|\@x| (|\expandafter|) is again used to ensure that the tail is expanded
into its component tokens before being passed as parameter.
\EndComment

\def \l@tt@rspace #1\@nd
    {\ifx  \@nd #1\@nd
           \let \n@xt = \relax
     \else
           \p@rtition #1\@nd
           \ifx \h@ad \sp@ce \hsss \h@ad \hsss
           \else
                \h@ad
                \ifx \t@il \@nd \else \hsss \fi
           \fi
           \@x \def \@x \n@xt \@x
                {\@x \l@tt@rspace \t@il \@nd}%
     \fi
     \n@xt
    }

\Comment
The operation of token-list partitioning is conceptually simple: one passes the
token list as parameter text to a macro defined to take two parameters, the
first simple and the second delimited; the first element of the list will be
split off as parameter-1, and the remaining material passed as parameter-2.
Unfortunately this na\"\i ve approach fails when the first element is a bare
space, as the semantics of <TeX> prevent such a space from ever being passed as
a simple parameter (it could be passed as a delimited parameter, but as one
does not know what token follows the space, defining an appropriate delimiter
structure would be tricky if not impossible).  The technique used here relies
upon the adjunct macro |\m@kespacexplicit|, which replaces a leading bare space
by a space protected by braces; such spaces may legitimately be passed as simple
parameters.  Once that operation has been completed, the re-constructed token
list is passed to |\p@rtiti@n|, which actually performs the partitioning as
above; again |\@x| (|\expandafter|) is used to expand |\b@dy| (the
re-constructed token list) into its component elements before being passed as
parameter text.
\EndComment

\def \p@rtition #1\@nd
    {\m@kespacexplicit #1\@nd
     \@x \p@rtiti@n \b@dy \@nd
    }

\def \p@rtiti@n #1#2\@nd
    {\def \h@ad {#1}%
     \def \t@il {#2}%
    }

\Comment
The operation of making a space explicit relies on prescience: the code needs
to know what token starts the token list before it knows how to proceed.
Prescience in <TeX> is usually accomplished by |\futurelet|, and this code
is no exception: |\futurelet| here causes |\h@ad| to be |\let| equal to
the leading token, and control is then ceded to |\m@kesp@cexplicit|.

The latter compares |\h@ad| with |\sp@cetoken| (remember the convolutions we
had to go through to get |\sp@cetoken| correctly defined in the first place),
and if they match (i.e.~if the leading token is a space), then |\b@dy|
(the control sequence through which the results of this operation will
be returned) is defined to expand to a protected space (a space surrounded by
braces), followed by the remainder of the elements; if they do not match
(i.e.~if the leading token is \stress {not} a space), then |\b@dy| is simply
defined to be the original token list, unmodified.  If the leading token
\stress {was} a space, it must be replaced by a protected space: this is
accomplished by |\pr@tectspace|.

The |\pr@tectspace| macro uses a delimited parameter structure, as do most
of the other macros in this suite, but the structure used here is unique,
in that the initial delimiter is a space.  Thus, when a token-list starting
with a space is passed as parameter text, that space is regarded as matching
the space in the delimiter structure and removed; the expansion of the macro
is therefore the tokens remaining once the leading space has been removed,
preceded by a protected space |{ }|.
\EndComment

\def \m@kespacexplicit #1\@nd
    {\futurelet \h@ad \m@kesp@cexplicit #1\@nd}

\def \m@kesp@cexplicit #1\@nd
    {\ifx \h@ad \sp@cetoken
          \@x \def \@x \b@dy
                     \@x {\pr@tectspace #1\@nd}%
     \else
          \def \b@dy {#1}%
     \fi
    }%

\@x \def \@x \pr@tectspace \sp@ce #1\@nd {{ }#1}

\Comment
The final step is to re-instate the category code of commercial-at.
\EndComment

\catcode `\@ = \the \@code

\Comment
Thus letter-spacing is accomplished.  The author hopes both that the code will
be found useful and that the explanation which accompanies it will be found
informative and interesting.
\EndComment

\usepackage[latin1]{inputenc}
\usepackage[english]{babel}

\usepackage[T1]{fontenc}
\usepackage{textcomp}
\usepackage{setspace}
\typearea[current]{last}
\usepackage[expansion=false]{microtype}
\usepackage{enumitem}
\setlist{itemsep=.1ex}

\ifpdfoutput{
  
  \usepackage[pdftex]{graphicx}
  \DeclareGraphicsExtensions{.pdf,.png,.jpg}
}{
  \usepackage{graphicx}
  \DeclareGraphicsExtensions{.eps.gz,.eps,.png,.jpg}
}
\usepackage[font=normalsize]{subfig}
\usepackage{rotating}
\usepackage{tabularx}
\usepackage{booktabs}
\usepackage{multirow}

\usepackage{tocvsec2}

\usepackage{makeidx}
\makeindex

\usepackage{amsmath}
\usepackage{amssymb}
\usepackage[liter]{unitsdef}

\usepackage{hyperref}
\hypersetup{%
  bookmarks          = true,
  bookmarksnumbered  = true,
  bookmarksopen      = true,
  bookmarksopenlevel = 1,
  linkcolor          = blue,
  breaklinks         = true,
  pdfauthor          = {Lutz Petersen},
  pdfsubject         = {Diploma thesis},
  pdftitle           = {Quantum Simulations in Ion Traps},
  pdfstartview       = {},
}
\ifpdfoutput{
  \hypersetup{colorlinks=true}
}{}
\renewcommand{\equationautorefname}{eq.}

\usepackage[figure,table]{hypcap}
\hypcapredef{sidewaysfigure}

\ifpdfoutput{
  \pdfcompresslevel=9
}{}

\newcommand{\be}{\begin{equation}}
\newcommand{\ee}{\end{equation}}

\newcommand{\benu}{\begin{enumerate}}
\newcommand{\eenu}{\end{enumerate}}
\newcommand{\bitem}{\begin{itemize}}
\newcommand{\eitem}{\end{itemize}}

\newcommand{\bt}{\begin{table}\begin{center}}
\newcommand{\et}{\end{center}\end{table}}
\newcommand{\btab}{\begin{tabular}}
\newcommand{\etab}{\end{tabular}}
\newcommand{\bfig}{\begin{figure}\centering}
\newcommand{\efig}{\end{figure}}

\newcommand{\down}{{\downarrow}}
\newcommand{\up}{{\uparrow}}

\usepackage{ifthen}
\usepackage{relsize}
\makeatletter
\newcommand{\ac}[1]{%
	\ifthenelse{\equal{\f@family}{cmr} \and \equal{\f@series}{m} \and \equal{\f@shape}{n}}%
	{\textsc{\MakeLowercase{#1}}}%
	{{\smaller\letterspace to 1.03\naturalwidth{#1}}}%
}
\makeatother

\newunit{\millikelvin}{\milli\kelvin}
\newunit{\mbar}{\millibar}
\newunit{\kayser}{\centimeter\unitsuperscript{-1}}
\newunit{\inch}{in}

\newunit{\mW}{\milliwatt}
\newunit{\microwatt}{\micro\watt}
\newunit{\mWpersqcm}{\milliwatt\unittimes\centimeter\unitsuperscript{-2}}

\newunit{\Hz}{\hertz}
\newunit{\THz}{\tera\hertz}

\newunit{\gauss}{G}
\newunit{\milligauss}{\milli\gauss}
\newunit{\microtesla}{\micro\tesla}
\newunit{\millihenry}{\milli\henry}
\newunit{\microhenry}{\micro\henry}

\newunit{\dB}{dB}
\newunit{\dBm}{dBm}

\newunit{\byte}{B}
\newunit{\kilobyte}{\kilo\byte}
\newunit{\megabyte}{\mega\byte}
\newunit{\MBit}{MBit}

\setMathOmega{$\Omega$}

\newcommand{\autoeqref}[1]{\hyperref[#1]{\equationautorefname~(\ref*{#1})}}

\begin{document}

\frontmatter
\settocdepth{none}

\addtolength{\oddsidemargin}{35pt}
\titlehead{
\begin{center}
  \textsc{\Large ludwig-maximilians-universität münchen} \\
  \textsc{\Large max-planck-institut für quantenoptik garching}
\vspace{45mm}
\end{center}
}
\title{\huge Quantum Simulations in Ion Traps \vskip 3mm\Large Towards Simulating the
Early Expanding Universe}
\author{\vspace{10mm}\\Diploma thesis\vspace{1.5mm}\\by Lutz Petersen}
\date{\vspace{13mm}December 2006}
\maketitle

\typearea[current]{last}

\chapter{Abstract}

This thesis provides an overview of an approach to quantum simulations
using magne\-sium-25 ions stored in a linear Paul trap as the carriers
of quantum information. Their quantum state is manipulated and read
out using ultraviolet laser beams. Several important steps towards
realising the first experiments have been undertaken, the most
striking of which is cooling the ions to their motional ground
state. We describe a first experiment simulating cosmological particle
creation in the Early Universe and discuss the expected results.

\chapter{Acknowledgements}

I would like to thank all the people that have, in one way or the other,
contributed to this thesis. First and foremost, I have to mention my
dear colleagues Axel, Günther, Hector, and Steffen who did a great job
not only in the laboratory but also by always making me feel in good
hands. Special thanks goes to my advisor Tobi for the great support
and perpetual encouragement, for the discussions about the freewill
and for the numerous weird stories about dogs and the like. Thanks
also to the quasi-group member and technician Arnold Steyer who time
and again impressed not only me with his approach to everyday life.

A big ``thank you'' to Ralf Schützhold who developed the proposal for
this thesis. We greatly profited from his collaboration and
acknowledge that he never sniffed at the 101 questions of an
experimentalist.

I gratefully acknowledge Paul Pham, his support and assistance in
making the electronics do what it is supposed to do. Thanks to Markus
Wewer of LOT Oriel and Axel Lucas of B.\,Halle for the good collaboration.
Many more people have helped us in the lab, especially all the amiable
colleagues at \ac{MPQ} who tried to satisfy even the most exceptional
desires. Although they are too numerous to be listed here in
completion, I'd like to mention Helmut Brückner and Tom Wiesmeier for
their excellent electronics support, Thomas Strobl and his crew for
their commitment to providing the best manufacturing of required
laboratory items, and Thomas Rieger as one representative of the Rempe
group helping us out with laboratory equipment whenever necessary.

Finally, I'd like to thank my parents for their support and affection
at all times. And last but not least thanks to my boyfriend Bastian
for reminding me that there is a life besides doing physics in the lab.

\ifpdfoutput{
  \cleardoublepage\phantomsection
  \currentpdfbookmark{\contentsname}{toc}
}{}
\tableofcontents

\chapter{Introduction}

The idea of quantum simulations was first formulated by Richard
Feynman \cite{feynman:simulation}. He recognised that a calculation to
predict the behaviour of quantum systems can only be performed
by utilising another quantum system. That way, quantum mechanical
properties like superposition states or entanglement are inherently
included in the simulation system.

One might be tempted to think that a state-of-the-art classical
computer would be able to calculate the time evolution of a complex
quantum system. A simple estimation elucidates that this is not the case
even for comparably small quantum systems: The state of a system of
multiple---not necessarily interacting---particles is described by the
coefficients of the respective product Hilbert space. For $n=200$
two-level particles there are $2^n = 1.61 \cdot 10^{60}$ coefficients,
a number that corresponds to the estimated amount of \textit{all
protons in the universe}. Too many parameters for any classical
computer even in the future.

The proposed quantum computer addresses this problem by consisting
itself of interacting qubits. 200 qubits will suffice to describe the
state of 200 two-level particles. With a universal set of quantum
manipulation operations it can in principle cope with many unsolved
quantum problem, e.\,g. the challenge of a prime factor in polynomial
time \cite{shor} that would render most classical encryption
algorithms vulnerable \cite{nielsen}. However, to realise a universal
quantum computer, an appreciable amount of around 1000 logical qubits
is required, each logical qubit requiring an overhead of about
100 ancilla qubits to allow for fault-tolerant operation
\cite{arda}. Although controlling and manipulating $10^5$ qubits with
sufficient fidelities appears to be technically feasible \cite{arda},
even the most optimistic estimations for the realisation of a
universal quantum computer are of the order of 10\,--\,20 years.

In our group, we try to implement a ``shortcut'' on the way towards a
quantum computer via the approach of quantum simulations in ion traps,
which follows Feynman's original proposal even closer. Take a quantum
system (denoted~\ac{A}) whose initial conditions and interaction
terms, i.\,e. its Hamiltonian, can be manipulated in a controlled
way. If this Hamiltonian matches the one of another system of interest
(denoted~\ac{B}), the time evolution as well as all properties
observed and deduced at system~\ac{A} can be analogously transferred
to system~\ac{B}. This is why we call system~\ac{A} an
\textit{analogous quantum simulator}. It opens the possibilities to
simulate specific systems beyond the scope of classical
computation. Examples include quantum spin systems such as high-$T_C$
superconductors whose physics is still far from being understood
\cite{porras}. The simulations may also help to better understand
effects such as quantum relativistic and gravitational issues
\cite{garay:sonic,lamata:dirac,bermudez:dirac}.

Another motivation for the realisation of a quantum simulator is to
actually ``observe'' phenomena in experiment that have so far only been
predicted by quantum theory. The first experiment to be implemented in our
quantum simulator addresses the particle creation process in the Early
Universe according to a proposal by Schützhold et al.
\cite{schuetzhold} in collaboration with our group: Matter as we
observe it nowadays shows aggregations (planets, stars, galaxies)
rather than being distributed evenly over space. According to the
standard model of cosmology the seeds for the required structure
formation are provided by quantum fluctuations during the very early
moments of the universe. One possible process is cosmological particle
creation during a non-adiabatic expansion of space-time
\cite{birrell}. We can identify the expansion of space-time with a
non-adiabatic change of the ions' trapping potential in our
laboratory. This will ``create'' pairs of phononic excitations
corresponding to particle pairs in the universe, see for example
\cite{dowling:unruh,walls:squeezed,schleich-wheeler}. Thus, the system
to be simulated is the universe itself (!), whereas even a single ion
can serve as a simulation system allowing for the investigation of
some of the most interesting properties.

Various systems can be exploited in terms of quantum simulators;
examples include neutral atoms, molecules, quantum dots and many more
\cite{arda}. In our laboratory, we implement a technique first
proposed by Cirac and Zoller \cite{cirac-zoller}: Laser-cooled ions
stored in a linear Paul trap under ultra-high vacuum serve as the
carriers of quantum information. By shining onto the ions with laser
beams we can initialise, manipulate, and determine their internal
electronic quantum state in a controlled way and with high
fidelities. In addition, the ions may conditionally interact among
each other by coupling their electronic state to a collective
vibrational mode of motion in the trap. As coupling to the environment
is very weak for such system, observed coherence times are long
compared to the duration of the required operations \cite{wineland:bible}.

We use two distinct hyperfine states of magnesium-25 ions to implement
a qubit where the coupling between the two levels is provided via a
two-photon stimulated Raman transition. Readout occurs by fluorescently
scattering photons in a \textit{cycling transition} allowing for
high-accuracy state-sensitive detection by means of a photomultiplier
or a \ac{CCD} camera. The \ac{NIST} group in Boulder, Colorado, are
currently implementing a similar system and we are grateful to profit
from their former experience with beryllium-9 ions.

Several important steps towards our first experiment have been
realised during a year's work on this thesis. As the initial
theoretical proposal \cite{dowling:unruh} posed severe problems to an
experimental realisation, we started a collaboration with the
quantum theory group of Ralf Schützhold\footnote{\href{mailto:schuetz@theory.phy.tu-dresden.de}{schuetz@theory.phy.tu-dresden.de}}
at the Technical University of Dresden. For the author it was most
interesting to contribute to a theoretical proposal and to
develop an experimental technique for its realisation
\cite{schuetzhold}. Progress achieved with the apparatus in the
laboratory includes the ability to trap ions, to drive coherent
transitions between their internal electronic states (Rabi flopping), to
couple their electronic state to vibrational modes, to measure their
vibrational state, and to efficiently initialise their quantum state
by cooling them close to the vibrational ground-state.

The group had to develop strong laser sources \cite{friedenauer} at
ultraviolet wavelengths to access the transitions of magnesium-25
ions. They are stored within a specially designed trapping apparatus
driven at parameters that have not been well investigated before.
Starting from the successful loading of the first ion, a lot of work
has been spent on the assembly and optimisation of the related laboratory
components and their operation. Let me focus on my significant
contributions: Theoretical considerations on disturbing micromotion of
ions in the trap as well as their experimental realisation
(\autoref{ch:micromotion}) allowed for a better understanding of
the trap apparatus, whose parameters lie in an intermediate regime
between the resolved-sideband and the non-resolved-sideband case
(natural linewidth of the observed transition is comparable to the
trapping frequency). Compensation of micromotion turned out to be
particularly important for a change of the trapping potential by more
than one order of magnitude needed to realise \cite{schuetzhold}. 
Further activities concerning the trap included the design and
fabrication of magnetic field coils intended for providing a
quantisation axis and required highly stable Zeeman shifts of the
involved qubit levels (\autoref{sec:magnetic_field}). External
magnetic field noise has to be minimised using an active compensation
system, which was home-built and tested. As for the laser
apparatus, works included the modification and optimisation of
Doppler-free saturation spectroscopy units
(\autoref{sec:iodine_spectroscopy}) as well as the setup and
optimisation of the laser beamline (infrared, visible, and
ultraviolet wavelengths) including 15 single- and double-pass
acousto-optical modulators (\autoref{sec:beamline}). Electronics
for locking our laser systems to an iodine signal were developed and
thoroughly tested for maximum reliability. The system containing our
single photon detection instruments (photomultiplier and \ac{CCD}
camera) had to be optimised for maximum sensitiveness thus improving
the signal-to-noise ratio by up to two orders of magnitude
(\autoref{ch:detection}). Tedious debugging and resoldering
sessions allowed to eventually put our home-built experimental control
system into operation. Digital electronics with a programmable
processor clocked at $100\MHz$ renders this system a real-time tool
controlling all the laser operation cycles during the proposed and
future quantum simulation experiments (\autoref{sec:paul-box}).

Apart from the hands-on activities appreciable amount of time has been
dedicated to software development, debugging, and documentation, which
should keep the system adaptable to new requirements. Greater software
projects encompassed the development of a compiler for fast control of
our experimental control system, the development of visualisation and
data acquisition software for our \ac{CCD} camera
(\autoref{sec:auge}), and a respective network protocol for
communication with our experimentation software system.

This thesis is divided into three parts: The first part gives an
overview of the theoretical models used within the field of quantum
simulations. We will discuss the two-level system as a description of
qubit dynamics and fluorescence phenomena (\autoref{ch:two-level})
and review our implementation of quantum simulations using
magnesium-25 ions (\autoref{ch:mg25}). The second part contains
a tour through our lab where we introduce the ion trap apparatus
(\autoref{ch:trap}), the laser apparatus (\autoref{ch:laser})
as well as our detection\,/\,data acquisition (\autoref{ch:detection})
and experimental control systems (\autoref{ch:control}). In the
third part we present experimental data. Among the results are: a
proposal of the implementation and measurement of \cite{schuetzhold}
(\autoref{ch:universe}), a review of methods for the compensation
of micromotion from an experimentalists' view
(\autoref{ch:micromotion}), and the worldwide first data on
coherent interactions using ${}^{25}\text{Mg}^+$ ions like Rabi
flopping and ground-state cooling experiments
(\autoref{ch:flopping}).

\mainmatter
\settocdepth{section}

\part{Theoretical considerations}
\chapter{Optical properties of a two-level system}
\label{ch:two-level}

In the context of quantum simulations we will be dealing with
couplings between electronic states that can be described by the model
of an effective two-level system. In particular, the two phenomena of
resonance fluorescence and Rabi oscillations can be understood using
the same set of equations.

We will denote the lower state with $\left|a\right>$ and the upper
state with $\left|b\right>$. We further assume the following conditions:
\benu
\item The two states are separated by an energy difference of $\hbar
\omega_0$.
\item The system interacts with an externally applied classical
electromagnetic field $E(t) = E_0 \cos(\omega t + \delta)$.
\item The population in the upper state has a limited lifetime $\tau$
and can decay into the lower state by spontaneous emission
characterised by the natural linewidth $\Gamma = 1/\tau$.
\eenu

\section{Optical Bloch equations}
\index{Bloch equations}

The time evolution of the two-level system is described by the time
evolution of its density matrix $\hat{\rho}$---or more precisely: by
the time evolution of the four density matrix elements $\rho_{ij}$,
where $i$ and $j$ encode either state $\left|a\right>$ or state
$\left|b\right>$. These equations are known as the \textit{optical
Bloch equations} (see for example \cite{loudon,scully}):
\be
\begin{split}
\dot{\rho}_{bb} &= -\frac{i \Omega}{2} \left(\rho_{ab} e^{i \Delta t} - \text{c.\,c.}\right) - \Gamma \rho_{bb} \text{,} \\
\dot{\rho}_{ab} &=  \frac{i \Omega}{2} e^{-i \Delta t} \left(1 - 2\rho_{bb}\right) + (i\Delta - \frac{\Gamma}{2}) \rho_{ab} \text{,} \\
\dot{\rho}_{ba} &=  \dot{\rho}_{ab}^* \text{,} \\
\dot{\rho}_{aa} &= -\dot{\rho}_{bb} \text{,}
\end{split}
\label{eq:bloch1}
\ee
where
\be
\Omega = \frac{\left| \left<a|\hat{x}|b\right> \right| E_0}{\hbar}
\label{eq:rabi_frequency}
\ee
is the Rabi frequency and
\be
\Delta = \omega_0 - \omega
\ee
is the detuning of the radiation field frequency $\omega$ with respect
to the system's transition frequency $\omega_0$. In principle, we can
distinguish between two different parameter regions:
\benu
\item \textit{Strong coupling} where the lifetime of the upper state
is much larger than the typical duration of a Rabi oscillation period,
$\Gamma \ll \Omega$, and
\item \textit{Weak field limit} where the upper state decays much
faster compared to the Rabi frequency, $\Gamma \gg \Omega$.
\eenu

\subsection{Rabi oscillations}

In the Rabi oscillation region the upper state is long-lived enough
for Rabi oscillations to occur: The radiation field will transfer the
population from $\left|a\right>$ to $\left|b\right>$ and vice
versa. Because of the limited lifetime of the upper state we will
observe a damped oscillation, see \autoref{fig:rabi1}. The
equilibrium population is close to $1/2$, which is due to the
saturation of the transition, see \autoref{sec:power_broadening}
below.

\bfig
\subfloat[$\Gamma = \Omega/5$]{
  \includegraphics{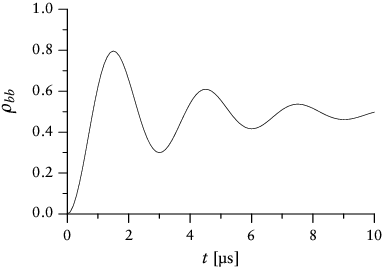}
  \label{fig:rabi1}}
\hfill
\subfloat[$\Gamma = 0$]{
  \includegraphics{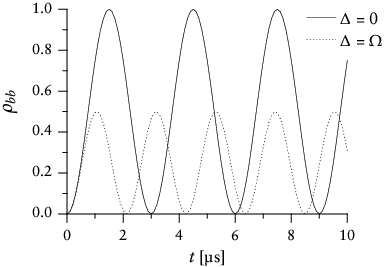}
  \label{fig:rabi2}}
\caption{Example Rabi flopping curves of the upper state population
$\rho_{bb}$ for $\Omega = 2\pi/3\MHz$ and different parameters for the
decay rates $\Gamma$. Unless noted otherwise, a detuning of $\Delta =
0$ has been chosen.}
\label{fig:rabi}
\efig

In the case that decay of the upper state can be neglected, the
\hyperref[eq:bloch1]{optical Bloch equations~(\ref*{eq:bloch1})} give a perfect
sinusoidal oscillation, see \autoref{fig:rabi2}. Both the oscillation
frequency and the oscillation amplitude depend on the detuning
$\Delta$; the former which we will call the effective Rabi frequency
is given by
\be
\Omega_\text{eff} = \sqrt{\Omega^2 + \Delta^2} \text{,}
\label{eq:effective_rabi}
\ee
whereas the peak-to-peak amplitude $\rho_{bb,0}$ of the
oscillation is determined by the ratio
\be
\rho_{bb,0} = \frac{\Omega^2}{\Omega_\text{eff}^2} \text{.}
\ee
The duration $\tau_\pi = \pi/\Omega$ it takes to transfer all the
population from one state to the other is called the $\pi$ flopping
duration. The reason for this naming is clear: During the $\pi$
flopping duration the phase of a Rabi oscillation is shifted by
$\pi$. Analogously, a $\pi$ pulse is an electric field pulse that has
the length of the $\pi$ flopping duration. Another important pulse
duration is the $\pi/2$ flopping duration. Starting from an
eigenstate, i.\,e. $\left|a\right>$ or $\left|b\right>$, it
distributes the probabilities like evenly among the two states.

\subsection{Weak field limit}

The weak field limit is characterised by a very small Rabi frequency
compared to the linewidth of the transition. As such a small Rabi
frequency is determined by a weak incident electric field, we call
this region the weak field limit. Here, the population of the upper
state reaches its equilibrium before any significant Rabi oscillations can
occur, see \autoref{fig:weak_field}. Resonance fluorescence phenomena
can often be described in the weak field limit.

\bfig
\includegraphics{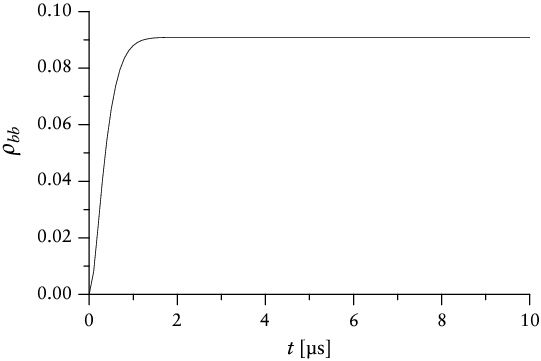}
\caption{Development of the upper-state population $\rho_{bb}$ over
time for $\Omega = 2\pi/3\MHz$, $\Gamma = 3\Omega$, and $\Delta = 0$.}
\label{fig:weak_field}
\efig

\section{Saturation effects}

\subsection{Power broadening}
\label{sec:power_broadening}

For intense incoming radiation fields, the effective linewidth of the
system's transition broadens, i.\,e. resolution of the transition
decreases with larger field intensities. To understand this, we have a
closer look at the steady-state solution of the \hyperref[eq:bloch1]{optical Bloch} 
\hyperref[eq:bloch1]{equations~(\ref*{eq:bloch1})}. We can deduce the population in the upper
state $\left|b\right>$ after Rabi oscillations have damped out:
\be
\rho_{bb}(t\rightarrow\infty) = \frac{\Omega^2}{4} \frac{1}{\Delta^2 + (\Gamma/2)^2 + \Omega^2/2} \text{.}
\label{eq:bloch_steady}
\ee
Thus, the upper state population $\rho_{bb}(\infty)$ is
Lorentzian-shaped in dependence of the detuning $\Delta$. The
effective linewidth of this distribution is given by
\be
\Gamma_\text{eff} = \sqrt{\Gamma^2 + 2\Omega^2} \text{.}
\ee
We see that the effective linewidth increases as the Rabi frequency
$\Omega$ increases. As the Rabi frequency is proportional to the
electric field amplitude $E_0$, \autoeqref{eq:rabi_frequency}, we
observe an increasing full width at half maximum~(\ac{FWHM}) with
increasing field intensities $I = \epsilon c E_0^2$ (where $\epsilon$
characterises the absolute permittivity of the optical medium).

In order to better characterise the power broadening effect we
introduce the \index{saturation intensity} saturation intensity
$I_\text{sat}$. It is defined as the intensity of a resonant incident
electromagnetic field causing the
\hyperref[eq:bloch_steady]{steady-state upper state
population~(\ref*{eq:bloch_steady})} to reach the value of $1/4$,
i.\,e.
\be
\Omega^2 = \frac{\Gamma^2}{2} \text{,}
\ee
which can also be written as \cite{curtis:thesis}
\be
I_\text{sat} = \frac{\hbar \Gamma \omega_0^3}{12\pi c^2} \text{.}
\ee
In the limit of infinite intensities the upper-state population
$\rho_{bb}$ takes the value of $1/2$. This result is independent
of $\Gamma$ or $\Delta$.

As an example, we investigate a fluorescent transition with a natural
linewidth of $\Gamma = 2\pi \cdot 43\MHz$ and plot $\rho_{bb}$ in
dependence of $\Delta$, \autoref{fig:power_broadening}. The
linewidth, i.\,e. the \ac{FWHM}, broadens to $\Gamma^\prime =
74.5\MHz$ as the incident intensity is increased from virtual zero to
twice the saturation intensity.

\bfig
\includegraphics{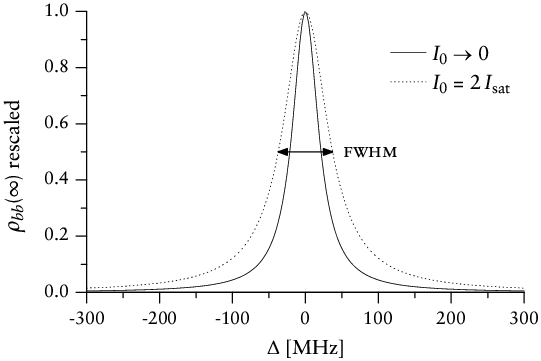}
\caption{Steady-state upper-state population $\rho_{bb}$ as a function
of the detuning $\Delta$. The transition lines have been rescaled to
$1.0$ at their respective maximum for better comparability.}
\label{fig:power_broadening}
\efig

\subsection{Attenuation}
\label{sec:attenuation}

\bfig
\includegraphics{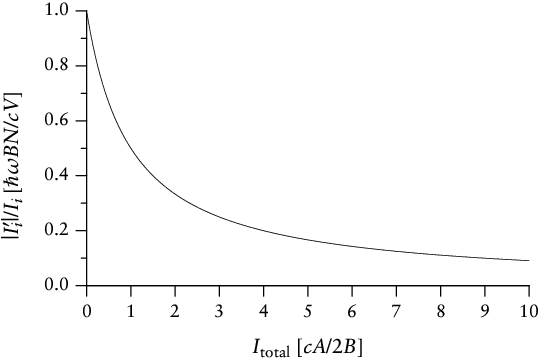}
\caption{Relative optical attenuation $|\text{d}I_i/\text{d}z|/I_i$ of a beam
travelling through a region with overall intensity $I_\text{total}$.}
\label{fig:attn}
\efig

The effect of attenuation of a laser beam in a saturated optical
medium plays an important role in e.\,g. saturation spectroscopy. To
understand the relevant processes consider the very simple model of a
two-level system with its associated \index{Einstein coefficients}
Einstein equations \cite{loudon}
\be
\begin{split}
\frac{\text{d}N_1}{\text{d}t} &= A N_2 - B \rho N_1 + B \rho N_2 \text{,} \\
\frac{\text{d}N_2}{\text{d}t} &= -\frac{\text{d}N_1}{\text{d}t} \text{,}
\end{split}
\label{eq:einstein}
\ee
where $N_1$ and $N_2$ denote the population in the lower and upper
levels respectively, $A$ and $B$ are the Einstein coefficients
related to spontaneous emission and absorption\,/\,stimulated
emission, and $\rho$ is the average energy density of the radiation
field.

In the steady state, neither $N_1$ nor $N_2$ change, that is
\be
N_2 = N \frac{1}{2 + \frac{A}{B \rho}} \text{.}
\label{eq:n2_steady_state}
\ee
We can estimate the optical attenuation, i.\,e. the change in energy
density $\rho$ over time or, equivalently, the change in intensity $I$
over the propagation distance $z$. It is proportional to the number of
photons absorbed minus the number of photons emitted by stimulated
emission:
\be
\begin{split}
\frac{\text{d}I}{\text{d}z} &= -\frac{\hbar\omega}{c V} \left(B I N_1 - B I N_2\right) \\
                            &= -\frac{\hbar\omega B I N}{c V} \left(1 - \frac{2 B I}{c A + 2 B I}\right)
\end{split}
\ee
where $V$ denotes some normalisation volume, $c$ is the speed of light
and the last equality has been derived using
\autoeqref{eq:n2_steady_state}.

We may also derive an expression for the optical attenuation in the
case where there are several optical sources, i.\,e. overall intensity
is equal to the sum of the individual intensities. As the underlying
process of optical attenuation is a single particle
process (absorption and emission of photons), attenuation will be
proportional to the respective intensity
\be
\frac{\text{d}I_i}{\text{d}z} = \frac{\text{d}I}{\text{d}z} \cdot \frac{I_i}{I} \text{,}
\ee
which means that the relative optical attenuation is the same for each
contributing source:
\be
\begin{split}
\frac{\text{d}I_i/\text{d}z}{I_i} = -\frac{\hbar\omega B N}{c V} \left(1 - \frac{2 B I_\text{total}}{c A + 2 B I_\text{total}}\right) \text{.}
\end{split}
\label{eq:attn}
\ee
\hyperref[fig:attn]{Fig.~\ref*{fig:attn}} illustrates the meaning of \autoeqref{eq:attn}:
The relative optical attenuation decreases when overall intensity is
increased, that is when the transition becomes more and more
saturated. In the limit of infinite overall intensities, the system
does not attenuate at all---it becomes transparent, which is also known
as \index{optical bleaching} optical bleaching.

\chapter{Magnesium-25 as a qubit system}
\label{ch:mg25}

\section{Electronic structure of magnesium-25}

We use trapped magnesium-25 ions~(${}^{25}\text{Mg}^+$) as the
carriers of quantum information. These ions have one valence electron
which can populate distinct energy levels, see \autoref{fig:mg25}. The
nuclear spin of $I = 5/2$ is responsible for energy splittings due to
the \index{hyperfine levels} hyperfine interaction: For a total
electronic spin of $J = 1/2$ there are two hyperfine sublevels $F \in
{2,3}$, fine structure levels with $J = 3/2$ split into four sublevels
$F \in {1,2,3,4}$.

The two hyperfine levels of the $3S_{1/2}$ manifold are used to store
quantum information in a qubit where the lower qubit state is
$\left|\down\right> = 3S_{1/2} \left|F=3, m_F=3\right>$ and the upper
state is $\left|\up\right> = 3S_{1/2} \left|F=2, m_F=2\right>$. Any
manipulation or state discrimination occurs by laser interaction that
couples these qubit levels to states of the $P$ manifold.

The $3S_{1/2} \leftrightarrow 3P_{3/2}$ transition lies at
$279.635\nanometer$ whereas the $3S_{1/2} \leftrightarrow
3P_{1/2}$ transition, separated from the former by the fine structure
splitting of $2\pi \cdot 2.746\THz$, lies at $280.353\nanometer$.
Both transitions are dipole-allowed and have radiative linewidths of
$2\pi \cdot 43\MHz$ each. In contrast, transitions between hyperfine
states of the $3S_{1/2}$ manifold are not dipole-allowed. Moreover,
these states are separated by a hyperfine splitting of merely
$1.789\GHz$, which is why direct radiative decay processes can be
neglected throughout.

Degeneracy of the hyperfine levels of the same $F$ manifold is broken
by applying a magnetic quantisation field $B$ leading to \index{Zeeman
energy shift} Zeeman energy shifts between different $m_F$ levels
\be
\Delta E = g_F m_F \mu_B B
\ee
where $g_F$ is the Landé factor and $\mu_B$ is the Bohr magneton. The
Zeeman splitting of the $3S_{1/2}$ hyperfine levels is of particular
importance for quantum simulation experiments as these hyperfine
levels will be used to store quantum information in a qubit. Their
Landé factors are $-1/3$ and $1/3$ respectively
\cite{ludsteck:thesis}. We would usually apply a magnetic field of
about $B = 5.6\gauss = 560\microtesla$ introducing energy splittings of
$\Delta E/\,\hbar = m_F \cdot 2\pi \cdot 2.6 \MHz$.

\bfig
\includegraphics{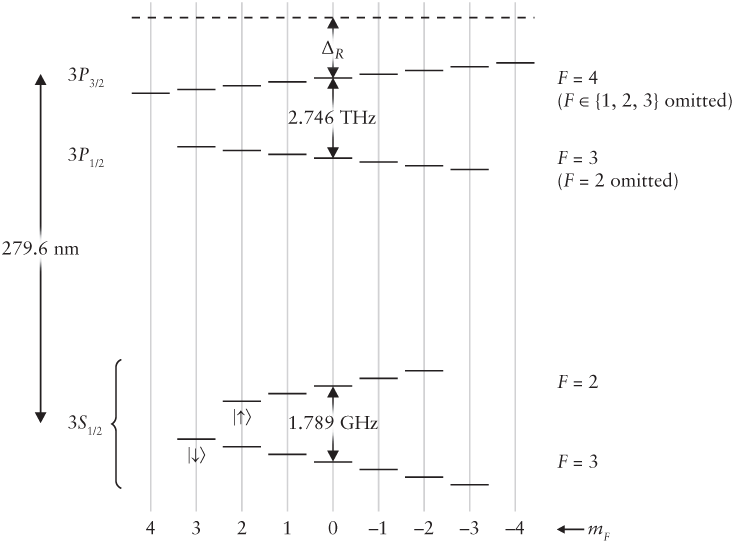}
\caption{Energy levels of ${}^{25}\text{Mg}^+$ (not to scale). Note
that we have omitted hyperfine levels of the $P$ manifold which are
not relevant to our quantum simulation experiments. $\left|\down\right>$
and $\left|\up\right>$ denote the two states of our qubit system,
$\Delta_R$ is the detuning of the virtual Raman level with respect to
the $3P_{3/2}$ level.}
\label{fig:mg25}
\efig

\section{Motional states}
\index{motional quantum number}

In our setup the magnesium ions are confined in a three-dimensional
harmonic potential using a combination of \ac{AC} and \ac{DC} electric
fields, see \autoref{ch:trap}. The motion of a single ion in the
storage potential can be described by the model of the quantum
mechanical oscillator. It is characterised by equidistant energy
levels separated by $\hbar \omega$ where $\omega$ is the resonance
frequency of the potential. We will denote its energy
eigenstates---the Fock states---by numbers ranging from zero to
infinity. Throughout this thesis, we will characterise the state of an
ion by its qubit (electronic) state (which is a linear combination of
$\left|\down\right>$ and $\left|\up\right>$) as well as by its
motional state (which is a linear combination of $\left|0\right>$,
$\left|1\right>$, $\left|2\right>$, etc.).

\section{Qubit state initialisation and detection}
\label{sec:state_detection}

\begin{figure}[b]\centering
\includegraphics{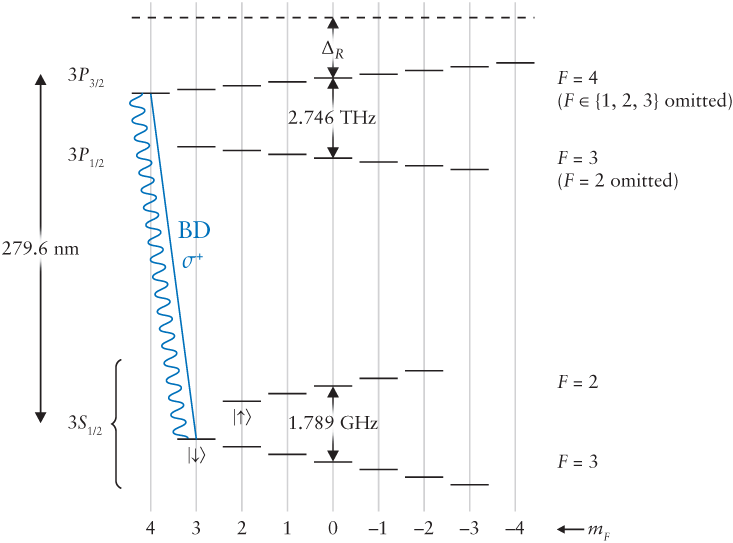}
\caption{Cycling resonance fluorescence transition used to detect
electronic population in the $\left|\down\right>$ state of our qubit
system (lengths not to scale).}
\label{fig:mg25-bd}
\efig

All our quantum simulation experiments begin by optically pumping the
electronic population into the $\left|\down\right>$ state. This is
achieved by shining onto an ion with two near-resonant laser beams,
\index{Blue Doppler laser} the Blue Doppler~(\ac{BD}) laser driving a
closed cycling transition $3S_{1/2} \left|F=3\right> \leftrightarrow
3P_{3/2} \left|F=4\right>$ and \index{Red Doppler laser} the Red
Doppler laser tuned to the $3S_{1/2} \left|F=2\right> \leftrightarrow
3P_{1/2} \left|F=3\right>$ transition. (The names of the two laser
beams are due to the fact that they can be used to Doppler cool
\cite{haensch:doppler} the ions, see below.) Both laser beams are
polarised $\sigma^+$ with respect to the magnetic quantisation
field. They thus transfer electronic population to one of the $3P$
states illustrated in \autoref{fig:mg25} increasing the $m_F$ value
at the same time. Subsequently, the population will decay into one of
the lower states by resonance fluorescence where the probability of
decay into $3S_{1/2}\left|F=3\right>$ is about twice as high as the
probability of decay into $3S_{1/2}\left|F=2\right>$. During resonance
fluorescence $m_F$ may change according to the dipole rule $\Delta m_F
= 0, \pm 1$. This way, all electronic population is eventually pumped
into the $\left|\down\right>$ state.

In order to discriminate ions in the $\left|\down\right>$ state from
ions in the $\left|\up\right>$ state we use the transition
$\left|\down\right> \leftrightarrow 3P_{3/2} \left|F=4, m_F=4\right>$,
see \autoref{fig:mg25-bd}. \index{cycling transition} The alert reader
will have noticed that this transition is driven by the \ac{BD}
laser. After electronic population has been shifted from
$\left|\down\right>$ to $3P_{3/2} \left|F=4, m_F=4\right>$ it can only
decay back into $\left|\down\right>$ (due to dipole rules) at the same
time emitting resonance fluorescent light. The \ac{BD} transition can
thus be excited over and over again, scattering many photons if the
ion is actually in the $\left|\down\right>$ state. This is called a
\textit{cycling transition}. Its properties are listed in
\autoref{tab:cycling}. In contrast, if the ion is in the
$\left|\up\right>$ state, the \ac{BD} laser cannot drive a resonant
transition to the $3P_{3/2}$ level. Off-resonant excitation is however
possible (although improbable: the transition linewidth of $43\MHz$
makes up for only $2.4\%$ of the hyperfine splitting between
$\left|\down\right>$ and $\left|\up\right>$) and effectively limits
the detection time: Once an off-resonant excitation has occurred,
there is a finite probability of decay into $\left|\down\right>$
thereby entering the cycling transition. Thus even an initially dark
ion will eventually fluoresce. In practice, we use detection durations
of $20\microsecond$\,--\,$50\microsecond$ to minimise the influence of
these effects (causing a loss of contrast).

\index{Blue Doppler transition}
\bt
\btab{ll}\toprule
transition frequency $\omega_{0,\text{BD}}$ & $2\pi \cdot 1.0720841(12)\cdot10^{15}\Hz$ \\
natural transition linewidth $\Gamma$       & $2\pi \cdot 43\MHz$ \\
saturation intensity $I_\text{sat}$         & $250\mWpersqcm$ \\ \bottomrule
\etab
\caption{\protect\ac{BD} cycling transition properties.}
\label{tab:cycling}
\et

Using the \hyperref[eq:bloch1]{optical Bloch equations~(\ref*{eq:bloch1})} and the parameters
listed in \autoref{tab:cycling} we may calculate the population in
the upper $3P_{3/2} \left|F=4, m_F=4\right>$ state for a system
initially in the ground state $\left|\down\right>$. As we can see from
\autoref{fig:fluorescence_bd} the weak field limit is a very good
approximation to our situation. With a steady-state upper state
population of $0.125$ we expect a fluorescence signal of $0.125 \cdot
2\pi\Gamma = 2\pi \cdot 34\MHz$, i.\,e. $3.4\cdot10^7$ scattered photons per
second.

\bfig
\includegraphics{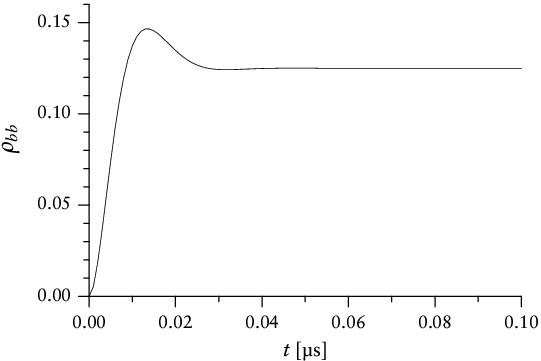}
\caption{Population in the detection state $3P_{3/2} \left|F=4,
m_F=4\right>$ for the \protect\ac{BD} fluorescent detection transition
in magnesium (not to scale). Typical experimentation values were used for this
plot---intensity is $I = (2/3)I_\text{sat} = 167\mWpersqcm$ and the
laser is red-detuned with respect to the resonance frequency by
$\Gamma/2 = 2\pi \cdot 21.5\MHz$.}
\label{fig:fluorescence_bd}
\efig

\section{Qubit state manipulation}

\subsection{Two-photon stimulated Raman transitions}
\label{sec:tps-raman}
\index{two-photon stimulated Raman transition}

In magnesium-25, the two qubit levels are energetically separated by
$\hbar \cdot 2\pi \cdot 1789\MHz$, lying in the radio
frequency~(\ac{RF}) range. In order to transfer electronic population
between the two qubit levels, we use a two-photon stimulated
Raman transition: Two laser beams shining onto an ion at the same time
drive the transition from $\left|\down\right>$ to $\left|\up\right>$ and
vice versa via \index{virtual Raman level} a shared virtual Raman
level $\left|\text{ram}\right>$, which is blue-detuned with respect to
the $P_{3/2}$ level by $\Delta_R = \text{several tens of
\unitsignonly{GHz}}$, see \autoref{fig:mg25-raman}. In a simple
picture we may consider the two-photon stimulated Raman process as the
absorption of a photon from one beam followed by a stimulated
emission into the other. During the two-photon stimulated Raman
process we alter the $m_F$ value by one, yet shine onto the ion with
two lasers. Therefore only one of the lasers is polarised $\sigma^+$
while the other must be polarised $\pi$ with respect to the magnetic
quantisation field axis. 

\begin{figure}[t]\centering
\includegraphics{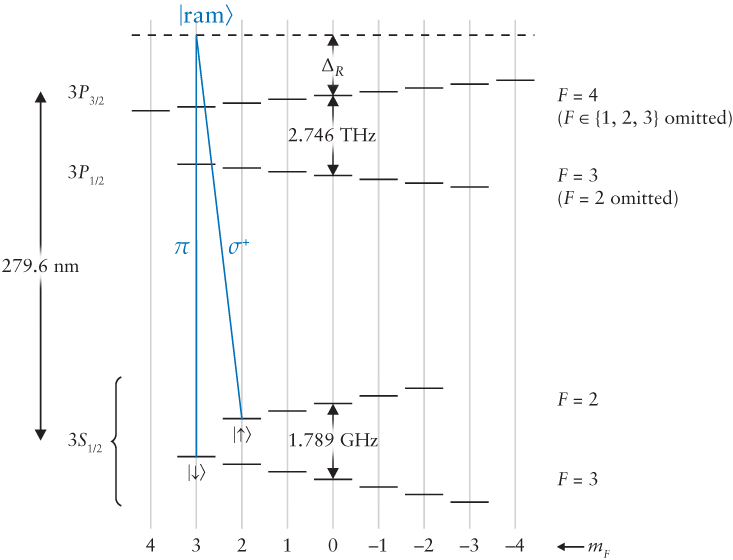}
\caption{Two-photon stimulated Raman transition used to transfer
electronic population between the $\left|\down\right>$ and
$\left|\up\right>$ states (lengths not to scale). Note that one of the
Raman beams has to be polarised $\sigma^+$ while the other has to be
polarised $\pi$ with respect to the magnetic quantisation field axis.}
\label{fig:mg25-raman}
\efig

The scheme of two-photon stimulated Raman transitions has several
advantages compared to driving the transition directly:
\bitem
\item The Rabi frequency can be adjusted by altering the intensity of
the two Raman lasers as well as by detuning the virtual Raman level,
see below.
\item The frequency difference between the two lasers, i.\,e. the
actual driving frequency, as well as their relative phase, can be
adjusted in a very accurate manner ($\Delta\omega \approx 2\pi \cdot
1\Hz$) using acousto-optic modulators in one or both of the laser
beams.
\item Small drifts of the Raman laser frequencies do not affect the
strength of the driven transition as their difference frequency
remains unchanged.
\item Although the effective driving frequency is in the \ac{RF}
range, the ion still experiences the strong electric field gradients
imposed by the \ac{UV} laser fields. These field gradients are
necessary to couple the electronic and motional states of the ion, see
\autoref{sec:motion_coupling}.
\eitem
In reality, the two Raman beams will of course couple to all the
states of the $3P$ manifold. These states may noncoherently decay into
$3S$ by \index{spontaneous emission} spontaneous emission, which is a
disadvantage of the two-photon stimulated Raman scheme. The
probability for spontaneous emission may however be significantly
reduced by choosing a large detuning $\Delta_R$. In return, we need
correspondingly large laser intensities of the order of
$10^5\mWpersqcm$ for the Raman beams. We use fibre lasers to provide
these powers, see \autoref{ch:laser}.

\index{adiabatic elimination} The coupling of both Raman lasers to the
$3P$ manifold can be adiabatically eliminated
\cite{heinzen:adiabatic,wineland:bible,kielpinski:thesis,king:thesis}
and we may describe the coupling by an effective two-level system as
we did in \autoref{ch:two-level}. \index{Raman Rabi frequency!parallel
geometry} The Raman Rabi frequency for a two-photon stimulated Raman
transition is given by \cite{wineland:bible}
\be
\Omega_R = \frac{2 \Omega_\down \Omega_\up}{\Delta_R}
\label{eq:coprop_rabi_frequency}
\ee
where $\Omega_\down$ and $\Omega_\up$ are the Rabi frequencies for the
coupling of $\left|\down\right> \leftrightarrow 3P$ and
$\left|\up\right> \leftrightarrow 3P$ transitions respectively. They
can be calculated from \autoeqref{eq:rabi_frequency} taking into
account that all states of the $3P$ manifold have to be considered for
the transition matrix element $\left<a|\hat{x}|b\right>$. Using
the Raman laser beam intensities $I_\down$ and $I_\up$ we may also
conclude
\be
\Omega_R \propto \frac{\sqrt{I_\down I_\up}}{\Delta_R} \text{.}
\label{eq:coprop_rabi_intensity}
\ee
Note that these equations for the Rabi frequency are only strictly
valid for co-propagat\-ing (=\,parallel) Raman beams. Other beam
configurations introduce a coupling between the electronic and
motional states, which changes the Rabi frequency~(see below).

\subsection{\texorpdfstring{\protect\ac{AC}}{AC} Stark shifts}
\label{sec:stark_shift}
\index{Stark shift}

Each Raman laser (coupling one of the qubit levels to the $3P$
manifold) causes an energy shift of the respective qubit level. This
effect can be attributed to the intense applied laser field---or more
precisely: to the strong alternating electric field, which is why it
is called an \ac{AC} Stark shift. The energy shifts can be determined
by \cite{kielpinski:thesis}
\be
\Delta E_i = \hbar \frac{\Omega_i^2}{\Delta_R}
\ee
where $i = \down,\up$.

The \ac{AC} Stark shifts are not exactly equal for the two qubit
states of ${}^{25}\text{Mg}^+$. This introduces problems as the
energy gap between the $\left|\down\right>$ and $\left|\up\right>$
states then becomes dependent on the applied Raman laser
intensity. This means we would have to adapt the frequency difference
of the Raman beams to their intensity. Fortunately, we can reduce the
influence of \ac{AC} Stark shifts by sufficiently large detunings
$\Delta_R$.

\section{Coupling qubit and motional states}
\label{sec:motion_coupling}

In the dipole approximation we may describe the coupling between an
ion and a laser beam by the interaction Hamiltonian
\be
\hat{H}_\text{int} = -\hat{\mathbf{p}} \cdot \mathbf{E}(\hat{\mathbf{x}}, t)
\ee
where $\hat{\mathbf{p}}$ is the electric dipole
moment operator of the ion and $\mathbf{E}$ is the electric field
vector at position $\hat{\mathbf{x}}$. The
coupling between $\hat{\mathbf{p}}$ and
$\hat{\mathbf{x}}$ can be approximated by
expanding the electric field in a power series:
\be
\hat{H}_\text{int} = -\hat{\mathbf{p}} \cdot \left[\mathbf{E}(0,t) + (\hat{\mathbf{x}}\cdot\mathbf{\nabla})\mathbf{E}|_{\mathbf{x}=0} + \mathcal{O}(\mathbf{x}^2)\right]
\ee
The essential term is the gradient term. It introduces the coupling
between the electronic~(qubit) and the motional states.

The direction of motion in which coupling occurs is determined by the
difference of the wavevectors of the two Raman beams
$\Delta\mathbf{k}$ \cite{kielpinski:thesis}. In principle, we may
choose any direction by appropriately adjusting the angles between
the two beams. Throughout this thesis however we will restrict
ourselves to coupling in the trap's axial direction $\hat{z}$ (also
see \autoref{fig:trap_west}): Both beams propagate perpendicular to
each other where the difference of their wavevectors points in the
axial direction so that $k_\text{eff} = |\Delta\mathbf{k}| =
\sqrt{2}k_1 = \sqrt{2}k_2$ as the moduli of their wavevectors are
virtually equal.

When these two laser beams shine onto the ion, they drive Rabi
oscillations between the states $\left|\down,n\right>$ and
$\left|\up,n^\prime\right>$ where $n$ and $n^\prime$ are not
necessarily equal motional states (in contrast to the co-propagating
case of \autoref{sec:tps-raman}). \index{Raman Rabi
frequency!orthogonal geometry} The Raman Rabi frequency is now given
by \cite{wineland:bible}
\be
\Omega_{n^\prime,n} = \Omega_R e^{-\eta^2/2} \sqrt{\frac{n_<!}{n_>!}} \eta^{|n^\prime-n|} L_{n_<}^{|n^\prime-n|}(\eta^2)
\label{eq:ortho_rabi_frequency}
\ee
where $\Omega_R$ is the co-propagating Raman Rabi frequency from
\autoeqref{eq:coprop_rabi_frequency}, $n_<$~($n_>$) is the
smaller~(greater) of $n$ and $n^\prime$, $L_n^\alpha$ is the generalised
Laguerre polynomial, $\eta$ is the so-called Lamb-Dicke parameter
defined by
\be
\eta = k_\text{eff} z_0
\ee
and $z_0$ is the zero-point wavepacket spread of the ion in the axial
direction $\hat{z}$. For our trap $\eta \approx 0.3$. Note that
$\Omega_{n^\prime,n} = \Omega_{n,n^\prime}$.

Three cases are of special interest:
\bitem
\item $n^\prime = n$. In this case $\Omega_{n,n} = \Omega_R
e^{-\eta^2/2} L_n(\eta^2)$ where $L_n$ is the~(usual) Laguerre
polynomial. As the motional state is unchanged by this transition it
is called the \textit{carrier}.
\item \index{sideband!transition} $n^\prime = n-1$. The motional state
is reduced by one, which is why this transition is called the first
\textit{red sideband} transition. If $n=0$ there is no lower motional
state that could be reached, so the red sideband cannot be driven
anymore.
\item $n^\prime = n+1$. The motional state is increased by one. We
call such a transition the first \textit{blue sideband} transition.
\eitem
Mathematically, things become a lot easier in the Lamb-Dicke
limit which is characterised by a wavepacket spread in the axial
direction that is much less than the effective laser wavelength, i.\,e.
\be
\left<\psi_\text{motion} \mid k_\text{eff}^2 z^2 \mid \psi_\text{motion}\right> \ll 1 \text{.}
\label{eq:lamb-dicke}
\ee
(This criterion is not strictly met for ions in our trap.) In this
limit the Laguerre polynomials $L_n^\alpha$ converge rather quickly and thus
yield simpler expressions for \autoeqref{eq:ortho_rabi_frequency}.
Interestingly, the Hamiltonian for a red sideband transition in the
Lamb-Dicke limit can be written as a Jaynes-Cummings Hamiltonian. This
type of Hamiltonian \cite{shore:jcm} actually describes the time
evolution of a two-level atom coupled to a single mode of the
quantised radiation field\footnote{Several cavity-QED experiments
implement this Hamiltonian directly.}.

The rather large \index{Lamb-Dicke factor} Lamb-Dicke factor $\eta
\approx 0.3$ of our trap makes it possible to drive not only first
sideband transitions $n \leftrightarrow n\pm1$ but also second
sideband transitions $n \leftrightarrow n\pm2$. The Rabi frequencies
for these transitions are smaller than their first sideband
counterparts ($\Omega_{2,0} = 0.21 \Omega_{1,0}$) which requires
longer $\pi$ pulses, but this is still in the feasible range of our
experimental setup. Smaller $\eta$ would further increase the $\pi$
time of second sideband transitions, eventually rendering this type of
transitions inaccessible (due to decoherence occuring at similar
timescales).

\section{Cooling the \texorpdfstring{ions'}{ions\textquoteright} motion}
\label{sec:cooling}

Our quantum simulation experiments will rely on an efficient
initialisation of the ions' motional state. As an example consider the
simulation of cosmological particle creation~\cite{schuetzhold}. The crucial
measurement will be to detect a change in the population of the
motional state $n=2$, ideally from zero to some finite value. This is
only possible if the better part of the population has previously been
transferred to the motional ground-state $n=0$.

As the ions usually carry a considerable amount of kinetic energy
after having been loaded into the trap, they have to be cooled. In our
setup we use two methods for cooling both driven by laser transitions
to eventually reach the motional ground-state.

\subsection{Doppler cooling}
\label{sec:doppler_cooling}

Consider a magnesium ion travelling towards our slightly red-detuned
\ac{BD} laser source (frequency $\omega$) with velocity $v$. Due to
the first-order relativistic Doppler shift it will experience a
shifted frequency
\be
\omega^\prime\ \dot{=}\ \omega \left(1 + \frac{v}{c}\right)
\ee
where $c$ is the speed of light. If $\omega^\prime$ matches the actual
\ac{BD} transition frequency, one photon is absorbed thereby
transferring negative momentum
\be
p_1 = -\frac{\hbar \omega}{c}
\label{eq:doppler_momentum}
\ee
onto the ion. The subsequent fluorescent emission can either occur by
stimulated emission (in which case the ion re-gains a momentum of $p_2
= -p_1$; the net momentum transferred vanishes) or spontaneously. For
isotropic spontaneous emission the mean transferred net momentum is
thus determined by \autoeqref{eq:doppler_momentum}. As long as the ion
can absorb counterpropagating photons from the \ac{BD} laser, it will
thus slow down, losing energy.

The optimum conditions for this process of Doppler cooling obviously
depend on the velocity of the ion. For faster ions we will have to
red-detune the \ac{BD} laser by greater amounts than for ions that
have already been cooled. \index{temperature!Doppler cooling} The
minimum temperature $T_D$  that can be reached using Doppler cooling
is the \textit{Doppler cooling limit} determined by
\cite{haensch:doppler}
\be
T_D = \frac{\hbar \Gamma}{2 k_B} \approx 1\millikelvin 
\ee
where $\Gamma = 2\pi \cdot 43\MHz$ is the radiative linewidth of the
\ac{BD} transition and $k_B$ is the Boltzmann constant. It represents
a lower limit on the temperature, which is reached for a red-detuning
with respect to the resonance frequency of $\Delta = \Gamma/2$
\cite{wineland:cooling}. \index{Blue Doppler laser!intensity}
Laser intensity is kept below the saturation intensity at about $(2/3)
I_\text{sat}$ in order to avoid power broadening effects, see
\autoref{sec:power_broadening}.

The occupation of motional states after Doppler-cooling follows a
Maxwell-Boltz\-mann distribution. The stronger the confinement of the
ions, the more population will reside in the lower motional states, see
\autoref{fig:doppler-occupation}.

\bfig
\includegraphics{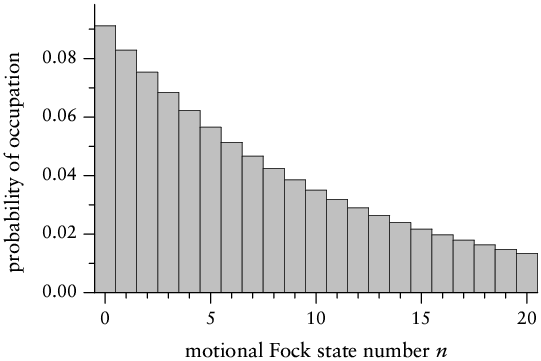}
\caption{Occupation probabilities of the motional Fock states
characterised by their number $n$ for a Doppler-cooled ion at a
temperature of $T_D = 1\millikelvin$. We assumed a harmonic potential
characterised by $\omega_z = 2\pi \cdot 2\MHz$. The mean
motional quantum number for this distribution is $\overline{n} = 10$.}
\label{fig:doppler-occupation}
\efig

\subsection{Resolved-sideband cooling}
\label{sec:sb_cooling}

It is desirable to initialise our qubit in a way that most of the
population resides in one particular motional state. The Rabi
frequency will then be well-defined even for an orthogonal Raman beam
geometry, see \autoref{sec:motion_coupling}. The method of choice
is to cool as much population as possible into the motional ground state
$\left|\down,n=0\right>$. If we calculate the motional occupation
numbers for an ion that has been pre-cooled to the Doppler limit $T_D =
1\millikelvin$ we however find that the probability for an ion to be
in the motional ground state is only about $9\%$, see
\autoref{fig:doppler-occupation}.  For proper state initialisation we
however require probabilities of $>90\%$. This means that we have to
transfer at least the occupation of the motional states
$\left|1\right>$ thru $\left|24\right>$ into the ground state
$\left|0\right>$.

We use the method of resolved-sideband cooling for this purpose
\cite{wineland:cooling, diedrich:sb_cooling, monroe:sb_cooling}.
Starting with a red sideband $\pi$-pulse optimised for $n=24$ we
first drive the transition $\left|\down,24\right> \rightarrow
\left|\up,23\right>$ and then optically pump the population into
$\left|\down,23\right>$, not changing the motional state of the ion,
see \autoref{fig:sb_cooling}.

\bfig
\subfloat[First step: Red sideband $\pi$-pulse using a two-photon
stimulated Raman transition.]{
  \includegraphics{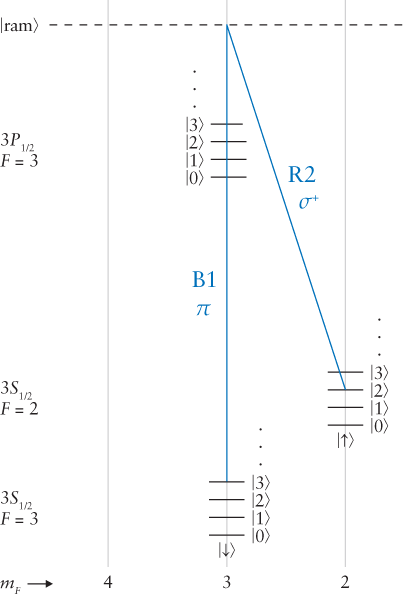}}
\hfill
\subfloat[Second step: Optical pumping into the ground state
$\left|\down\right>$ using the Red Doppler beam.]{
  \includegraphics{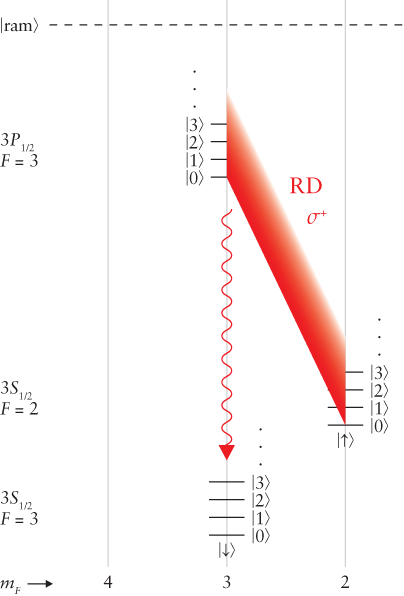}}
\caption{Resolved-sideband cooling of ${}^{25}\text{Mg}^+$. The
figures illustrate the cooling process from $n=3$ to $n^\prime=2$.}
\label{fig:sb_cooling}
\efig

\noindent The optical pumping process uses the Red Doppler~(\ac{RD}) beam to
transfers the population into the $P_{1/2} \left|F=3, m_F=3,
n=23\right>$ state which then decays into $\left|\down,23\right>$ by
spontaneous emission. As some part of the $P_{1/2}$ population might
however decay into the $S_{1/2} \left|F=3, m_F=2\right>$ state (see
\autoref{fig:mg25}), we subsequently shine onto the ion with the
repumper beam that resonantly couples $S_{1/2} \left|F=3, m_F=2\right>
\leftrightarrow P_{1/2} \left|F=3, m_F=3\right>$. It clears the
population in $\left|F=3, m_F=2\right>$ state by optically pumping it
into $\left|\down\right>$. Both the optical pumping beams are enabled
for a duration allowing for several scattering events to occur. This
ensures that even population which is scattered back into its initial
state is eventually transferred into $\left|\down\right>$. In the end,
most of the population initially in $\left|\down,24\right>$ will have
been transferred to $\left|\down,23\right>$.

In the \hyperref[eq:lamb-dicke]{Lamb-Dicke limit~(\ref*{eq:lamb-dicke})} the recoil energy due
to the spontaneous emission processes is small compared to the energy
of the motional quanta. As mentioned above, we may thus neglect
excitations of the motional spectrum in first order.

The two-photon stimulated Raman pulses applied to transfer $n=24$ into
$n^\prime=23$ will also affect the other motional states $n<24$. As their Rabi
frequencies $\Omega_{n-1,n}$ will be different from $\Omega_{23,24}$,
we cannot concurrently drive these transitions with high
efficiencies. Instead, population in these other motional states will
somehow distribute over $n$ and $n-1$. Anyhow, the important thing
about this iteration $n=24 \rightarrow n^\prime=23$ of sideband
cooling is the fact that the motional state $n=24$ will be depleted
afterwards. It will thus not be populated by subsequent sideband
cooling iterations. Iterating the sideband cooling processes over the
lower motional states ($n=23 \rightarrow n^\prime=22$, $n=22
\rightarrow n^\prime=21$, etc.) we will eventually transfer most of
the population initially occupying the motional states $0 \leq n \leq
24$ into $n^\prime=0$.

As shown in \autoref{ch:flopping} we obtain mean motional quantum
numbers of $\overline{n} = 0.6$ after resolved-sideband cooling. For
even lower values we would have to increase the ions' confinements as
described in \cite{monroe:sb_cooling}.

\part{Assembly of the experimental apparatus}
\chapter{Trap apparatus}
\label{ch:trap}

The magnesium ions used for experimentation throughout are stored
inside a quadru\-pole trap operated in the radio frequency~(\ac{RF})
range. The trap itself is kept inside an ultra-high vacuum~($p <
10^{-10}\mbar$) so as to isolate the ions from noise, collisions, or
other perturbing signals of the surrounding environment thus allowing
for rather long quantum state lifetimes and precise quantum state
manipulation.

\section{Linear Paul trap}

\begin{figure}[t]\centering
\includegraphics[width=\textwidth]{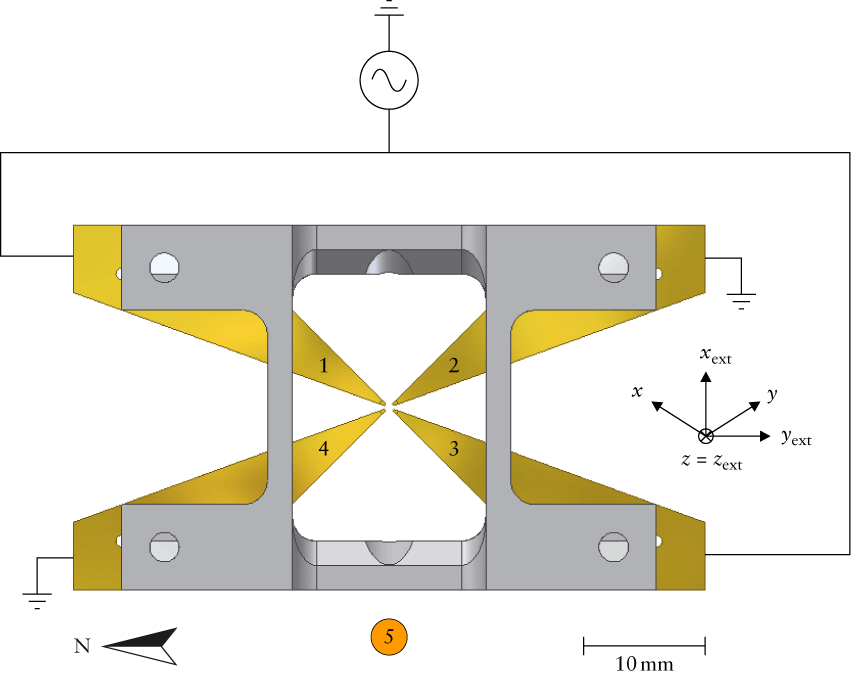}
\caption{Western view of the employed ion trap. $x$, $y$, and $z$
denote the non-orthogonal trap-inherent directions, while
$x_\text{ext}$, $y_\text{ext}$, and $z_\text{ext}$ represent an
external orthogonal coordinate system. Part no.~1--4: electrodes. Part
no.~5: compensation electrode.}
\label{fig:trap_west}
\efig

For our experiments we use a segmented linear Paul trap,
\autoref{fig:trap_west}. In a rough view the trap consists of
four electrodes where two opposing ones~(2 and 4) are grounded and the
other two~(1 and 3) are connected to an \ac{RF} voltage. The
associated \ac{RF} quadrupole potential radially confines the
positively charged ions to the trap axis $\hat{z}$, which lies
parallel to the electrode surfaces.

\begin{figure}[t]
\includegraphics[width=\textwidth]{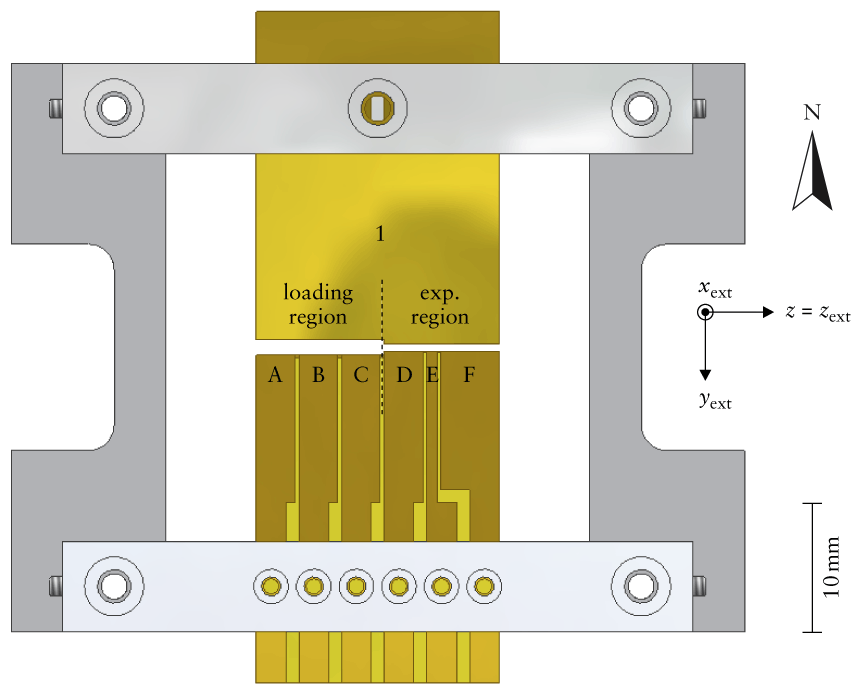}
\caption{Top view of the employed ion trap. $x_\text{ext}$,
$y_\text{ext}$, and $z_\text{ext}$ represent the external orthogonal
coordinate system.}
\label{fig:trap_top}
\efig

If we take a closer look on the \ac{RF}-grounded electrodes we can see
that they are each segmented into six independent sub-electrodes,
\autoref{fig:trap_top}. Any three opposing pairs of the 2x6
sub-electrodes are able to provide a confining potential along the
trap axis by keeping them at suitable static~(\ac{DC}) voltages. As long
as the outer sub-electrode pairs are kept at a higher voltage than the
middle one, the ions will align nearby the middle electrode. In
conjunction with the radial \ac{RF} potential, ions can be confined or
shuttled to arbitrary positions along the trap axis.

The trap can be divided into two regions characterised by the distance
$R$ of the electrodes from the trap centre: The loading region~($R =
800\micrometer$) and the storage or experimentation region~($R =
400\micrometer$). We use the loading region solely to ionise
atoms from a thermal atomic beam and load them into the trap. That
way we can avoid the contamination of the electrodes in the
experimentation region with magnesium atoms. The bigger size of the
loading regions allows for a good penetration by the atomic beams
whereas the smaller size of the experimentation region is able to
provide a stronger confinement. By appropriately altering the \ac{DC}
voltages we can shuttle the ions along the axis and transfer them from
the loading region into the experimentation region with near 100\%
efficiency. All experiments will be carried out in the experimentation
region.

\subsection{Radial confinement}
\label{sec:radial_confinement}

Radial confinement to the trap axis is provided by an \ac{RF}
potential. This method of confining positively charged particles using
alternating electric fields \cite{paul:traps} has first been described
by Wolfgang Paul (Nobel prize in 1989).

\noindent We will be storing ions of mass $m$. For this purpose, we
apply a trapping voltage $V_0 \cos \Omega_\text{RF} t$ to electrodes 1
and 3 of our trap. The two remaining electrodes 2 and 4 (more precisely:
the respective \ac{DC} subelectrodes) are kept at a constant \ac{DC}
voltage $U_0$. As mentioned before, the minimum distance from the
electrode surface to the trap centre is denoted by $R$. The equations
of motion for such a system each have the type of a Mathieu
differential equation which can be solved analytically for the radial
directions $\hat{x}$ and $\hat{y}$ \cite{major:traps}. The first order
solution is
\be
\begin{split}
x(t) &= x_0 \cos(\omega_x t + \phi_x) \left(1 + \frac{q}{2} \cos \Omega_\text{RF} t\right) \text{,} \\
y(t) &= y_0 \cos(\omega_y t + \phi_y) \left(1 - \frac{q}{2} \cos \Omega_\text{RF} t\right) \text{,}
\end{split}
\label{eq:radial_motion}
\ee
where $x_0$, $y_0$, $\phi_x$, and $\phi_y$ are determined by the initial
conditions. We introduced the dimensionless parameters
\be
\begin{split}
a &= \frac{4 e U_0}{m \Omega_\text{RF}^2 R^2} \text{,} \\
q &= \frac{2 e V_0}{m \Omega_\text{RF}^2 R^2} \text{.}
\end{split}
\ee
$\omega_i$ are the so-called \index{secular frequency} secular
frequencies defined by
\be
\begin{split}
\omega_x &= \frac{\Omega_\text{RF}}{2} \sqrt{\frac{q^2}{2} + a} \text{,} \\
\omega_y &= \frac{\Omega_\text{RF}}{2} \sqrt{\frac{q^2}{2} - a} \text{.}
\end{split}
\ee
As we can see, the ions' motion in each of the two radial directions
can be divided into two parts (see \autoref{fig:trap_motion}):

\bitem
\item Secular motion with angular frequency $\omega_i$, also called
macromotion. The associated frequencies determine the effective depth
of our radial potentials
\[
\begin{split}
\Phi_x &= \frac{1}{2} m \omega_x^2 x^2 \text{,} \\
\Phi_y &= \frac{1}{2} m \omega_y^2 y^2 \text{.}
\end{split}
\]
\item \index{micromotion} Micromotion with the driving frequency
$\Omega_\text{RF}$. As the ion is displaced from the trap axis, the
distances to the opposing trap electrodes are not any more
equal. Therefore the ion's motion is driven with the \ac{RF}
frequency.
\eitem

\bfig
\includegraphics{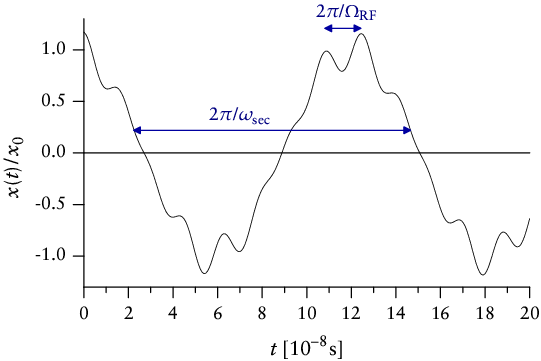}
\caption{First order solution for the radial $\hat{x}$ motion of a
single ${}^{25}\text{Mg}^+$ ion in the experimentation region of our
trap. The following parameters were used to produce this figure: $U_0
= -10\volt$, $V_0 = 1000\volt$, $\Omega_\text{RF} = 56\MHz$. This
yields $a = 7.78\cdot10^{-3}$, $q = 0.389$. Once an ion has been
cooled, the oscillation amplitude is very small: $x_0 < 0.1\micrometer$.}
\label{fig:trap_motion}
\efig

\noindent We use a common frequency generator (Hewlett-Packard~HP8640B)
connected to an \ac{RF} amplifier (Amplifier Research~10W1000) to
generate the trapping \ac{RF} voltages. They deliver a
\index{RF power and voltage@\ac{RF} power} maximum output power of
$44\dBm$ or $27\watt$. We further amplify the voltage via a
high-\ac{Q} helical resonator transformator connected to the
chamber~($Q_\text{total} = 90$) with a fixed driving frequency of
$\Omega_\text{RF} = 56\MHz$. This setup allows for \ac{RF} voltages up
to $1100\volt$. Associated secular frequencies range up to $8.2\MHz$.

During the first two hours of operation, the electrodes will heat up
due to resistive losses. As a result, the trap geometry changes
slightly and a portion of the applied \ac{RF} power is backreflected
into the \ac{RF} amplifier. We manually optimise the injection
efficiency by adjusting the position of the coil inside the helical
resonator. Special care must be taken not to exaggerate the injected
\ac{RF} power as highly temperated electrodes will decrease the
lifetime of stored ions. The micromotion compensation settings (see
\autoref{ch:micromotion}) are also highly sensitive to a thermal
deformation of the trap.

\subsection{Axial confinement}
\label{sec:axial}

Axial confinement of the ions is provided by applying \ac{DC} voltages
to the sub-electrode segments. The depth of this potential well and
thereby its resonance frequency was estimated using the commercially
available software package \ac{SIMION}. For the experimentation region
(see \autoref{fig:trap_top}) where sub-electrodes \ac{D} and \ac{F} 
are connected to a voltage of magnitude $U_\text{axial}$ and
sub-electrode \ac{E} is kept at ground we can calculate
\be
\omega_z = \sqrt{\frac{2 e U_\text{axial}}{m z_0^2}}
\label{eq:axial_freq}
\ee
where $z_0$ is the ``width'' of the harmonic potential. \ac{SIMION}
determined a value of $z_0 = 4.1\millimeter$. We applied a
low-amplitude \ac{RF} voltage to the central sub-electrode \ac{E} and
thereby resonantly excited the ion motion, which can be observed using
our camera, \autoref{fig:ion_excited_axial}. That way we rather estimate
$z_0 = (1.40\pm0.05)\millimeter$.

\bfig
\subfloat[Focused image of an ion.]{
  \includegraphics{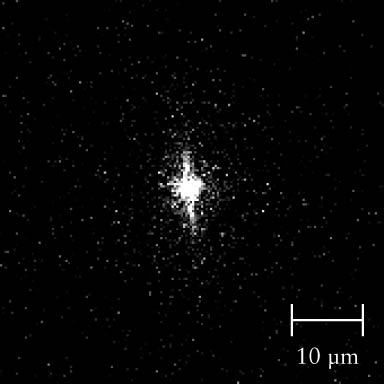}}
\hfill
\subfloat[Ion excited in the axial direction.]{
  \includegraphics{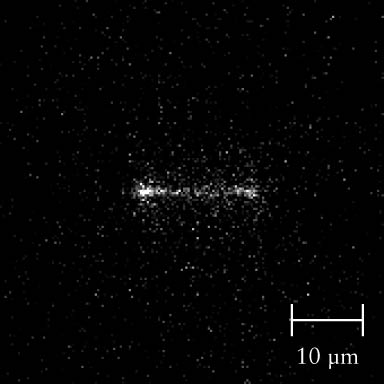}}
\caption{\protect\ac{CCD} camera images of an ion at a \protect\ac{BD}
intensity of $I = (2/3)I_\text{sat}$. Exposure time was $100\millisecond$.}
\label{fig:ion_excited_axial}
\efig

It is desirable to make the axial confinement as strong as
possible. This will improve the efficiency of Doppler pre-cooling (see
\autoref{sec:doppler_cooling}) and reduce the probability of
motional excitation by thermal effects. Normally, we would operate at
an \index{axial confinement} axial confinement of $2\pi \cdot 2\MHz$.

Due to the lack of damping, ions in the trap can be excited very
efficiently. In order to reduce unwanted excitation by injection of
external noise all of the \ac{DC} sub-electrodes have been equipped
with \index{low-pass filter} low-pass filters. The cutoff frequencies
are $2\pi\cdot5\MHz$ each, with the exception of the filter for the
central sub-electrode which has a cutoff frequency of
$2\pi\cdot2\MHz$. That way higher frequency noise is banned from
entering the trap, ramping the axial potential as required for
simulating cosmological particle creation is however still possible at
slew rates of $40\volt/\unitsignonly{\microsecond}$.

\subsection{Mutual interference of confinements}
\label{sec:trap_confinement_interference}

From an experimentator's point of view it is desirable to have
completely decoupled confinements in the radial and axial
directions. This is however not true for the design of our trap.

\noindent Without any additional radial confinement (\ac{RF} supply switched
off), \ac{DC} voltages applied to subelectrodes~\ac{D} and~\ac{F} will
cause a focusing effect in the $\hat{z}$ direction \textit{and} in the
$\hat{y}$ direction (see \autoref{fig:trap_west} for the coordinate
axes). Both confinements can be associated positive frequencies
$\omega_x$ and $\omega_y$. Due to Poisson's equation $\Delta\Phi = 0$
for the potential $\Phi$ there must however be a net defocusing effect
in the $\hat{x}$ direction, equivalent to an imaginary frequency $i
\omega_x$. The stronger the radial confinement $\omega_z$, the greater
will be the frequency splitting of the radial frequencies
$\Delta\omega_r = \omega_x-\omega_y$.

The defocusing effect in the $\hat{x}$ direction is compensated by a
sufficiently strong \ac{RF} radial confinement. As we have seen in
\autoref{sec:radial_confinement}, the axial confinement may give
rise to two different secular frequencies whose splitting
\be
\omega_x-\omega_y \approx \Omega_\text{RF} \frac{a}{\sqrt{2} q} \text{for $a \ll q^2$}
\ee
depends on the voltage $U_0$ applied to the central \ac{DC}
subelectrode, e.\,g. subelectrode \ac{E} for the experimentation
region.

If an ion's $z$ position is not centred in the region between two
opposing \ac{DC} subelectrodes, it will oscillate with the \ac{RF}
frequency. This effect can be attributed to the asymmetric geometry of
our trap: The \ac{DC} subelectrodes are separated by slits whereas the
\ac{RF} electrodes aren't. An axial off-position will thus induce a
net oscillating force on the ion. See \autoref{ch:micromotion}
for further micromotion effects.

The optimum $z$ position (where the ion is not oscillating) is
furthermore dependent on the strength of the applied \ac{RF}
potential. This effect is probably caused by slight changes in trap
geometry due to \ac{RF} dependent heating of the electrodes.

\subsection{A word on trap geometries}
\label{sec:heating}

Several research groups have demonstrated the fabrication and
operation of ion traps featuring electrode\,--\,ion distances of
$100\micrometer$ and below \cite{nist:microfabrication}. Obviously,
these designs are suitable for incorporation into larger arrays of ion
traps. They however usually suffer from relatively high motional
\index{heating rate} heating rates $\propto R^{-4}$. Motional
coherence times for the stored ions thus diminish for shrinking trap
sizes. For our single trap setup we chose a rather large trap geometry
of $R = 400\micrometer$ in order to minimise these heating effects. We
expect $\gamma$ to be smaller than $0.01$ excited quanta per
millisecond at an axial confinement of $\omega_z = 2\pi \cdot
2\MHz$. As the durations of typical experiments are of the order of
$1\millisecond$, heating should not be an issue.

\subsection{Compensation electrode}

The trap apparatus is completed by one compensation electrode
\index{compensation electrode} consisting of a piece of copper wire
($3\millimeter$ in diameter) which is located beneath the trap
axis, see \autoref{fig:trap_west}. This electrode can be used to
compensate for imperfect external \ac{DC} fields in the up-down
direction, see \autoref{ch:micromotion}. By appropriately tuning
the \ac{DC} voltages applied to the sub-electrodes we can as well
create compensation fields in the horizontal directions---radially by
differential voltages applied to an opposing subelectrode pair and
axially by appropriate voltages of neighbouring subelectrode pairs.

\section{Vacuum chamber}
\index{vacuum}

To minimise ion loss due to collisions with residual gas particles, we
have to provide a vacuum of better than $10^{-10}\mbar$. Evacuation to
the ultra-high vacuum level was attained as follows:
\benu
\item Evacuation using an oilfree molecular turbopump system (Pfeiffer
Vacuum) in combination with an oilfree scrollpump for
forepumping for about 24 hours. According to an electronic full-range
gauge (rangeability: atmospheric pressure down to $10^{-9}\mbar$) we
obtained pressures of $10^{-7}\mbar$.
\item Baking out the vacuum chamber. Using a home-built brick oven,
the vacuum chamber was baked out at $150\celsius$ for a duration of
approximately seven days, all the time being connected to the turbopump
system. Pressure inside the chamber eventually decreased back to
$10^{-7}\mbar$.
\item Running the ion getter pump in conjunction with the titanium
sublimation pump while the chamber cooled down. At room temperature, a
pressure level of some $10^{-10}\mbar$ was obtained. These
measurements were carried out using an \index{ion gauge} ion gauge
system (Granville-Phillips model~350).
\item Titanium sublimation. The titanium sublimation pump was run
several times, which again lowered the pressure to several
$10^{-11}\mbar$.
\item Installation of the vacuum chamber on an optical table.
\eenu
We secured the ion getter pump against power failure using an
uninterruptible power supply~(\ac{UPS}). Power interruptions of up to
two hours can be bypassed that way. The titanium sublimation pump is
run about once a month for a few minutes. Every run precipitates
titanium over the inner surface of a double-walled cylinder which is
connected to the getter pump housing and can be cooled using liquid
nitrogen.

Our ion gauge system is calibrated for a nitrogenic atmosphere, which
certainly differs from the composition of the residual gas inside the
chamber. Furthermore, at pressures of $10^{-10}\mbar$ and below, we
operate the gauge close to its sensitivity limit. The indication of
the gauge will thus probably differ from the actual pressure inside
the chamber. In practice, we use the ion lifetime inside our trap to
judge pressure. While this may not provide a quantitative measure of
the pressure itself, it is the quantity of interest that matters with
quantum simulations. Depending on the temperature of the electrodes we
attained ion lifetimes between two hours (at maximum \ac{RF} power on
the electrodes) and ten hours and above (at moderate \ac{RF} settings).

The vacuum chamber is equipped with a total of twelve ports, see
\autoref{fig:chamber}. Port~\ac{C} is the connector for the helical
resonator, port~\ac{D} acts as a feedthrough for the \ac{DC} voltages
applied to the trap subelectrodes. Two windows serve as observation
ports where the upper port~\ac{A} is designated for electronic observation
systems (camera, photomultiplier) and port~\ac{B} can be used to have
a look inside the chamber. All other eight ports have considerably
smaller $1$-inch windows coated for maximum transmission at
$280\nanometer$. They are used exclusively to shine into the
trapping volume using \ac{UV} laser beams from different directions:
\bitem
\item Four windows serve beams from the north-eastern~(\ac{E}),
south-eastern~(\ac{F}), south-western~(\ac{G}), and
north-western~(\ac{H}) directions. All of these beams propagate
parallel to the optical table layer.
\item Another four windows serve beams from the eastern~(\ac{J},
\ac{K}) and western~(\ac{L}, \ac{M}) directions. The propagate at
angles of $20\degree$ relative to the optical table layer.
\eitem
All windows are made from fused silica which is transparent to \ac{UV}
radiation and does not exhibit any birefringence. An exception is port
\ac{A} which has a sapphire crystal window. It is part of the
diffraction limited optics for the \ac{CCD} camera and as such
required by the objective attached to port \ac{A} for optimum image
quality, see \autoref{sec:camera}.
\bfig
\subfloat[Plan view.]{
\includegraphics{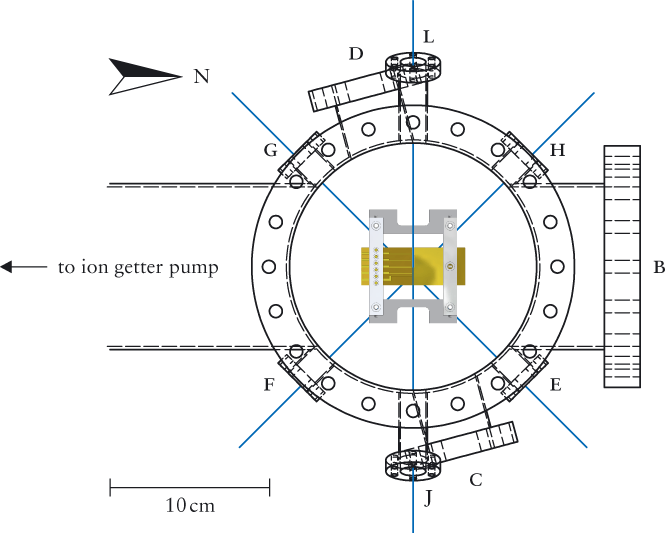}} \\
\subfloat[Eastern view.]{
\includegraphics{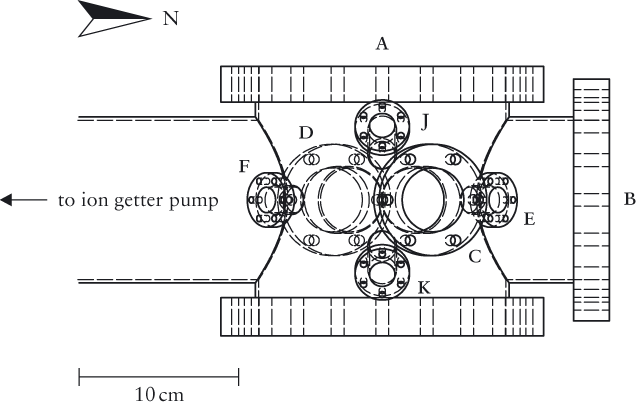}}
\caption{Sketches of the vacuum chamber. Laser beams entering the
chamber are drawn in blue colour.}
\label{fig:chamber}
\efig

\section{Ion source}
\label{sec:ionisation}

\begin{figure}[t]\centering
\includegraphics[width=\textwidth]{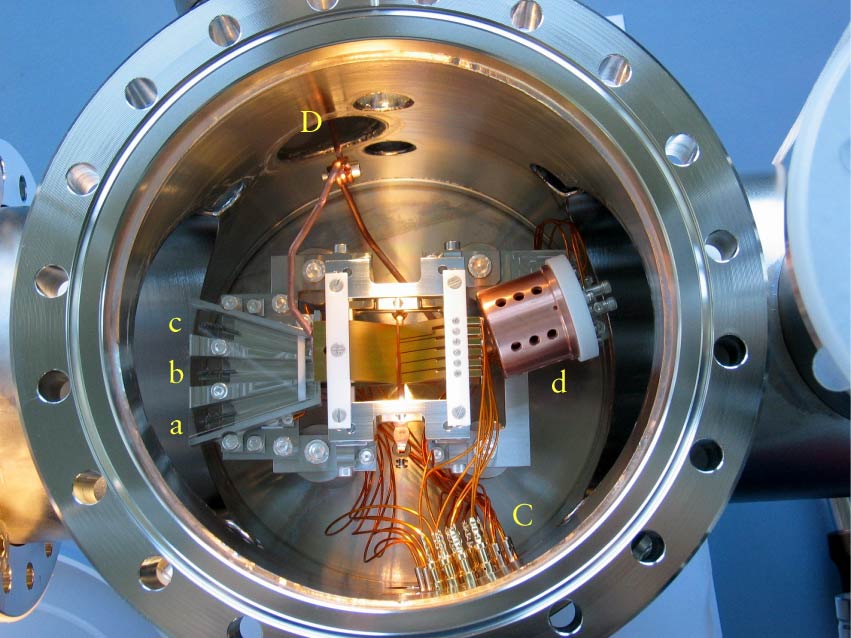}
\caption{Top view of the whole trap apparatus. Ports are denoted with
uppercase letters. Legend: a,~b,~c\,=\,atomic ovens; d\,=\,electron gun.}
\label{fig:trap_apparatus_top}
\efig

To store ions inside a Paul trap, they have to be ionised inside the
active trapping region near the centre of the trap. Our trap apparatus
features three independent atomic ovens emitting their atoms into the
centre of the loading region where we photo-ionise them (laser beam
through port~\ac{L}), see \autoref{fig:trap_apparatus_top}. In
addition, we have an electron gun for backup\,/\,emergency purposes.
All ovens consist of tantalum tubes (length $l \approx 10\millimeter$,
diameter $d \approx 1\millimeter$) whose rear halves are filled with a
granulate of magnesium. We fixed the ovens in their respective positions
using two tantalum wires soldered to the tubes. These also serve as
current entries, actually heating the ovens. Oven \ac{A} contains a
natural composition of magnesium isotopes, i.\,e. 79\%
${}^{24}\text{Mg}$, 10\% ${}^{25}\text{Mg}$, and 11\%
${}^{26}\text{Mg}$, while the other two ovens~\ac{B} and~\ac{C}
emit isotopically enriched ${}^{25}\text{Mg}$ atoms.
The ovens are arranged such that oven~\ac{B} emits its atoms
perpendicular to the direction of the ionising beam and ovens~\ac{A}
and~\ac{C} share an angle of $15\degree$ relative to oven~\ac{B}. This
setup allows us to selectively load ${}^{24}\text{Mg}^+$,
${}^{25}\text{Mg}^+$ or even ${}^{26}\text{Mg}^+$. We supply currents
between $2.1\ampere$ and $2.4\ampere$ to load ions into the trap with
a rate of between one and fifty ions per minute.

\index{photo-ionisation|main} Photo-ionisation of the magnesium atoms
is implemented by a two-photon absorption process \cite{aarhus}:
First, one electron of the $3s$ level is shifted to the $3p$ level
($3s^2 \rightarrow 3s3p {}^1P_1$) by resonant excitation at
$285\nanometer$. The same wavelength can be used to eventually ionise
the magnesium atom by transferring the excited electron into the
continuum \cite{madsen}. We use a frequency-doubled dye laser to
accomplish this task. The absolute photo-ionisation frequencies can be
determined using Doppler-free iodine saturation spectroscopy (see
\autoref{sec:iodine_spectroscopy}) in front of the frequency doubling
stage, i.\,e. at $570\nanometer$ and are summarised in
\autoref{tab:mg_ionisation}. Doppler shifts have to be taken into
consideration for ovens~\ac{A} and~\ac{C}. For the oven currents used
in our setup, \index{temperature!atomic oven} we can estimate
temperatures of $T \approx 600\kelvin$.

\bt
\btab{ccccc}\toprule
\multirow{2}*{oven} & \multirow{2}*{Mg isotope}         & typical oven current &       $3s^2 \rightarrow 3s3p {}^1P_1$ frequency \\
                    &                                   & [$\unitsignonly{\ampere}$] & [$2\pi \cdot 10^{15}\Hz$] \\
\cmidrule(r){1--1}
\cmidrule(lr){2--2}
\cmidrule(lr){3--3}
\cmidrule(lr){4--4}
\cmidrule(l){5--5}
A                   & ${}^{24}\text{Mg}$                & $2.1$                      & $1.050811538$ \\
B                   & ${}^{25}\text{Mg}$                & \textit{not yet measured}  & \textit{not yet measured} \\
C                   & ${}^{25}\text{Mg}$                & $2.2$                      & $1.050811834$ \\ \bottomrule
\etab
\caption{Resonant $3s^2 \rightarrow 3s3p {}^1P_1$ transitions for
photo-ionising magnesium as determined from the Doppler-free
spectroscopy setup. With the oven currents indicated we are able to
load a single ion into the trap on a $60\second$ timescale.}
\label{tab:mg_ionisation}
\et

The process of photo-ionisation has several advantages over the
electron gun method:
\bitem
\item In contrast to an electron beam a photo-ionising laser beam is
bundled and will usually not hit the electrode surfaces. Thus, vacuum
is affected less.
\item A photo-ionising laser beam will not charge the electrode
surfaces, which makes micromotion compensation
(\autoref{ch:micromotion}) more stable.
\item We may selectively load particular isotopes of magnesium by
properly adjusting the ionisation frequency.
\item The two-photon ionisation process of magnesium also scatters
some fluorescent light, which is why the atomic magnesium beam can be
visualised using the photo-ionising laser beam. Therefore, we can easily
detect whether the atomic ovens are working and whether the other
laser beams are correctly positioned.
\eitem
One drawback of using photo-ionisation for magnesium is the
unavailability of diode or fibre lasers at the required wavelength of
$285\nanometer$. We thus use a frequency-doubled dye laser for this
purpose.

\section{Magnetic field}
\label{sec:magnetic_field}

As described in \autoref{ch:mg25} controlled manipulation of
the electronic population of ${}^{25}\text{Mg}^+$ requires
discrimination of different $m_F$ levels of the states involved. We
have to apply a magnetic field to introduce an appropriate Zeeman
splitting between the otherwise degenerate $m_F$ levels. For
${}^{25}\text{Mg}^+$ the Zeeman splitting of the qubit shelves
($S_{1/2}$, $\left|F=2\right>$ and $\left|F=3\right>$) is about
$0.5\MHz/\unitsignonly{\gauss}$.

\subsection{Field-generating coils}

\begin{table}[b]\begin{center}
\btab{ll}\toprule
inner coil radius               & $60\millimeter$ \\
outer coil radius               & $100\millimeter$ \\
distance between coils          & $479\millimeter$ \\
number of windings per coil     & $336$ \\
resistance per coil             & $1.1\ohm$ \\
inductance per coil             & $7\millihenry$ (estimated) \\
calculated field at trap centre & $1.68\gauss/\unitsignonly{\ampere}$ \\
actual field at trap centre     & $0.932\gauss/\unitsignonly{\ampere}$ \\
operating current               & $6.000\ampere$ \\ \bottomrule
\etab
\caption{Characteristics of the two field-generating coils. The actual
field at the trap centre was determined via Rabi flopping, see
\autoref{sec:shelving}.}
\label{tab:magnetic_z}
\et

Due to the geometry of our vacuum chamber it is not advantageous to
create a spatially homogeneous magnetic field using Helmholtz's
configuration. We have to employ coils whose mutual distance $d =
479\millimeter$ is greater than their mean radius $\overline{r} =
100\millimeter$, see \autoref{fig:magnetic_photo}. This configuration
leads to the required field strengths in the centre (see
\autoref{tab:magnetic_z}), but shows a quadratic field dependence on
the position between the coils, see \autoref{fig:magnetic_z}. In the
$10\millimeter$ vicinity around the centre, relative deviations of the
magnetic field still remain below $10^{-4}$. \hyperref[tab:magnetic_z]{Tab.~\ref*{tab:magnetic_z}}
summarises the relevant parameters.

\bfig
\includegraphics{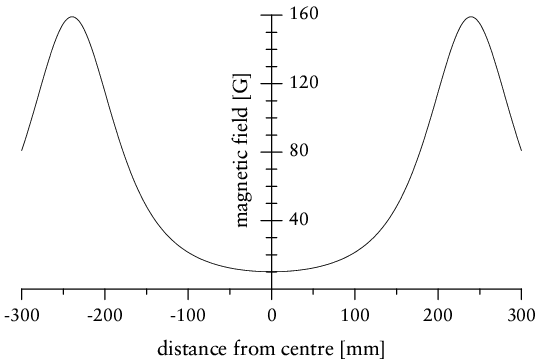}
\caption{Calculated magnetic field strength induced by the two coils
of \autoref{tab:magnetic_z} at a current of $I = 6\ampere$.}
\label{fig:magnetic_z}
\efig

The great discrepancy between the calculated and actual field strengths
at the centre of our trap~($44\%$) could be attributed to ferromagnetic
parts in the vacuum chamber or the trap itself. Most probably, the
material of the vacuum chamber (non-magnetic stainless steel) contains
some ferromagnetic inclusions. The conversion of non-magnetic stainless
steel into its ferromagnetic variant is well-known and can occur during
cold deformation and welding processes. Our vacuum chamber had to be
re-welded several times to eventually seal it. That might have caused
ferromagnetic inclusions which effectively weaken the magnetic field
in their vicinity.

Normally, we operate the coils at a current of $I = 6\ampere$ which is
generated by two low-noise Hewlett-Packard HP6264B power supplies in
constant-current mode (ripple and noise $< 5\milliampere$,
drift $< 8\milliampere$). The coils' windings are attached on a copper
mount to improve heat flow to the outside. After about three hours, the
coils will have reached their \index{temperature!magnetic field coil}
steady-state temperature of $60\celsius$. For higher currents, the
coils' temperature rises quickly and adversely affects optical
components in the vicinity. Water-cooling the coils would probably
improve this situation.

\bfig
\includegraphics{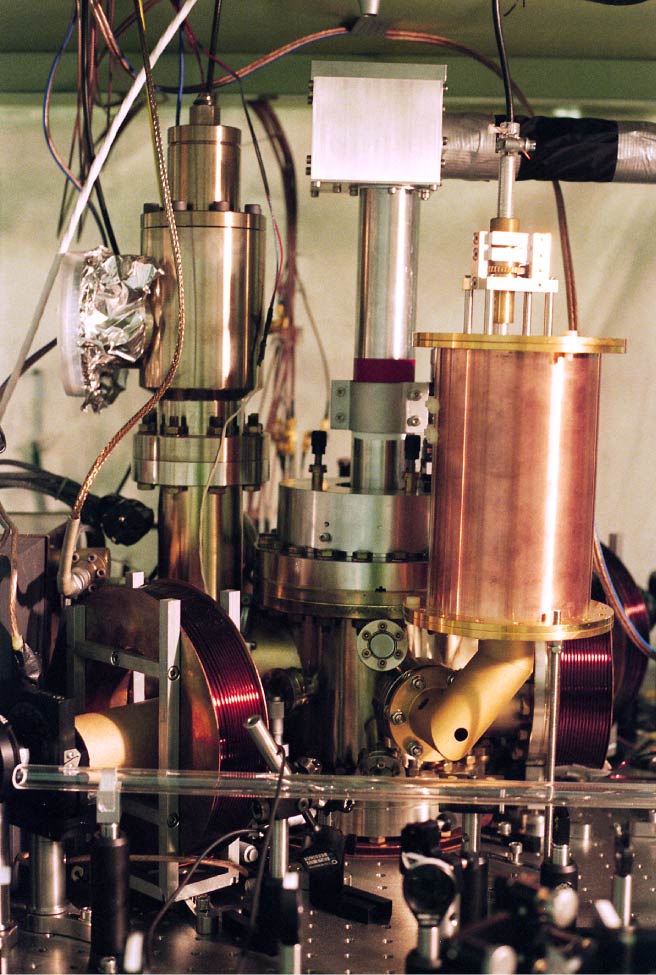}
\caption{Eastern view of the vacuum chamber. The southeastern
field-generating coil is clearly visible at the bottom-left part of the
image. The northwestern field-generating coil appears blurred while
the spatial compensation coils below and north of the chamber are easy to
spot.}
\label{fig:magnetic_photo}
\efig

\subsection{Spatial compensation coils}

\bt
\btab{lll}\toprule
Parameter                                        & northern coil & lower coil \\
\cmidrule(r){1--1}
\cmidrule(lr){2--2}
\cmidrule(l){3--3}
mean radius
[$\unitsignonly{\millimeter}$]                   & $81.3$        & $90.7$ \\
distance from trap centre
[$\unitsignonly{\millimeter}$]                   & $160.2$       & $119.0$ \\
number of windings                               & $60$          & $20$ \\
resistance [$\unitsignonly{\ohm}$]               & $0.3$         & $0.2$ \\
inductance (estimated)
[$\unitsignonly{\millihenry}$]                   & $0.2$         & $0.02$ \\
calculated field at trap centre
[$\unitsignonly{\gauss}/\unitsignonly{\ampere}$] & $0.430$       & $0.309$ \\
operating current
[$\unitsignonly{\ampere}$]                       & $3.20$        & $2.95$ \\ \bottomrule
\etab
\caption{Characteristics of the spatial compensation coils.}
\label{tab:magnetic_spatial}
\et

Being our quantisation axis the magnetic field has to point exactly
into the same direction as the wavevector of the beams entering the
trap along this axis. Otherwise, it would be impossible to polarise
these beams purely $\sigma^\pm$ with respect to the field axis. This
is however important for efficiently driving the cycling detection
transition, see \autoref{sec:state_detection}. Two spatial
compensation coils will tilt the magnetic field axis as needed. They
are located north and below the vacuum chamber respectively. Their
characteristics are summarised in \autoref{tab:magnetic_spatial}.

\subsection{Temporal compensation coil}
\label{sec:magnetic_temporal}

Temporal changes of the magnetic field strength will cause fluctuating
Zeeman shifts and thus reduce the fidelity of operations on quantum
states relying on fixed energy levels. Therefore, it is
desirable not to have variations in the magnetic field greater than
$10^{-4}$. We stabilise the magnetic field against external noise and
fluctuations of our power supplies using a feedback-regulated
system: The signal of a miniature high-precision \index{fluxgate
sensor} fluxgate sensor (Stefan Mayer Instruments FLC\,100) is
processed by a bandpass filter~($3\Hz < f < 3\kHz$) and a high-current
operational amplifier~($I_\text{max} = 5\ampere$) supplying the
compensation coil. This coil consists of only a few additional
windings so as to minimise its inductivity and thus maximise its
regulation bandwidth. It is wound directly on one of the
field-generating coils. Characteristics are listed in
\autoref{tab:magnetic_temporal}.

\bt
\btab{ll}\toprule
radius                          & $100\millimeter$ \\
distance from trap centre       & $239.5\millimeter$ \\
number of windings              & $5$ \\
resistance                      & $0.4\ohm$ \\
inductance                      & $1.5\microhenry$ (estimated) \\
calculated field at trap centre & $18.0\milligauss/\unitsignonly{\ampere}$ \\
regulation bandwidth            & DC\,--\,$150\kHz$ \\ \bottomrule
\etab
\caption{Characteristics of the temporal compensation coil.}
\label{tab:magnetic_temporal}
\et

\chapter{Laser apparatus}
\label{ch:laser}

All laser beams used throughout experimentation have frequencies lying
in the ultraviolet (\ac{UV}) range. We can identify four main laser
sources serving different purposes in a simulation experiment:
\benu
\item Photo-ionisation, see \autoref{sec:ionisation}.
\item Qubit manipulation. In order to drive the transition between the
two qubit states, we use a two photon stimulated Raman
transition. It involves a virtual level blue-detuned by several tens
of \GHz with respect to the $3P_{3/2}$ level in order to reduce
spontaneous emission effects, see \autoref{sec:tps-raman}.
Therefore, the respective laser source must be able to provide
intensities of the order of $10^5\mWpersqcm$.
\item State detection and Doppler cooling. The state-detecting laser
drives a closed cycling transition which is also used for
Doppler-cooling the ions. We will refer to this laser source as the
Blue Doppler (\ac{BD}) laser.
\item State correction and sideband cooling. Erroneous state
manipulations can partly be corrected using an additional resonant
transition which is also used for sideband cooling. Providing similar
properties as the \ac{BD} laser, we will analogously call this laser
source the Red Doppler (\ac{RD}) laser.
\eenu
Laser frequencies of the resonant transitions have to match the
respective transition frequencies within a range of $\pm1\MHz$. We use
Doppler-free iodine saturation spectroscopy to adjust and/or lock the
lasers to the required frequencies.

\section{Laser sources}

\subsection{Dye laser}

We generate $285\nanometer$ light required for the \index{photo-ionisation}
photo-ionisation process by second harmonic generation (\ac{SHG}) of a
\index{dye laser} dye laser (Coherent model~899) operated at
$570\nanometer$. The required frequency is determined roughly using a
wavemeter ($\pm500\MHz$) and fine-adjusted using a Doppler-free iodine
spectroscopy signal.

We use rhodamine-19 dissolved in ethylene glycol as a dye
solution. For proper jet formation, the solution is cooled down to
$9\celsius$. Under normal conditions, i.\,e. when the dye has already
bleached out and lost its initially high gain (this is the case after
about one week), we pump the dye laser with $4\watt$ of
frequency-doubled solid state laser light (Coherent Verdi V10) at
$532\nanometer$. We are able to operate the laser using these
parameters for about two months.

\subsection{Fibre lasers}

All three remaining laser systems consist of specially designed fibre
laser systems (Koheras Boostik models) near $1120\nanometer$ (\ac{IR})
which are frequency-doubled twice to generate wavelengths near
$280\nanometer$ (\ac{UV}). These fibre lasers feature very narrow
linewidths of $2\pi \cdot 200\kHz$ essential for coupling light into
the \ac{SHG} resonators (see below). As an active gain medium they use
an optical waveguide doped with ytterbium. In our systems, the fibres are
attached to a metal plate and have a piezoelectric transducer
(\ac{PZT}) at one end. The output frequency can be precisely adjusted
by either tuning the \index{temperature!fibre laser} metal plate
temperature (on the timescale of a minute) or tuning the \ac{PZT}
voltage (bandwidth of $20\kHz$); both mechanisms alter the fibre
length. Technical specifications are summarised in \autoref{tab:fibre_lasers}.

\begin{table}[b]\begin{center}
\btab{lcccc}\toprule
\multirow{2}*{laser source} & frequency       & max. output & temperature                                             & \ac{PZT} tuning \\
                            & range [THz]     & power [W]   & tuning [$\text{GHz}_\text{IR}/\unitsignonly{\celsius}$] & [$\text{MHz}_\text{IR}/\text{V}$]         \\
\cmidrule(r){1--1}
\cmidrule(lr){2--2}         
\cmidrule(lr){3--3}
\cmidrule(lr){4--4}
\cmidrule(l){5--5}
Raman                       & $267.98-268.10$ & $1.90$      & $2.3$                                                   & $+670$ \\
\ac{BD}                     & $267.97-268.08$ & $0.90$      & $2.7$                                                   & $-17$  \\
\ac{RD}                     & $267.28-267.39$ & $1.27$      & $2.8$                                                   & $+17$  \\ \bottomrule
\etab
\caption{Technical specifications of the used fibre laser sources.}
\label{tab:fibre_lasers}
\et

Reliable ``turn-key'' operation of the fibre lasers is estimated at
three years, which is the lifetime of the employed pumping
diodes. Whereas we would observe very stable and reliable operation
during the first year, two of our three fibre laser systems failed
afterwards and had to be shipped to the manufacturer for minor repair
works. The \ac{BD} system, which had been in use most heavily, first
failed after one and a half years. It had to be repaired twice which
decreased stable output power from $1.44\watt$ to $0.90\watt$. In one
case, a wavelength division multiplexer (\ac{WDM}) broke, in the other
case a pumping diode failed. However, this system is supposed to
regain the initial performance after an upgrade of its pre-amplifier
section. The \ac{RD} system broke after two years, the reason was a
bleached beamsplitter. We stress that we did not run the fibre lasers
continuously but rather for a maximum of twelve hours per day.

The fibre laser driven \ac{UV} sources will drive coherent transitions
between different energy levels of magnesium. As fluctuations in phase
or amplitude will diminish the fidelity of these transitions, they
should be kept at a minimum. Concerning the stability of the intensity
after the two \ac{SHG} stages (see below) we observe 4\% drops in
amplitude on a microsecond timescale, and 7\% drops less than once a
minute \cite{friedenauer}, which could hardly be achieved using
conventional dye laser setups.

\section{Second harmonic generation}
\label{sec:shg}
\index{SHG@\ac{SHG}|see{second harmonic generation}}
\index{second harmonic generation}

For second harmonic generation~(\ac{SHG}) we use resonators designed
as bow-tie cavities with sputtered mirrors featuring reflectivities of
$99.98\%$ \cite{friedenauer}. The length of these resonators is actively
stabilised using a mirror mounted onto a piezoelectric
crystal~(\ac{PZT}) in combination with the locking technique of Hänsch
and Couillaud \cite{haensch_couillaud}. We will distinguish two
different types of second harmonic generation:
\benu
\item \textit{\ac{SHG} of visible green light from infrared (\ac{IR})
light}. \ac{IR} light from one of our fibre laser systems first passes
an optical isolator (transmittivity $\approx 90\%$, isolation $\approx
30\dB$) which prevents damage of the fibre lasers by back reflection.
Subsequently, it is coupled into the \ac{SHG} resonator using a
modematching telescope. The \ac{IR} light is frequency-doubled with
the help of a lithium triborate (\ac{LBO}) crystal cut at $90\degree$
angles and antireflectance coated for both the fundamental and
harmonic wavelengths. Non-critical type~\ac{I} phase matching is
achieved using a \index{temperature!crystal|(} crystal oven
temperature of $94\celsius$. The folding angle of the resonator is
kept small ($10\degree$) to minimise astigmatism of the fundamental
wave. In the end this system yields a maximum conversion efficiency of
$52.7\%$.

For about two hours after the fibre laser systems have been
powered up, their \index{polarisation drift} \ac{IR} polarisation
will drift, which causes the frequency-doubled output power to
vary. After two hours, we can still observe polarisation drifts, but
these are caused by changes in the ambient temperature (power
alterations of $5\%$ per $1\kelvin$ \textit{behind} the resonator). We
reduce these effects by stabilising the temperature of the fibre
lasers' Peltier elements.

\item \textit{\ac{SHG} of ultraviolet (\ac{UV}) light from visible green
light.} \ac{UV} light is eventually generated using a second \ac{SHG}
resonator with a beta barium borate (\ac{BBO}) crystal using critical
type~\ac{I} phase matching. Due to reported problems with
the adhesion of antireflectance coatings on \ac{BBO} at high
intensities we have had the front and back sides of our crystals cut at
Brewster's angles instead. The thereby induced astigmatism of the
fundamental wave after one ``run'' through the resonator is
compensated by choosing an appropriate folding angle of the resonator
($27.4\degree$). In order to prevent water from condensing on the
hygroscopic surface \index{temperature!crystal|)} we permanently heat
the crystal to $50\celsius$ and expose it to a flow of oxygen. Walkoff
of the harmonic wave inside the crystal results in an astigmatic
output beam which is corrected, i.\,e. projected into a proper
$\text{\ac{TEM}}_{00}$ mode, using a cylindrical telescope. We
obtained conversion efficiencies of up to $28.9\%$ (mostly due to
losses of the s-polarised harmonic wave exiting the Brewster-cut
crystal).
\eenu

\bfig
\includegraphics[width=\textwidth]{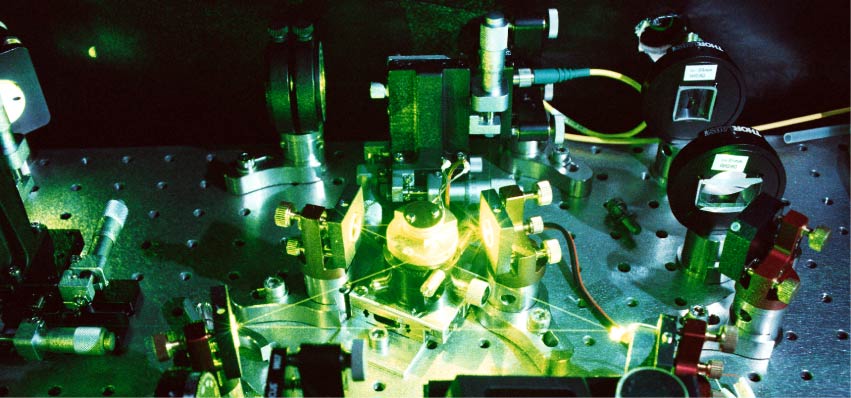}
\caption{\protect\ac{BBO} resonator for frequency-doubling our dye laser
source. The input coupler and the attached optical fibre can be seen in
the background. At the bottom left part of the resonator a small
mirror mounted on a \protect\ac{PZT} element keeps the resonator's optical
length constant. On the right hand side one can see the cylindrical
telescope used for correcting the astigmatism of the output beam.}
\label{fig:bbo_resonator}
\efig

\noindent In the Raman and \ac{BD} systems, we directly couple light emitted
from the first \ac{SHG} stage into the second one. Both stages reside
on a common breadboard enhancing portability. The \ac{RD} system is
spatially split: Output of the \ac{LBO} resonator is fed into the
\ac{BBO} resonator using a single-mode optical waveguide. Whereas the
first scheme provides high effective conversion efficiencies, it can be
quite subtle to re-adjust both \ac{SHG} resonators without touching
alignment of the output \ac{UV} beam. The second scheme does not
feature maximum conversion efficiencies due to losses at the
waveguide's incoupler (we achieve incoupling efficiencies about
$85\%$), but it can more easily be re-adjusted. In addition, it is
more flexible as we can couple light from other laser sources at
$560\nanometer$ (such as a dye laser) into the \ac{BBO} cavity without
many efforts---which is very handy should one fibre laser system
fail. We actually do use such a system for generating the beam
required for photo-ionisation, see \autoref{sec:photo_laser}.

\section{Beamlines}
\label{sec:beamline}

\bfig
\includegraphics{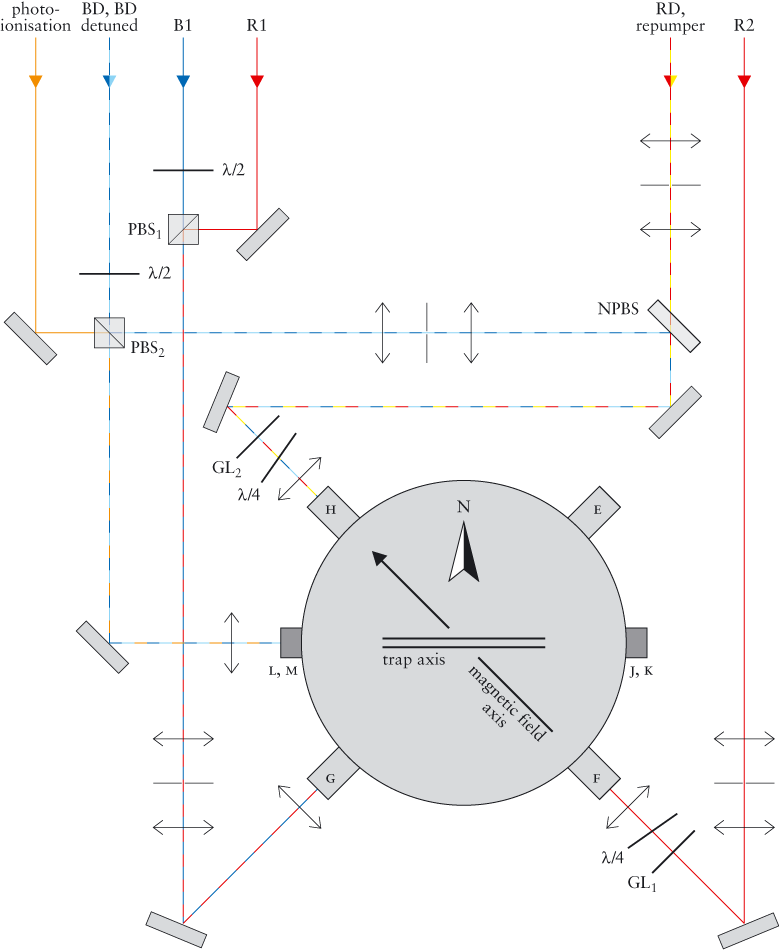}
\caption{Beamline setup used for experimentation. All input beams
(upper part of the figure) are s-polarised. Legend: PBS~= polarising beam
splitter, NPBS~=non-polarising beam splitter, $\text{GL}_i$ = Glan
Laser polariser.}
\label{fig:beamline}
\efig

An overview of the beamline setup is given in
\autoref{fig:beamline}. All initial beams (upper part of the figure)
leave their respective \ac{SHG} resonators with s-polarisation,
i.\,e. their electric field vector is perpendicular to the optical
table surface. For our first experiments we will consider the
following beams:
\bitem
\item Photo-ionising beam at $285\nanometer$. It enters the chamber
exclusively via port~\ac{L} (the upper port, \ac{M} is the lower
port).
\item \ac{BD} and \ac{BD} detuned beams. Using the $\lambda/2$
waveplate in combination with the polarising beam splitter
$\text{PBS}_2$ their power can be directed either towards port~\ac{L}
(p-polarised) or towards port~\ac{H} ($\sigma^+$-polarised due to the
$\lambda/4$ waveplate in front of port~\ac{H}) or a combination of
both.
\item \ac{RD} and repumper beams. They are superposed on the \ac{BD}
and \ac{BD} detuned beams at the non-polarising beam splitter
\ac{NPBS} which retains the s-polarisation of all beams. Therefore,
they also enter the chamber $\sigma^+$-polarised via port~\ac{H}.
\item Raman beam B$1$. It is p-polarised using a $\lambda/2$ waveplate
and enters the chamber via port~\ac{G}. Therefore, it is
$\pi$-polarised with respect to the magnetic field axis.
\item Raman beam R$1$. R$1$ is superposed on B$1$ at the polarising
beam splitter $\text{PBS}_1$, thus also directed towards
port~\ac{G}.
\item Raman beam R$2$. It is $\sigma^-$-polarised by the $\lambda/4$
waveplate in front of port~\ac{F}. However, in the framework of the
magnetic field axis, it is $\sigma^+$-polarised.
\eitem
Note the deviation from the perfect $90\degree$ angle for the
$\lambda/4$ waveplates. Although specified for $280\nanometer$---the
wavelength used for all the beams passing through---these waveplates
have to be inclined in order to maximise their
efficiency\footnote{Tolerances of $5\degree$ lie within the specifications
of the manufacturer (B.\,Halle, Berlin)}.

The combination of beams entering the chamber via port~\ac{L} is used for
photo-ionising magnesium atoms and loading the respective ions into
the trap. It thus points to the centre of the loading region. All
other beams are focused into the centre of the experimentation
region.

All beams entering the chamber are focused by a ``final'' lens (focal
length $f = 200\millimeter$) so that the focal point lies in the
middle of the trap volume. That way intensity as well as spatial
resolution is increased while scattered light causing an increased
background signal is reduced at the same time. Furthermore, we use
these lenses to position the respective beams inside the trap volume
with an accuracy of $\pm3\micrometer$.

Wherever possible, we ``housed'' the beams with transparent tubes so
as to minimise \index{beam steering} beam steering and phase
fluctuation effects caused by fluctuations (either temporal or
spatial) of the air's refractive index. Such fluctuations are normally
omnipresent in the lab: Various thermal sources will heat the air in
their vicinity thus causing air turbulences. The same is true for
people passing by close to the optical tables: They will generate wake
vortices of the air behind them. The transparent housings protect the
air inside from these fluctuations introduced from the outside.

In order to further still the air on the optical tables, we house the
complete optical apparatus with flow boxes on the top and transparent
windows at the sides. The flow boxes will provide a laminar flow of
dust-filtered air as well as a slight overpressure inside the
housing. Dust particles still inside the housing should be swept
outside that way.

\subsection{Photo-ionising beam}
\label{sec:photo_laser}

$570\nanometer$ light from our dye laser system is coupled into the
respective resonator by means of an optical waveguide. We attain
waveguide coupling efficiencies, i.\,e. optical power after the
waveguide in relation to the power before the waveguide, of around
$85\%$. There are no switching elements in the beamline---we simply
block the photo-ionising beam by means of a beamdump once enough ions
have been loaded. \hyperref[tab:photo]{Tab.~\ref*{tab:photo}} lists the relevant beam
parameters.

\bt
\btab{ll}\toprule
operating power                                               & $0.5\mW$\,--\,$2.0\mW$ \\
waist at trap centre                                          & $\sim 70\micrometer$ \\
frequency (${}^{24}\text{Mg} \rightarrow {}^{24}\text{Mg}^+$) & $1.050810699 \cdot 10^{15}\Hz$ \\
frequency (${}^{25}\text{Mg} \rightarrow {}^{25}\text{Mg}^+$) & $1.050811443 \cdot 10^{15}\Hz$ \\ \bottomrule
\etab
\caption{Characteristics of the photo-ionising laser beam.}
\label{tab:photo}
\et

\subsection{Raman beams}
\label{sec:raman}

\begin{table}[b]\begin{center}
\btab{lcc}\toprule
\multirow{2}*{Raman beam} & \multirow{2}*{max. power [mW]} & waist at trap \\
                          &                                & centre [\unitsignonly{\micrometer}] \\
\cmidrule(r){1--1}
\cmidrule(lr){2--2}
\cmidrule(l){3--3}
B$1$                      & 1.2                            & 33 \\
R$1$                      & 48                             & 35 \\
R$2$                      & 48                             & 35 \\ \bottomrule
\etab
\caption{Raman beam parameters.}
\label{tab:raman}
\et

\begin{sidewaysfigure}\centering
\includegraphics{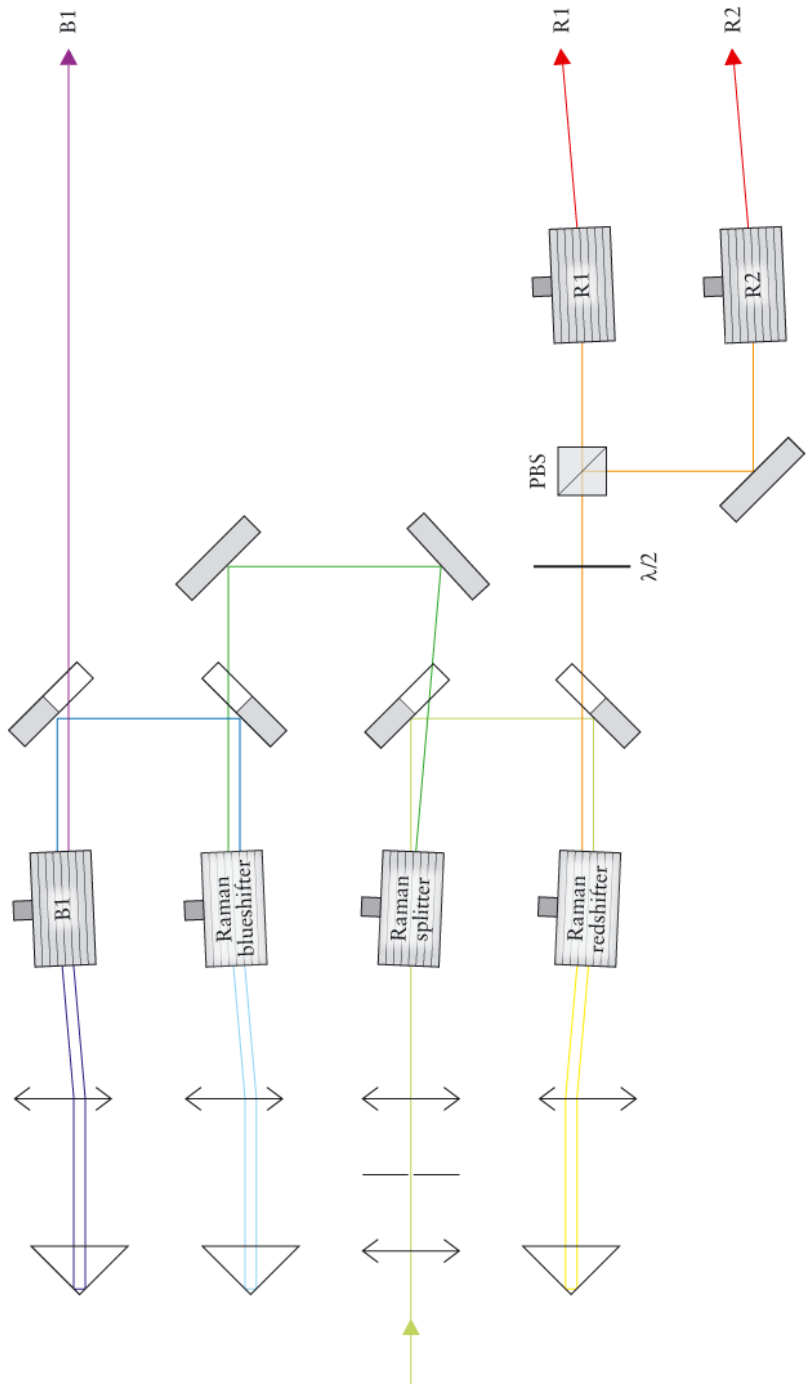}
\caption{Setup used to generate the three Raman beams. Legend: PBS~=
polarising beam splitter. Note that the beam offsets induced by the
double-pass prisms actually occur in the vertical direction
(perpendicular to the drawing plane). This avoids a parallel offset when
an \protect\ac{AOM}'s frequency is altered.}
\label{fig:raman}
\end{sidewaysfigure}

The Raman beam setup is shown in \autoref{fig:raman}. From an initial
$325\mW$ \ac{UV} beam, we generate three individual beams that can be
switched on and off and controlled with respect to their frequency,
phase and amplitude using acousto-optical modulators~(\ac{AOM}s): A
blueshifted B$1$ beam, and two redshifted beams named R$1$ and R$2$. The
frequency difference between the blueshifted and redshifted beams can
be precisely adjusted using the driver frequency of the Raman
blueshifter. This is actually not an acousto-optical modulator
(\ac{AOM}) but an \index{acousto-optical deflector} acousto-optical
\textit{deflector} featuring constant deflection
efficiencies~($\pm5\%$) over a wide range of driving frequencies
($200\MHz$\,--\,$240\MHz$). With the other \ac{AOM}s operating at
$220\MHz$ we can thus tune the frequency difference between $1720\MHz$
and $1800\MHz$. The driving powers for the B$1$, R$1$ and R$2$
\ac{AOM}s determine their diffraction efficiencies and thus optical
powers of the respective beams which can be varied from $0\%$ to about
$85\%$ of the initial optical power. The $\lambda/2$ plate is used to
additionally adjust the relative intensities of the R$1$ and R$2$
before starting experimentation.

\hyperref[tab:raman]{Tab.~\ref*{tab:raman}} contains a summary of the utilised beam
parameters. For our experiments, only the frequency difference of B$1$
and R$1$\,/\,R$2$ matters. As in addition all Raman beams couple to a
virtual level, which is blue-detuned by at least $80\GHz$ with respect
to the $3P_{3/2}$ level, frequency drifts of $\pm20\MHz$ induced by
changes of the ambient temperature ($\pm2\celsius$) remain negligible.

Although the beam waists are small compared to the dimensions of our
ion trap, we do observe a considerable amount of stray light when the
beams shine into the trap volume with many milliwatts of power.
However, these beams will not be enabled during detection; thus
contrast of our measurements is not affected.

\subsection{Blue Doppler beams}

The \ac{BD} beam is detuned by half the linewidth ($\Delta =
\Gamma/2$) of the \ac{BD} transition and attenuated to an intensity of
$I = (2/3)I_\text{sat}$ in order to maximise the Doppler cooling
efficiency. As such, it would actually suffice in order to Doppler
cool the ions and conduct detection in quantum simulation
experiments. However, due to collision processes with the residual gas
in the vacuum chamber ions stored in the trap will from time to time
gain so much velocity that the \ac{BD} beam alone will not be able to
re-cool them on a millisecond timescale before the start of each
experiment. This is where the detuned Blue Doppler beam (\ac{BD}
detuned) comes into play. \index{Blue Doppler detuned} With a
frequency that is redshifted $660\MHz$ with respect to the \ac{BD}
frequency and an \index{Blue Doppler detuned!intensity} intensity of
$I \approx 6I_\text{sat}$, it will even cool hotter ions (see
\autoref{tab:bd}).

\begin{table}[b]\begin{center}
\btab{lcccc}

\toprule
\multirow{2}*{\ac{BD} beam} & max.       & operating                         & waist at trap                       & frequency      \\
                            & power [mW] & power [\unitsignonly{\microwatt}] & centre [\unitsignonly{\micrometer}] & [$10^{15}\Hz$] \\
\cmidrule(r){1--1}
\cmidrule(lr){2--2}
\cmidrule(lr){3--3}
\cmidrule(lr){4--4}
\cmidrule(l){5--5}
\ac{BD}                     & $2.0$      & $5.5$                             & $47$                                & $1.072085218$  \\
\ac{BD} detuned             & $2.0$      & $100$                             & $47$                                & $1.072084558$  \\ \bottomrule
\etab
\caption{\protect\ac{BD} beam parameters calibrated for ${}^{25}\text{Mg}^+$.}
\label{tab:bd}
\et

\begin{sidewaysfigure}\centering
\includegraphics{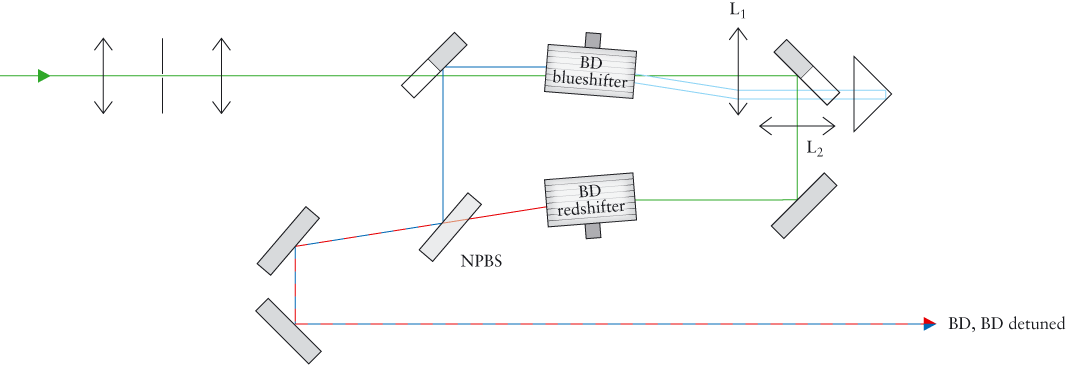} \\
\includegraphics{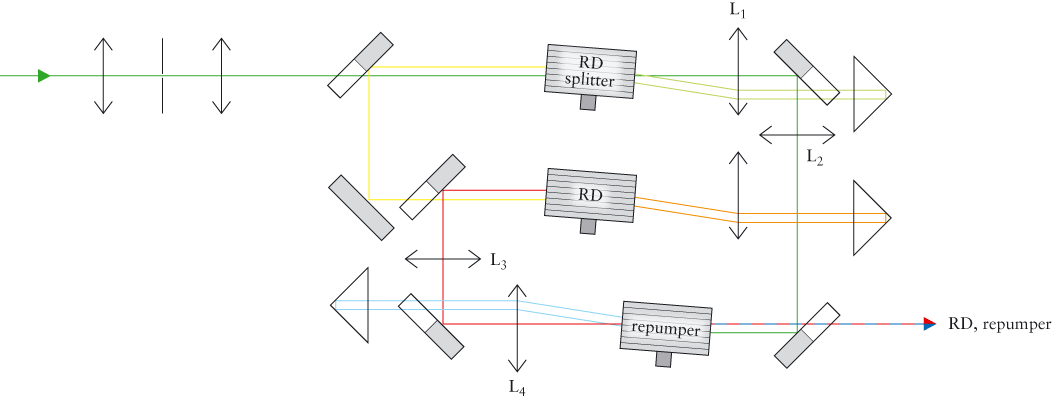}
\caption{Setups used to generate the two Blue Doppler and the two Red
Doppler beams. Legend: NPBS~= non-polarising beam splitter. Note that the beam offsets
induced by the double-pass prisms actually occur in the vertical
direction (perpendicular to the drawing plane). This avoids a parallel
offset when an \protect\ac{AOM}'s frequency is altered.}
\label{fig:bdrd}
\end{sidewaysfigure}

\hyperref[fig:bdrd]{Fig.~\ref*{fig:bdrd}} shows the setup for the two Blue Doppler
beams. They are created from an initial $90\mW$ of \ac{UV} beam
power. Note that lens $\text{L}_2$ is necessary to re-collimate the
undeflected beam, as it passes through lens $\text{L}_1$ only once. In
practice, the undeflected beam's waist can be adjusted by altering the
focal length ratio of $\text{L}_1$ and $\text{L}_2$.

The \ac{AOM}s introduce a frequency difference of $660\MHz$ between
the two generated beams. We also use the \ac{BD} and \ac{BD} detuned
\ac{AOM}s to switch the respective beams on and off during
experimentation. By manually tuning variable \ac{RF} attenuators in
the \ac{AOM} driving circuits we can adjust the optical powers of the
two beams. It is sufficient to conduct these adjustments once before
experimentation starts.

\subsection{Red Doppler beams}

The Red Doppler beam generation setup is shown in \autoref{fig:bdrd}.
We generate a redshifted Red Doppler (\ac{RD}) and a blueshifted
repumper beam, separated by a frequency difference of $1789\MHz$. Note
that the repumper \ac{AOM} is operated at $454\MHz$ to achieve
this. Initial \ac{UV} beam power is $90\mW$. In analogy to the \ac{BD}
setup, we use two re-collimating lenses $\text{L}_2$ and $\text{L}_3$
to correct for the non-doublepassed beams.

Both the RD and repumper beams can be switched on and off using the
respective \ac{AOM}s. As in the case of the \ac{BD} setup, we adjust
their optical powers once before starting an experiment using manual
variable attenuators.

\bt
\btab{lcccc}\toprule
\multirow{2}*{\ac{RD} beam} & max.       & operating                         & waist at trap                       & frequency      \\
                            & power [mW] & power [\unitsignonly{\microwatt}] & centre [\unitsignonly{\micrometer}] & [$10^{15}\Hz$] \\
\cmidrule(r){1--1}
\cmidrule(lr){2--2}
\cmidrule(lr){3--3}
\cmidrule(lr){4--4}
\cmidrule(l){5--5}
\ac{RD}                     & $1.2$      & $5$                               & $45$                                & $1.069338736$  \\
repumper                    & $1.2$      & $5$                               & $78$                                & $1.069340525$  \\ \bottomrule
\etab
\caption{\protect\ac{RD} beam parameters calibrated for ${}^{25}\text{Mg}^+$.}
\label{tab:rd}
\et

\section{Doppler-free iodine spectroscopy}
\label{sec:iodine_spectroscopy}

Whereas we can measure a laser's wavelength using a wavemeter
(accuracies of $10^{-7}$\,--\,$10^{-8}$) we still have to
determine---and in some cases lock---the frequency of our laser systems with a
resolution of $\pm1\MHz$. This is achieved using Doppler-free iodine
spectroscopy. Transition lines of iodine are well-known
\cite{iodine_atlas} and can be calculated to a high precision using
additional data on hyperfine splittings, which makes them ideal for use
as a frequency standard. We use Toptica's
IodineSpec\footnote{\href{http://www.toptica.com/}{http://www.toptica.com/}} for this purpose.

We use saturation spectroscopy in conjunction with a signal-enhancing
lock-in amplifier whose modulation signal drives an \ac{AOM}
\cite{haensch:saturation}. Other spectroscopic techniques are
available, e.\,g. polarisation spectroscopy
\cite{haensch:polarisation, pearman:polarisation} which may reduce the
experimental effort for some purposes. In our setup however, an
\ac{AOM} is necessary anyway to provide the required frequency shifts,
which is why we stick with saturation spectroscopy.

\subsection{Principles of Doppler-free saturation spectroscopy}

The simplest method of saturation spectroscopy uses two beams, one
reference beam that freely propagates (through vacuum or air) and a
probe beam which passes through the material and is attenuated on its
way according to \autoeqref{eq:attn} where $i = 1$ and $I_\text{total}
= I_1$. Subtraction of the respective intensities will yield a measure
of the transition strength, i.\,e. of how efficiently the transition
could be driven. As a function of the beams' frequency this will give
a spectroscopic signal.

\bfig
\includegraphics{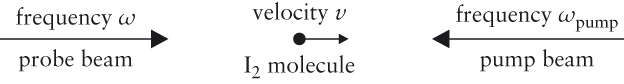}
\caption{Schematic of Doppler-free iodine spectroscopy.}
\label{fig:doppler-free_labels}
\efig

If the optical medium used for spectroscopic measurements is a gas (as
is the case with iodine), linewidths of the optical transitions will
be considerably Doppler-broadened~($\sim1\GHz$) at room temperatures
decreasing the frequency resolution. Dopp\-ler-free spectroscopy
overcomes this limitation by employing three beams in total:
\benu
\item the reference beam (frequency $\omega$, intensity $I$) which
freely propagates through the iodine cell,
\item the probe beam (frequency $\omega$, intensity $I$ as well)
which also propagates through the iodine cell, most commonly parallel
to the reference beam, but is superimposed by
\item the couterpropagating pump beam (frequency
$\omega_\text{pump}$, intensity $I_\text{pump}$). This one features a
very strong intensity suited to optically bleach the medium.
\eenu

\bfig
\includegraphics{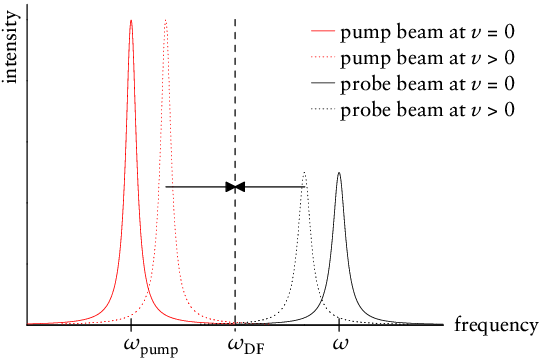}
\caption{Intensity profiles as seen by an iodine molecule. Solid
curves: $v = 0$. Dotted curves: $v > 0$.}
\label{fig:doppler-free}
\efig

\noindent This setup is illustrated in figs.~\ref{fig:doppler-free_labels} and
\ref{fig:doppler-free}. An iodine molecule moving with velocity $v$ is
hit by the probe beam and the pump beam from opposite directions. If
the molecule were at rest, it would ``see'' a probe beam frequency of
$\omega$ and a pump beam frequency of $\omega_\text{pump}$. Now that
the molecule is moving at velocity $v$, the probe beam frequency is
redshifted in the frame of the molecule (dotted curve in
\autoref{fig:doppler-free}). Analogously, the pump beam frequency appears
blueshifted. At some particular velocity $v_\text{DF}$, associated
with the frequency $\omega_\text{DF}$, the intensity profiles of
probe and pump beams overlap. Only then will the pump beam ``talk''
to the same velocity class of molecules as the probe beam. If in
addition $\omega_\text{DF}$ happens to be the frequency of an iodine
transition, attenuation of the probe beam will be decreased according
to \autoeqref{eq:attn}. In other words: A decrease in the attenuation of
the probe beam will only occur for a special velocity class of
molecules; thus Doppler broadening is circumvented. As attenuation of
the (unpumped) reference beam is never decreased, we will observe a
non-zero difference in the intensities of the probe and reference
beams if $\omega_\text{DF}$ matches the frequency of an iodine transition.

As indicated in \autoref{fig:doppler-free} $\omega_\text{DF}$ lies in
the middle between $\omega$ and $\omega_\text{pump}$:
\be
\omega_\text{DF} = \frac{\omega + \omega_\text{pump}}{2} \text{.}
\label{eq:omega_df}
\ee
This is because as a first approximation the blue Doppler shift of the
probe beam equals the red Doppler shift of the pump beam.

\subsection{Experimental setup}

\bfig
\includegraphics{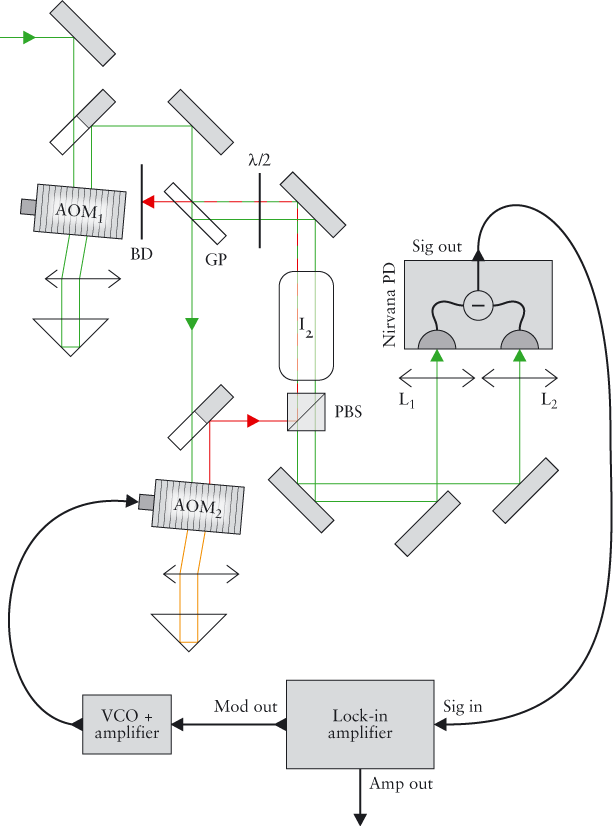}
\caption{Doppler-free iodine saturation spectroscopy setup. Legend:
$\text{AOM}_i$~= acousto-optic modulator, GP~= glass plate, PBS~=
polarising beam splitter, BD~= beam dump, VCO~= voltage-controlled
oscillator. Note that the beam offsets induced by the double-pass
prisms actually occur in the vertical direction (perpendicular to the
drawing plane). This avoids a parallel offset when an
\protect\ac{AOM}'s frequency is altered.}
\label{fig:iodine}
\efig

The experimental setup for iodine spectroscopising the once
frequency-doubled \ac{BD} laser (at $560\nanometer$) is shown in
\autoref{fig:iodine}. We operate $\text{AOM}_1$ at a frequency of
$210.5\MHz$ and $\text{AOM}_2$ at $80\MHz$. Probe and reference beams
are coloured green, the pump beam is painted red. In the end, both the
probe and the reference beam hit a \index{Nirvana photodiode}
differential photodetector (Newport Nirvana) which subtracts the
individually measured powers from each other. This type of
photodetector has one additional feature that greatly facilitates
operation: Low-frequency differences in the two signals are
automatically evened out. So even if powers in the two beams differ by
some \ac{DC} offset, the detector output signal is still zero.

The output signal of the photodetector is in general too weak to be
used directly. Furthermore, it has to be differentiated so as to
provide a zero-crossing which we use for locking (see below). We
achieve both these aims by phase-sensitive lock-in techniques. For
this purpose, we wobble the driving frequency of $\text{AOM}_2$---and
thus the pump beam frequency---with $10\kHz$. \index{lock-in
amplifier} The lock-in amplifier will phase-sensitively filter the
amplitude of a $10\kHz$ modulation in the photodetected signal thereby
enhancing the signal-to-noise ratio by several magnitudes.

Wobbling the pump beam frequency and thus $\omega_\text{DF}$
corresponds to a wiggle on the horizontal axis of
\autoref{fig:doppler-free}. As long as the centre frequency
$\left<\omega_\text{DF}\right>$ lies at the rising edge of an iodine
transition line, the Nirvana photodetector will accordingly measure a
$10\kHz$ modulation of the incoming probe intensity, which causes the
lock-in amplifier to output a negative voltage. If the centre
frequency lies at the falling edge, the intensity modulation of the
probe beam is phase-shifted by $\pi$, which causes the lock-in
amplifier to output a positive voltage. In between, when
$\left<\omega_\text{DF}\right>$ exactly matches the transition
frequency, the lock-in amplifier will output a voltage of zero. An
image of the curves obtained that way is shown in
\autoref{fig:iodine_signal}.

\bfig
\includegraphics{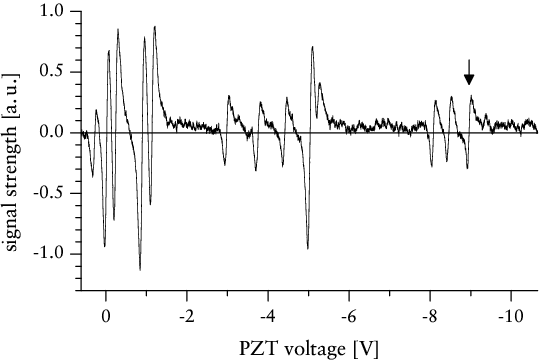}
\caption{Doppler-free iodine spectroscopic signal obtained from the
\protect\ac{BD} laser after the first \protect\ac{SHG} stage. The signal has been
averaged over four acquisitions with an initial beam power of
$5.2\mW$. The arrow indicates the line that the laser is locked
to. Note that the small overall positive offset ensures that the laser
will run to redder frequencies in case the lock is lost. That way,
heating the ions is prevented.}
\label{fig:iodine_signal}
\efig

The iodine spectroscopy setups attached to the dye laser and the
\ac{RD} laser are in principle equal to that of the \ac{BD}
laser. They differ in their particular setup of \ac{AOM}s: Firstly,
they both lack an initial frequency shifter $\text{AOM}_1$. Secondly,
$\text{AOM}_2$ in the dye laser iodine setup modulates both the
pump and probe beams thereby causing a Doppler free spectroscopic
frequency that is shifted by $160\GHz$ instead of only $80\GHz$. In
the \ac{RD} setup $\text{AOM}_2$ shifts to bluer frequencies by $91\MHz$.

\index{dispersive signal} The dispersive spectroscopic signal can be
used to comfortably lock the laser to one of the signal's
zero-crossings, i.\,e. one of the iodine transition frequencies. The
feedback loop of such a regulation simply consists of an op-amp adder
circuit (to be able to roughly adjust the frequency) combined with an
op-amp integrator circuit (to adjust the regulation feedback
speed). If the signal is below zero (left handside of a
zero-crossing), the laser's frequency is increased and vice versa,
thus locking the laser to the specific iodine transition.

During experimentation we iodine-lock the \ac{BD} and \ac{RD}
lasers. The dye laser feeding the photo-ionisation \ac{SHG} stage also
has an iodine apparatus attached which we use to adjust its
frequency. We do however not lock to this signal as frequency
requirements for the photo-ionisation process are not as strict as for
the \ac{BD} or \ac{RD} beams. \hyperref[tab:iodine]{Tab.~\ref*{tab:iodine}} gives an overview
over the parameters used.

\bt
\btab{lcccc}\toprule
\multirow{2}*{laser source}       & initial beam         & iodine                      & signal/noise          \\
                                  & power [mW]           & frequency [THz]             & ratio                 \\
\cmidrule(r){1--1}
\cmidrule(lr){2--2}
\cmidrule(lr){3--3}
\cmidrule(lr){4--4}
\cmidrule(l){5--5}
\ac{BD}                           & $5.2$                & $536.041888$                & $1.75:1$              \\
\ac{RD}                           & $12.0$               & $534.669899$                & $100:1$               \\
dye laser                         & \multirow{2}*{$7.9$} & \multirow{2}*{$525.405609$} & \multirow{2}*{$10:1$} \\
adjusted for ${}^{24}\text{Mg}^+$ &                      &                             &                       \\
dye laser                         & \multirow{2}*{$7.9$} & \multirow{2}*{$525.405757$} & \multirow{2}*{$10:1$} \\
adjusted for ${}^{25}\text{Mg}^+$ &                      &                             &                       \\ \bottomrule
\etab
\caption{Parameters used for the iodine spectroscopy setups throughout
experimentation. Unless noted otherwise, frequencies are calibrated
for ${}^{25}\text{Mg}^+$. Iodine frequencies were extracted from Toptica's
IodineSpec. The discrepancy between the dye laser's iodine frequencies
listed here and the frequencies listed in \autoref{tab:photo} is
due to the Doppler-shift of the not yet ionised magnesium atoms, see
\autoref{sec:ionisation}.}
\label{tab:iodine}
\et

Using the iodine multiplet between $536.041859\THz$ and $536.041888\THz$
we can lock the \ac{BD} laser to frequencies between $536.042299\THz$
and $536.042408\THz$ by appropriately adjusting the $\text{AOM}_1$
shift. As can be seen in \autoref{fig:iodine}, only the pump beam is
frequency-shifted twice. \hyperref[eq:omega_df]{Eq.~(\ref*{eq:omega_df})} shows that we could
span a greater frequency range for the laser lock if the
probe/reference beams were frequency-shifted as well, i.\,e. if the
they also passed through $\text{AOM}_2$. We however observed a
\textit{permanent} $10\kHz$ signature of the probe beam for this setup,
leading to an overall offset in the lock-in amplifier output which
prevented locking due to the lack of proper zero-crossings. We assume
that wobbling the frequency of $\text{AOM}_2$ introduced some steering
between the probe and the pump beams. The frequency range spanned by
the setup illustrated in \autoref{fig:iodine} will however suffice
for our applications.

\chapter{Visualisation and data acquisition}
\label{ch:detection}

The last step of every quantum simulation experiment is reading out
the final qubit state. We have to determine whether the ion in
question is bright (fluoresces) or dark when we shine onto it using
the \ac{BD} laser beam.

\section{Optics}

All detection equipment uses the vacuum chamber's port \ac{A}, which
is closed by a coated sapphire crystal window. It is required by the
attached objective for optimum imaging quality. At $280\nanometer$ it
transmits about $99\%$ of the incident intensity.

\index{objective} We use a specially designed high-quality F/$1$
objective (aperture angle $55.6\degree$, focal length
$19.3\millimeter$) with a magnification factor of $50$ (B.\,Halle,
Berlin). That way, we should be able to resolve ion-to-ion distances
of down to $0.35\micrometer$ (see \autoref{fig:objective_quality}).
Actually resolvable ion-to-ion distances depend on the focal setting
of the objective, we achieved a minimum of $2\micrometer$, see
\autoref{fig:objective_focal}.

\bfig
\subfloat[Ion centred below objective.]{
\includegraphics{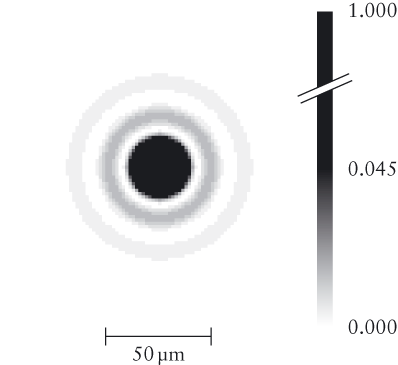}}
\hfill
\subfloat[Ion off by $50\micrometer$ from central position.]{
\includegraphics{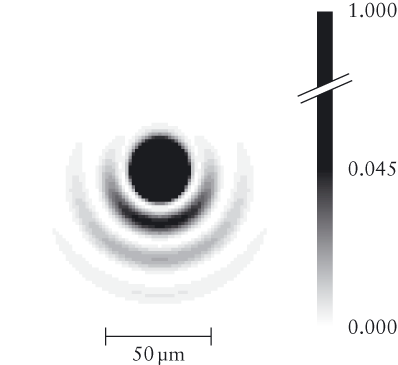}}
\caption{Image of a point-like fluorescing ion as it appears at the
magnifying end of the objective. What is actually drawn here are the
point spread functions (\protect\ac{PSF}) of the objective calculated
using the software package \protect\ac{ZEMAX}. The image contrast has
been greatly increased---all values above $0.045$ are represented by
black colour.}
\label{fig:objective_quality}
\efig

\bfig
\includegraphics{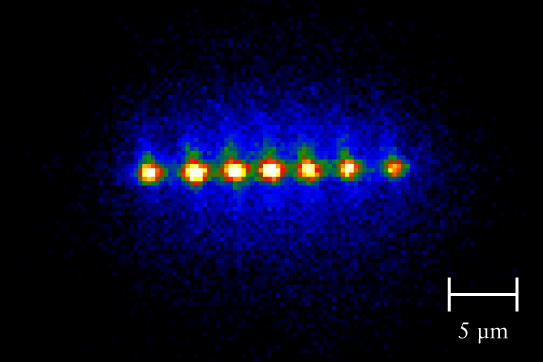}
\caption{Image of a chain of seven ions at an axial confinement of
$\approx 2\pi \cdot 2\MHz$. The spatial resolution for one ion can be
estimated at $2\micrometer$.}
\label{fig:objective_focal}
\efig

\bfig
\includegraphics{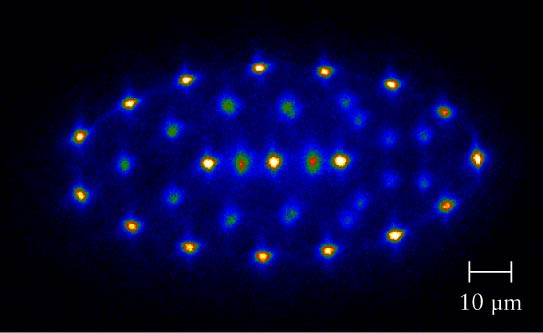}
\caption{Cigar-shaped Coulomb crystal of approx.~40 ions. The great
ion-to-ion distances are due to a relatively weak radial confinements
($\omega_\text{radial} \approx 2\pi \cdot 1\MHz$, $\omega_\text{axial}
\approx 2\pi \cdot 100\kHz$.}
\label{fig:ion_crystal}
\efig

\bfig
\subfloat[$\omega_\text{axial} < \omega_\text{radial}$.]{
  \includegraphics{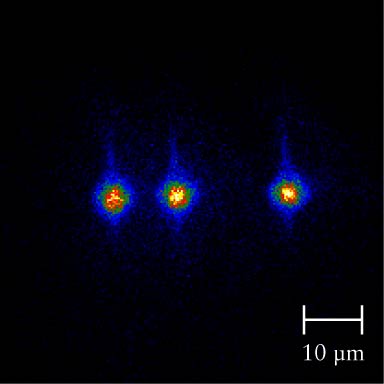}}
\hfill
\subfloat[$\omega_\text{axial} > \omega_\text{radial}$.]{
  \includegraphics{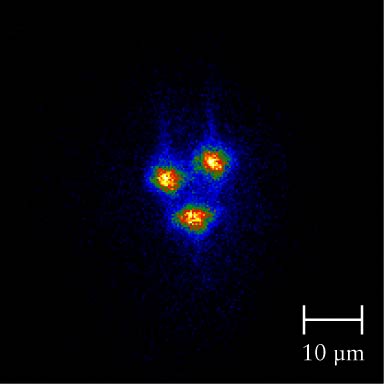}}
\caption{Coulomb crystal of four ions including one ``dark'' ion,
which is most probably magnesium hydride $\text{MgH}^+$ (resulting
from a collision of the magnesium ion with hydrogen from the residual
gas).}
\label{fig:dark_ion}
\efig

Fluorescing magnesium ions will emit $\sigma^\pm$ polarised light in
the cycling \ac{BD} transition, see \autoref{sec:state_detection}.
\index{emission characteristics} The emission characteristics is
determined by
\be
P(\theta,\phi) = 1 + \cos^2(\theta)
\ee
where $\theta$ and $\phi$ are the angles of spherical
coordinates. $\theta = 0$ equals the direction of the applied magnetic
quantisation axis so that the direction towards the detection port
\ac{A} of our chamber is given by $\theta = \pi/2$ (see
\autoref{fig:beamline}). Our objective thus catches $4.7\%$ of the
photons scattered. From this point of view it would have been
advantageous to observe the ions at an angle $\theta$ close to
zero. However, such a setup would block the horizontal laser beams
intersecting the trap axis under $45\degree$.

After having passed through the objective the beams are reflected on a
$45\degree$ mirror with a reflectivity of $99\%$ and enter
\index{detection switch box} a box with our detection instruments---a
\ac{CCD} camera and a photomultiplier (\ac{PMT}). By means of a
motor-driven flip mirror ($\text{reflectivity}>99\%$) we can direct
the light either onto the camera or onto the photomultiplier. See
\autoref{fig:cambox} for the assembly of the detection box. The whole
beamline between the objective and our detection instruments is
shielded against ambient light sources by either non-transparent metal
housings or---in the case of the detection box---by laser safety
curtain wrappings.

\bfig
\subfloat[Top view.]{
\includegraphics{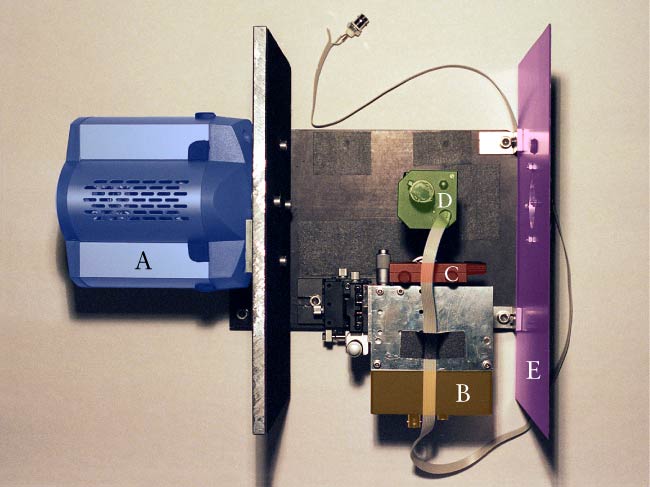}} \\
\subfloat[Front view.]{
\includegraphics{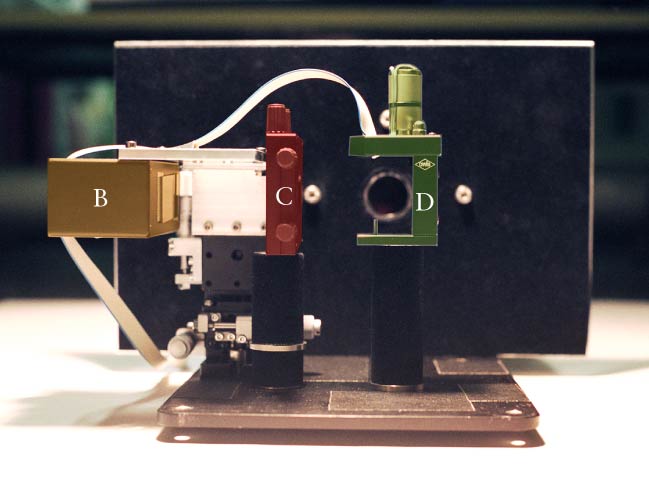}}
\caption{Assembly of the detection box consisting of camera (A),
photomultiplier (B), slit aperture (C), flip mirror (D), and entrance
aperture (E). All parts have been highlighted with distinct
colours. The photomultiplier is mounted on a linear translation stage
for optimum positioning.}
\label{fig:cambox}
\efig

Both the camera and the \ac{PMT} are equipped with glass filters
(Schott $\text{UG5}$, thickness $2\millimeter$) opaque at visible wavelengths
but transparent (transmission $88\%$) at $280\nanometer$. Altogether
we must consider photon losses of the order of $14\%$ on the way from
the ions to our detection instruments. \hyperref[tab:count_rate]{Tab.~\ref*{tab:count_rate}}
estimates the expected photon count rate.

\bt
\btab{ll}\toprule
fluorescent scattering rate & $33.8\MHz$ \\
objective captures $4.7\%$  & $\leadsto 1.59\MHz$ \\
$14\%$ of optical losses    & $\leadsto 1.37\MHz$ \\ \bottomrule
\etab
\caption{Calculation of the expected count rate behind all
optics. The fluorescent scattering rate was calculated using
\autoeqref{eq:bloch_steady} for a detuning of $\Delta = \Gamma/2 =
21.5\MHz$ and an intensity of $I = (2/3) I_\text{sat} =
167\mWpersqcm$ which equals a Rabi frequency
of $\Omega = \Gamma/\sqrt{3}$. These parameters for the
\protect\ac{BD} detection laser are used throughout during
experimentation.}
\label{tab:count_rate}
\et

\section{\texorpdfstring{\protect\ac{PMT}}{PMT}}
\index{PMT@\ac{PMT}|see{photomultiplier}}
\index{photomultiplier}

The most straightforward way of precisely measuring weak light sources
is to use a photomultiplier \ac{PMT}. We use a Hamamatsu~H8259 with a
quantum efficiency of about $14\%$ at $280\nanometer$ and a pulse-pair
resolution of $35\nanosecond$. Its photoactive window has a size
of $4 \times 20 \millimeter$. As we want to make sure that only a
minimal amount of stray light hits the \ac{PMT}, we have placed a
manual variable slit aperture in front of it. In order to prevent
saturation of particular regions on the photoactive window, we defocus
the ions' image thus distributing the intensity over a greater
region. Using standard operating intensities for the \ac{BD} detection
beam (see \autoref{tab:bd}) we get count rates of $210\kHz$ at a
\index{PMT@\ac{PMT}!signal-to-background} signal\,/\,background ratio
of about $1000:1$. Taking into account the \ac{PMT}'s quantum
efficiency this result fits the expected count rates
(\autoref{tab:count_rate}). Normally we would use an ``exposure
time'', i.\,e. the timespan that we count the \ac{PMT} clicks, of
$20\microsecond$ to $50\microsecond$, acquiring between $4$ and $10$
photons on average.

\section{\texorpdfstring{\protect\ac{CCD}}{CCD} camera}
\label{sec:camera}
\index{CCD camera@\ac{CCD} camera}

\bt
\btab{ll}\toprule
horizontal readout speed & $5\MHz$ \\
vertical shift speed     & $0.4\microsecond$ \\
vertical shift amplitude & $+2$ \\
\ac{EMCCD} gain          & $650$ \\
pre-amplifier gain       & $2.4$ \\
frame transfer mode      & \textit{off} \\ \bottomrule
\etab
\caption{Camera readout parameters used during experimentation.}
\label{tab:cam_params}
\et

\begin{table}[b]\begin{center}
\btab{ll}\toprule
count rate before the camera (\autoref{tab:count_rate}) & $1.37\MHz$ \\
$8\%$ losses through uncoated entrance window           & $\leadsto 1.26\MHz$ \\
quantum efficiency of $34\%$                            & $\leadsto 427\kHz$ \\
\ac{EMCCD} gain of $650$                                & $\leadsto 278\MHz$ \\
$23$ electrons per count (estimated)                    & $\leadsto 12.1\MHz$ \\
ion's image encompasses $20$ pixels                     & $\leadsto 603\kHz/\text{pixel}$ \\ \bottomrule
\etab
\caption{Camera count rate to be expected in theory for the parameters
listed in \autoref{tab:cam_params}.}
\label{tab:cam_count_rate}
\et

Using the \ac{CCD} camera as a detection instrument is a more
sophisticated way of observing stored ions as it allows to distinguish
single ions from each other. Our Andor iXon~DV887DCS-UVB has a photoactive
window of $8 \times 8 \millimeter$ which is very handy during the ion
loading process and allows for judgements on imaging quality.
The \ac{CCD} is equipped with an on-chip electron multiplication
facility (which is why the chip is also called an \ac{EMCCD} chip)
that can increase the signal generated by the released photoelectrons
by a factor up to $850$, depending on the actual readout
settings. Relevant readout parameters that we normally use during
experimentation are listed in \autoref{tab:cam_params}. With a
nominal quantum efficiency of $34\%$ at $280\nanometer$ the \ac{CCD}
is supposed to be more sensitive than the
\ac{PMT}. \hyperref[tab:cam_count_rate]{Tab.~\ref*{tab:cam_count_rate}} gives an estimation of the
expected camera count rate. In reality, the noise introduced by
readout or thermal effects might limit the sensitivity of the \ac{CCD}
chip \cite{andor:basics}. We thus measured the signal of a single
fluorescing ion using the parameters of tabs.~\ref{tab:count_rate}
and~\ref{tab:cam_params} for different horizontal readout speeds. The
result is illustrated in \autoref{fig:cam_signal}. At $5\MHz$
horizontal readout speed \index{CCD camera@\ac{CCD}
camera!signal-to-background} the count rate per pixel above ground is
about $75\kHz$ which differs by more than one order of magnitude from
our expected results. At the time of printing of this thesis, we were
trying to figure out what might cause such a low count rate in
collaboration with the manufacturer. In terms of count rate (not
normalized per pixel), the \ac{CCD} still outperforms the
\ac{PMT}---even at very short exposure times.

One drawback of using the \ac{CCD} camera is the inherent delay
between subsequent acquisitions caused by the \ac{CCD} readout
process. Thus, detailed timing information on timescales of
nanoseconds---needed for compensation of micromotion,
\autoref{ch:micromotion} for example---is only available using
the \ac{PMT}.

\bfig
\includegraphics{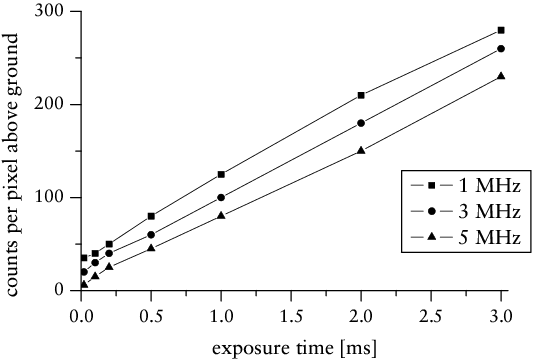}
\caption{Signal above ground of the \protect\ac{CCD} camera for
different horizontal readout speeds and different exposure
times. Intensity for the \protect\ac{BD} detection beam was adjusted
at $2/3$ of the saturation intensity. Note that there is no strict
linear dependence for very short exposure times. We assume that even
if the exposure time was set to zero, the \protect\ac{CCD} chip was
exhibited to photons for a finite duration. This issue was still
being investigated as of printing of this thesis.}
\label{fig:cam_signal}
\efig

\subsection{Noise and gain}

In principle, a \ac{CCD} chip has three sources of noise: thermal
noise (thermal release of photoelectrons), clock-induced charge
(resulting from shifting the electrons across the \ac{CCD} chip too
fast), and readout noise (caused by the electron-counting
sensor). Thermal noise can be reduced very efficiently to values below
$1 \text{e}^-/\text{pixel}/\text{second}$ by cooling the \ac{CCD} chip
to \index{temperature!CCD camera@\ac{CCD} camera} temperatures below
$-25\celsius$. Clock-induced charges may be avoided by lowering the
shift speeds transferring the electrons across the \ac{CCD}
chip. However, this also lowers the achievable frame rate (images per
unit time). Readout noise is an inherent property of the sensor and
cannot be reduced.

We measured the \ac{EMCCD} gain of our camera using the ion gauge
inside our vacuum chamber as a constant light source. For each camera
setting we acquired two images; one with the camera shutter opened and
one with the shutter closed. The signal can then be defined as the
difference in the count rates of the two images. Although \ac{EMCCD}
gain should depend only weakly on the \ac{CCD}'s temperature, we
measured considerably distinct values for different temperatures, see
\autoref{tab:emccd_temperature}. Maximum gain values throughout
improved with decreasing temperatures. Therefore we eventually
water-cooled the camera's Peltier cooling element to $18\degree$ by
connecting it to the chiller of our Verdi~V10 laser. We achieved
temperatures of down to $-90\celsius$, stable operation is however
only possible above $-85\celsius$.

\bt
\btab{lcc}\toprule
horizontal readout speed & max. gain at $-50\celsius$ & max. gain at $-80\celsius$ \\
\cmidrule(r){1--1}
\cmidrule(lr){2--2}
\cmidrule(l){3--3}
$1\MHz$                  & $155$                      & $410$ \\
$3\MHz$                  & $195$                      & $422$ \\
$5\MHz$                  & $173$                      & $647$ \\
$10\MHz$                 & $225$                      & $842$ \\ \bottomrule
\etab
\caption{Maximum \protect\ac{EMCCD} gains for different
\protect\ac{CCD} chip temperatures. Vertical shift speed was set at
$3.4\microsecond$ so as to avoid clock-induced charge noise.}
\label{tab:emccd_temperature}
\et

\subsection{Optimisation of image quality}
\index{objective!image optimisation}

If the objective above the vacuum chamber is misaligned in any way
(tilted or off-positioned), image quality and thus spatial resolution
is decreased. We conducted the following procedure to optimise the
imaging:
\bitem
\item Load one ion into the trap.
\item Defocus the ion's image starting from optimum focusing by
approaching the objective to the ion. This leads to a halo around the
bright spot associated with the ion, see \autoref{fig:defocused1}.
\item Centre the bright spot inside the halo by tilting the objective.
\item Defocus the ion's image starting from optimum focusing by moving
the objective away from the ion. For our objective this leads to a
bright ring, see \autoref{fig:defocused2}.
\item Move the objective perpendicular to the focal plane in order to
make the ring appear equally bright at all positions.
\eitem
On \autoref{fig:defocused2}, one can see an elliptical structure
centred on the circular one. Depending on the amplitude of micromotion
the intensity of the two structures changes: If micromotion is
virtually zero, the ellipse appears darker than the circle. With
increasing micromotion, the ellipse becomes lighter eventually
outshining the circle. The reason for this structure is not completely
understood. It might be related to polarisation of the scattered light
where birefringence of the sapphire crystal window could play a role.

\bfig
\subfloat[Distance between objective and ion is too small.]{
  \includegraphics{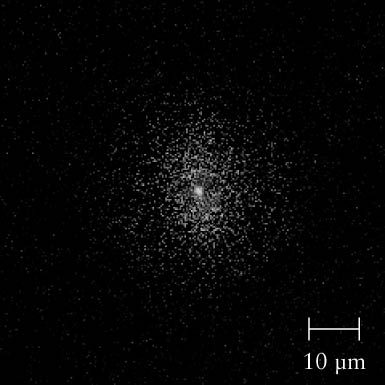}
  \label{fig:defocused1}
}
\hfill
\subfloat[Distance between objective and ion is too great.]{
  \includegraphics{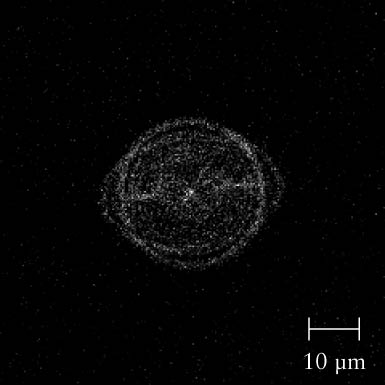}
  \label{fig:defocused2}
}
\caption{Defocused images of ions as they appear using the
\protect\ac{CCD} camera. Exposure time was $100\millisecond$, the
\protect\ac{BD} laser parameters were adjusted according to the
figures of \autoref{tab:bd}.}
\label{fig:defocused}
\efig

\section{Auge software}
\label{sec:auge}
\index{Auge software@\textit{Auge} software|main}

For future experiments involving more than one ion it
will be necessary to determine the state of each ion
independently. Obviously, such a discrimination is not possible using
the \ac{PMT}---it does not deliver any spatial information. For the
sake of spatially discriminating ions, we use the \ac{CCD} camera.

Although Andor bundles its camera with a powerful software for image
acquisition and analysis, this programme does not satisfy our requirements
(sorted by decreasing importance):
\bitem
\item Results of image analysis cannot be transferred to our computer
in charge of the experimental analysis in real-time.
\item Regions of interest, i.\,e. image subregions that are to be
analysed, are rectangular-shaped and can thus not be adapted to the
shape of an ion. This restriction leads to a loss of contrast since
pixels of the background signal will also be evaluated.
\item Regions of interest do not automatically adapt to slightly
changed image positions of the ions (which can be attributed to
thermal effects of the involved optics).
\item Performance of the Andor software is not suitable for the high
acquisition rates (more than $100$ images per second) needed for our
experiments.
\eitem
The author of this thesis has written a camera controlling and image
processing software from scratch in order to overcome these limitations.
This software, called \textit{Auge} for simplicity, runs on an
AMD~Athlon~64~3400+ under Windows. It has been written in C++ using
the standard Win32 API for maximum execution speed. For a schematic of
the application's components and their cooperation refer to
\autoref{app:auge}.

We can identify two modes of operation. The first is an
all-interactive mode actually displaying the camera image on screen.
The user can adjust the camera parameters such as exposure time,
electron multiplication or readout speeds. Furthermore, the software
lets you define elliptical regions of interest that each enclose the
image of one ion, see \autoref{fig:auge}. The second is a
remote-controlled mode where our experimentation software
\textit{Flocke} (see \autoref{sec:interplay}) takes over control
and in turn receives count rates inside the previously defined regions
of interest over the \ac{TCP/IP} network.

Communication between \textit{Flocke} and \textit{Auge} follows a
proprietary protocol called Auge Transfer Protocol (\ac{ATP}). It is a
client\,/\,server protocol where \textit{Auge} acts as the server and
\textit{Flocke} as the client directing commands at \textit{Auge}. See
\autoref{app:atp} for a detailed description.

\begin{figure}[t]\centering
\includegraphics{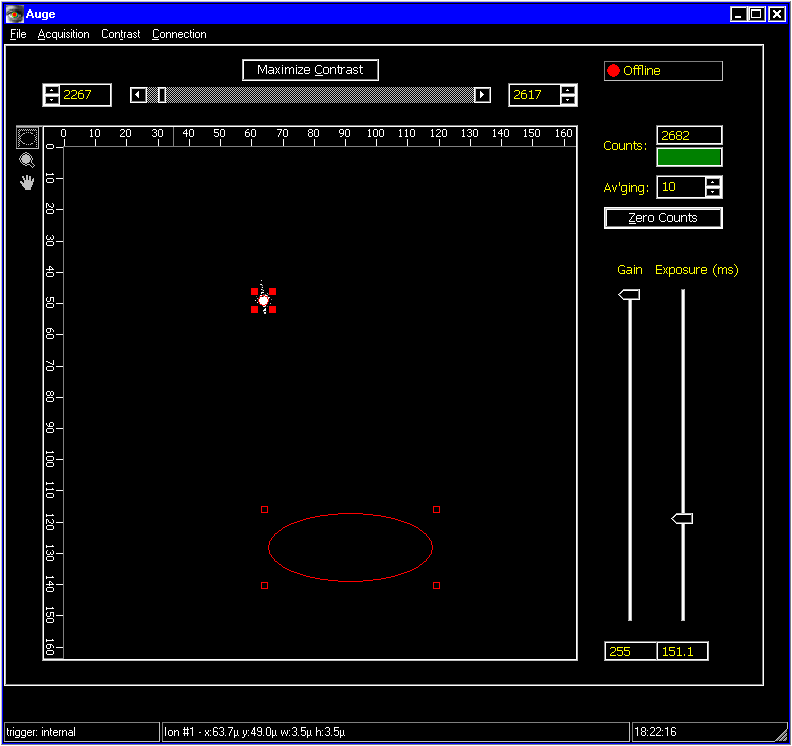}
\caption{Screenshot of \textit{Auge} in interactive mode.}
\label{fig:auge}
\efig

\noindent As one particular quantum simulation experiment (including the
process of state detection) has to be repeated several hundred
times to acquire one data point, it is desirable to minimize the time
elapsing between subsequent \ac{CCD} readouts. In interactive
visualisation mode where we read out the full \ac{CCD} frame this is at
least $57\millisecond$ or $17.5$ frames per second (\ac{FPS}) using
the parameters of \autoref{tab:cam_params}. The camera offers two
techniques to decrease readout times:
\bitem
\item Adjustment of the readout frame. Here, the camera will readout a
small rectangular portion of the whole image frame. \textit{Auge}
automatically calculates the minimum image frame that still encompasses
all the regions of interest. For a square of $5 \times 5$ pixels
(encompassing one ion) we attain about $580$\,FPS. In first order, the
duration for reading one frame scales linearly with its size.
\item \index{acquisition!row-wise} Row-wise acquisition. In this mode,
the \ac{CCD} chip is exposed several times before being read out, see
\autoref{fig:rowwise}. This is only feasible if several image rows fit
onto the \ac{CCD} chip concurrently. A rectangular frame of $512
\times 5$ pixels fits onto the chip $204$ times and leads to a maximum
frame rate of about $1700$\,FPS.
\eitem
As \textit{Flocke} takes over control, it transfers the desired frame
rate to \textit{Auge} which in turn switches to the most feasible
camera readout mode that still fulfils the frame rate desires.
Although row-wise acquisition allows for higher frame rates, it
involves additional effort to coordinate subsequent acquisitions. As
the \ac{CCD} is not cleared after every exposure we have to make sure
that we shift the \ac{CCD} photoelectrons as soon as acquisition time
is over. Let's refer to this acquisition as row no.\,1. Subsequently, we
will most probably perform operations involving fluorescence of the ions,
e.\,g. a Doppler re-cooling process, which we do not need to
acquire. Thus, the next row no.\,2, can be ignored. Only the
following row, no.\,3, will again contain relevant data. This
procedure effectively halves the frame rate.

As the \ac{CCD} is not cleared after every exposure, any background
signal that affects the entire \ac{CCD} frame is accumulated
\textit{within every row acquired}. This effect might be reduced by
properly positioning an aperture in front of the camera; this however
limits the significance of full frame acquisitions (which are very handy
during the ion loading process) .

Further problems arise due to the intermittent readout of the
\ac{CCD}. This readout takes about $120\millisecond$. Consequently, we
will have to suspend the experiment for at least $120\millisecond$
before we continue acquisition. Take the example given in
\autoref{fig:rowwise}: We take seven exposures without any waiting
times in between, then we will have to wait for the \ac{CCD} to be
read out and suspend experimentation for the meantime.

\bfig
\includegraphics[width=\textwidth]{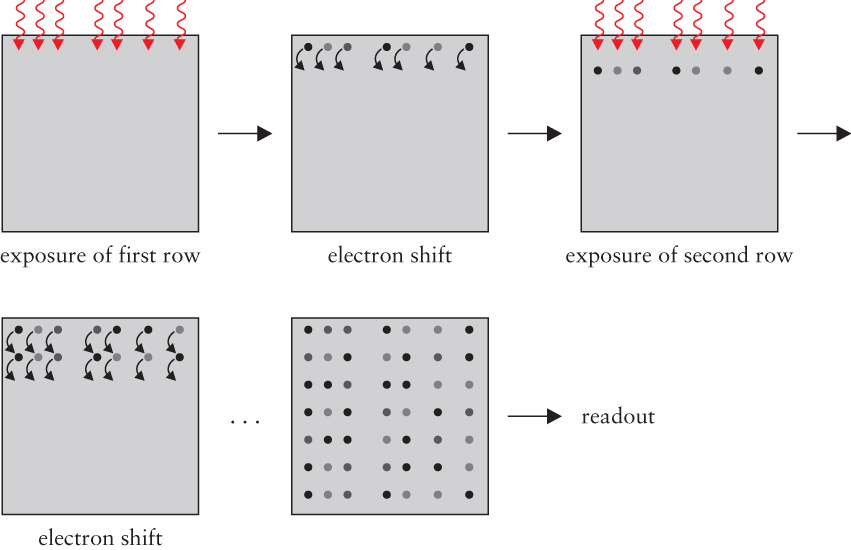}
\caption{Row-wise acquisition mode. In the case illustrated seven
acquisitions fit on the \protect\ac{CCD} chip at the same time.}
\label{fig:rowwise}
\efig

Due to the drawbacks involved with row-wise acquisition, it will be
advantageous to use the method of adjusting the readout frame for
realising our first experiments including \cite{schuetzhold}. This
mode is easier to operate and synchronize with the actual
experiment. However, it also shows some peculiarities. When the readout
frame is reduced (this occurs when we switch from the all-interactive
to Flocke-controlled mode), the \ac{CCD}'s dark count rate increases
at timescales of minutes. This behaviour might be associated with
heating effects inside the \ac{CCD}. With smaller readout frames,
i.\,e. higher frame rates, the chip temperature rises, eventually
leading to an increased dark count rate. At the time of printing of
this thesis, we were still investigating this issue in collaboration
with the camera's manufacturer.  This behaviour persisted even if we
water-cooled the \ac{CCD} to $-80\celsius$.

\chapter{Experimental control}
\label{ch:control}

An experimental control system used to conduct quantum simulation
experiments must satisfy the following requirements:
\bitem
\item High repetition rates. Any experiment dealing with measurements
of quantum states has to acquire considerable amounts of
\index{acquisition!statistics} statistical data (by conducting the
same experiment over and over again) in order to make statements on
superposition states. Thus, the duration for the acquisition of one
data set crucially depends on the repetition rate.
\item Good time resolution. Typical pulse durations for the laser
beams involved in a quantum simulation experiment are of the order of
$10\microsecond$. Therefore, an experimental control system must be
able to control associated components with a time resolution of better
than $100\nanosecond$.
\item Precise adjustment of the frequencies, phases, and amplitudes of
\ac{RF} output. This output will be used to drive acousto-optical
modulators~(\ac{AOM}s) that provide the required laser beam switching
and frequency-shifting capabilities.
\eitem
We meet these requirements using a home-built all-digital control box
which is clocked at $100\MHz$ and communicates with a self-written computer
programme. For simplicity, this box will be referred to as the
\textit{Paul box} \cite{pham:thesis,pham:manual}, according to its
manufacturer. It controls particular \ac{AOM}s and provides \ac{TTL}
outputs to trigger other devices in the laboratory. In addition, we
use what we will call the \textit{Jäger box}\footnote{Jäger
Computergesteuerte Messtechnik, \href{http://www.adwin.de/}{http://www.adwin.de/}} which controls
the \ac{DC} voltages at the trap electrodes and provides certain
measurement capabilities.

\section{\texorpdfstring{\protect\ac{AOM}}{AOM} control}
\index{acousto-optical modulator}

During experimentation all of our Raman beams (see
\autoref{sec:raman}) have to be controlled with respect to their
amplitude, frequency, and mutual phase. As these parameters are
subject to change during an experiment, we drive the respective
\ac{AOM}s using the Paul box frequency generation feature. Other beams
(\ac{BD} or \ac{RD} beams) will retain their frequencies and
amplitudes during an experiment, but will still have to be switched on
and off. Driving signals for the \ac{AOM}s of these beams can thus be
generated externally by voltage-controlled oscillators~(\ac{VCO},
MiniCircuits ZX series), appropriately attenuated using manual
variable or even fixed \ac{RF} attenuators, and eventually switched on
and off by \ac{RF} switches (MiniCircuits ZASW series). It is thus
sufficient to control these switches using the Paul box \ac{TTL}
output facility.

\index{acousto-optical modulator!driving power} Driving powers for the
\ac{UV} \ac{AOM}s (IntraAction ASM series) are $2.0\watt$ each, and
$3.5\watt$ for the deflector (Raman blueshifter). We use \ac{RF}
amplifiers (MiniCircuits ZHL series) with gains between $30\dB$ and
$40\dB$ to achieve these powers.

\section{Paul box}
\label{sec:paul-box}

We use the Paul box primarily to control all relevant \ac{AOM}s, some
of which must only be switched on and off while the driving input of
others must be controlled in amplitude, frequency, and phase. As the
\index{Paul box!time resolution} time resolution of Paul box commands
and actions is $10\nanosecond$, we also use it for accurate triggering
of other devices in the laboratory.

\subsection{Assembly}

\begin{sidewaysfigure}\centering
\includegraphics{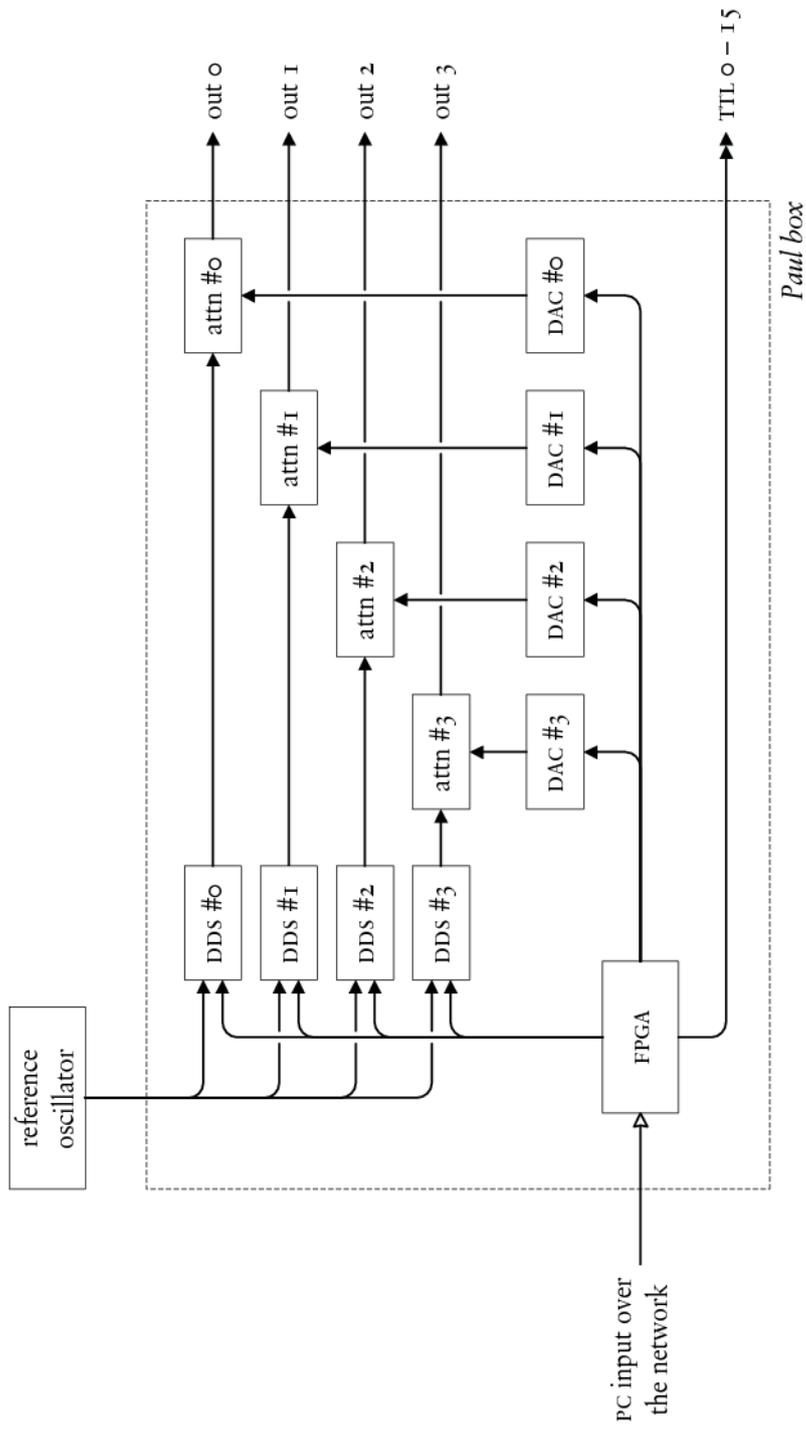}
\caption{Schematic of Paul box functionality. Legend:
\protect\ac{FPGA}\,=\,field programmable gate array (programmable processor),
\protect\ac{DDS}\,=\,direct digital synthesis board,
\protect\ac{DAC}\,=\,digital-to-analog converter,
attn\,=\,\protect\ac{RF} attenuator.}
\label{fig:paulbox}
\end{sidewaysfigure}

\index{DDS@\protect\ac{DDS}|see{Paul box direct digital synthesisers}}
An overview of the components included in the Paul box is given in
\autoref{fig:paulbox}. An externally generated reference frequency of
$f_0 = 1.2\GHz \pm 1\Hz$~(Hewlett-Packard HP8660C) is fed into four
\index{Paul box!direct digital synthesisers} direct digital
synthesisers~(\ac{DDS}) each of which can generate sinusoidal signals
with an arbitrary frequency $f < f_0/2$ and an arbitrary phase. The
output of each \ac{DDS} board can be individually varied in amplitude
using voltage variable \ac{RF} attenuators. Each attenuator is
connected to a digital-to-analog converter~(\ac{DAC}) which determines
its attenuation factor; the dynamic range at $220\MHz$ is $26\dB$, see
\autoref{fig:paulbox-attn}. Our box is equipped with four of these
universal frequency generation units. All the \ac{DDS} and \ac{DAC}
boards are in turn controlled by a microprocessor (field programmable
gate array, \ac{FPGA}) clocked at $100\MHz$ which receives its
commands over an Ethernet network. The \ac{FPGA} can also directly
control sixteen \ac{TTL} output signals and can be triggered by a
total of eight \ac{TTL} inputs, although the triggering feature has
not been required for the experiments conducted within the scope of
this thesis.

\bfig
\includegraphics{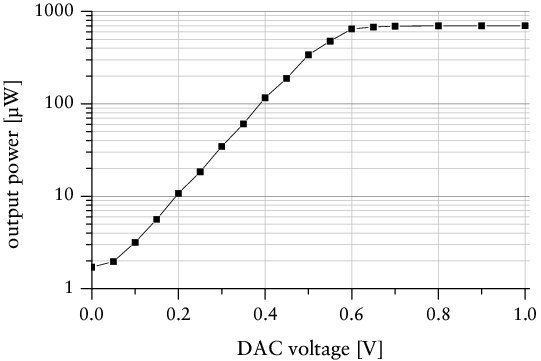}
\caption{Output \protect\ac{RF} power behind the Paul box's
voltage-variable attenuators at $220\MHz$. The remaining output power at a
\protect\ac{DAC} setting of zero can be switched off completely by
either switching off \protect\ac{DDS} output or by using
\protect\ac{RF} switches.}
\label{fig:paulbox-attn}
\efig

\noindent The Paul box was designed by Paul~T.~Pham\footnote{\href{mailto:ppham@cs.washington.edu}{ppham@cs.washington.edu}}
who assembled all the hardware on site at Garching. He has also
constructed similar boxes for the quantum simulation/computation
groups at Innsbruck, Austria and the University of Washington,
Seattle. Unfortunately, there was not enough time for Paul to
thoroughly test and debug his box before he had to return to the
\ac{US}. As a consequence, we spent considerable time to eliminate
particular bugs. Just a few examples: One of these bugs relates to the
fact that our box contains four \ac{DDS} boards instead of the two
\ac{DDS} boards included in all the boxes that had been built
before. The different signal delays in the connecting cables of the
\ac{DDS} boards destroyed temporal synchronisation between the
individual \ac{DDS} boards. Shortening the cable lengths eventually
solved this problem. Another bug was hidden on the commercially
available \ac{DDS} boards themselves. Their circuits contained an
erroneous connector that we had to remove by soldering. Tedious work
was also spent on debugging the \ac{DAC} boards. Originally they were
designed to operate with a differential \ac{EVENT} input. As this
proved no stable operation we re-soldered these boards to accept
unipolar \ac{EVENT}s. Anyway, after these repair works, the Paul box
proves to be a very stable and reliable piece of hardware. We stress
that we would not have been able to construct a similar box on our
own.

\subsection{\texorpdfstring{\protect\ac{FPGA}}{FPGA} software}

The \ac{FPGA} is responsible for the following tasks:
\bitem
\item Receiving input over the Ethernet network. The input is called a
\textit{pulse programme} containing directives for the \ac{DDS} boards,
\ac{DAC} boards, and \ac{TTL} outputs and is transferred to the box
inside a proprietary wrapper protocol called Pulse Transfer Protocol~(\ac{PTP}).
\item Executing the pulse programme. This part of the \ac{FPGA} software
is called Pulse Control Processor~\ac{PCP}. Apart from pulse commands it also
handles \texttt{WAIT} instructions, \texttt{LOOP}s and
subroutines. With the \ac{FPGA} being clocked at $100\MHz$ the
\ac{PCP} will process simple commands (such as \ac{TTL} switching)
within $10\nanosecond$.
\eitem
As the \ac{FPGA} and the \ac{DDS} boards are clocked by different
frequency sources, i.\,e. they are not synchronised, we observe a
\index{Paul box!jitter} max. $10\nanosecond$ jitter between the
execution of \ac{FPGA} commands and \ac{DDS} operation. For our
experiments however, this jitter will not measurably decrease
fidelities of quantum state manipulations.

The \ac{FPGA} software may be regarded as beta software (work in
progress), and as such it still contains a number of bugs. Thanks
to Paul's assistance we corrected almost all of them---at least those
whose presence avoided proper experimentation operation.

\subsection{\texorpdfstring{\protect\ac{PCP}}{PCP} compiler}
\index{PCP compiler@\ac{PCP} compiler|see{Paul box compiler}}
\index{Paul box!compiler}

Before being sent to the Paul box a pulse programme has to be translated
into a binary that is understood by the \ac{PCP}. Such a compilation
is most conveniently done on the computer communicating with the
box. While we were still testing and debugging the box hardware and
software, we used an already existing Python interface for this
purpose. Soon it became clear however that performance of the Python
compiler would not suffice to conduct our experiments (probably due to
heavy use of object allocation during the compilation process). We
eventually developed our own \ac{PCP} compiler written in \ac{C/C++}
to speed operations up. That way, compilation times for a typical pulse
programme with some $100\kilobyte$ decreased by a factor of about
$2000$.

Although being developed in \ac{C/C++} we have designed our compiler
as a Python module so that existing programmes---especially our graphical
user interface~(\ac{GUI}) pro\-gramme---could easily make use of
it. The compiler accepts a very simple human-readable pulse
description language (see \autoref{app:lxx}) that still
encompasses all the \ac{PCP} features.

\section{Jäger box}
\index{Jäger box}

The Jäger box---an ADwin-Gold system---features 8 \ac{DAC}s and 16
\ac{ADC}s for analog applications (both offering full-range
resolutions of $0.3\millivolt$), and a total of 32 digital in-/outputs
as well as 4 \ac{TTL} counters with a time resolution of $50\nanosecond$
for digital applications. In contrast to the Paul box it can handle
several processes at the same time in a multitasking environment. The
logic of these processes remains the same for every experiment, which
is why they are loaded once and then run forever (well, at least until
the box is switched off). Data exchange with a computer is then
carried out via Ethernet/UDP.

The status quo of our experimental setup is to have three processes
running on the Jäger box:
\bitem
\item A low-priority process controlling the \ac{DC} voltages at the
trap's sub-electrodes. By adjusting these voltages we can---by means
of software implementation---shift ions from the loading section into
the experimentation section of the trap and vice versa. Furthermore,
the axial confinement can be varied this way.
\item A high-priority process counting the pulses of a
photomultiplier~(\ac{PMT}) and outputting the count value upon the
arrival of an external trigger.
\item Another externally triggered process that measures the voltage
of a time-to-ampli\-tude converter~(\ac{TAC}) used for compensation of
micromotion (see \autoref{ch:micromotion}).
\eitem

\section{Interplay}
\label{sec:interplay}
\index{Flocke software@\textit{Flocke} software}

The heart of our experimental control system is a piece of software
called \textit{Flocke}\footnote{Obviously it had been snowing when the
programme was given its name.}. It has been written using Python
\cite{python:docs} and features a portable \ac{GTK} user interface
(see \autoref{fig:flocke}), result displaying and saving
capabilities. It runs under Linux on an AMD Athlon~XP 2800+ machine
connected to a $100\MBit$ network.

\begin{figure}[t]\centering
\includegraphics{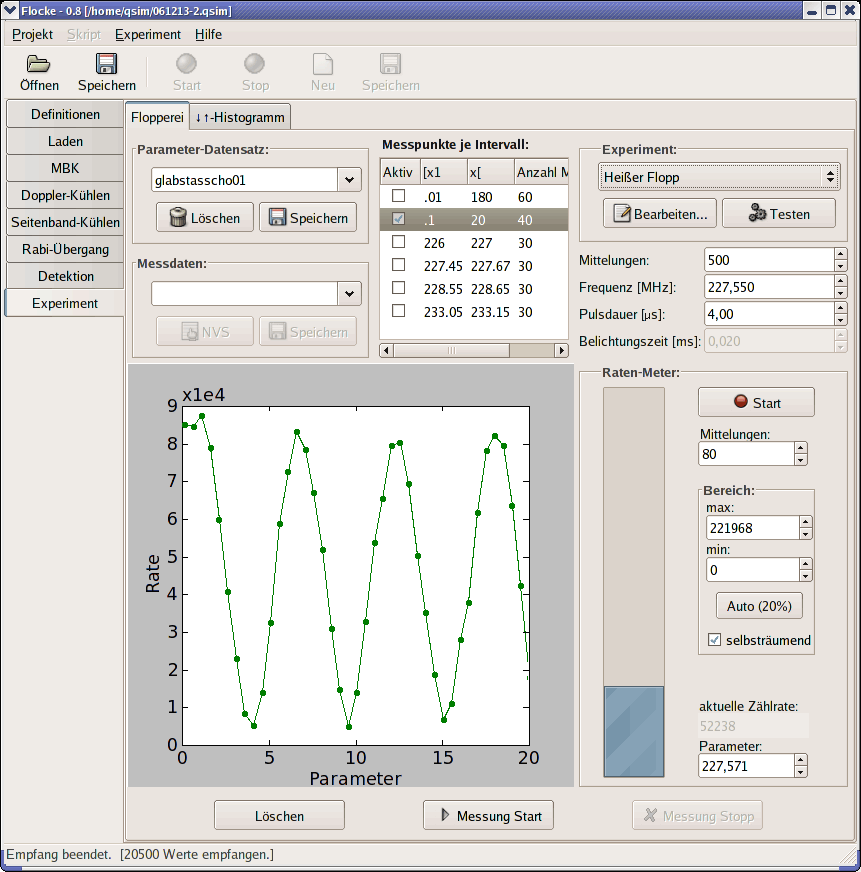}
\caption{Screenshot of \textit{Flocke} at work.}
\label{fig:flocke}
\efig

\hyperref[fig:interplay]{Fig.~\ref*{fig:interplay}} illustrates the
interplay between \textit{Flocke} and the other components controlling
and evaluating an experiment. On the input side we have a total of
three tools at hand to investigate ions in the trap. \index{CCD
camera@\ac{CCD} camera!interplay} The camera and
\index{photomultiplier!interplay} \ac{PMT} are primarily used to
measure the fluorescence of an ion expressed in a count rate. Their
detection processes thus have to be synchronised to the \ac{BD}
detection laser beam. The camera receives its acquisition triggers by
the Paul box, respective count rates are transferred to
\textit{Flocke} by the visualisation software \textit{Auge}. In the
case of \ac{PMT} acquisition, the Paul box sends its triggers to the
Jäger box which in turn returns the number of \ac{PMT} clicks that
have occurred between two subsequent trigger events. Both input
devices (including the visualisation software \textit{Auge}) are
described in more detail in \autoref{ch:detection}.

\begin{sidewaysfigure}\centering
\includegraphics{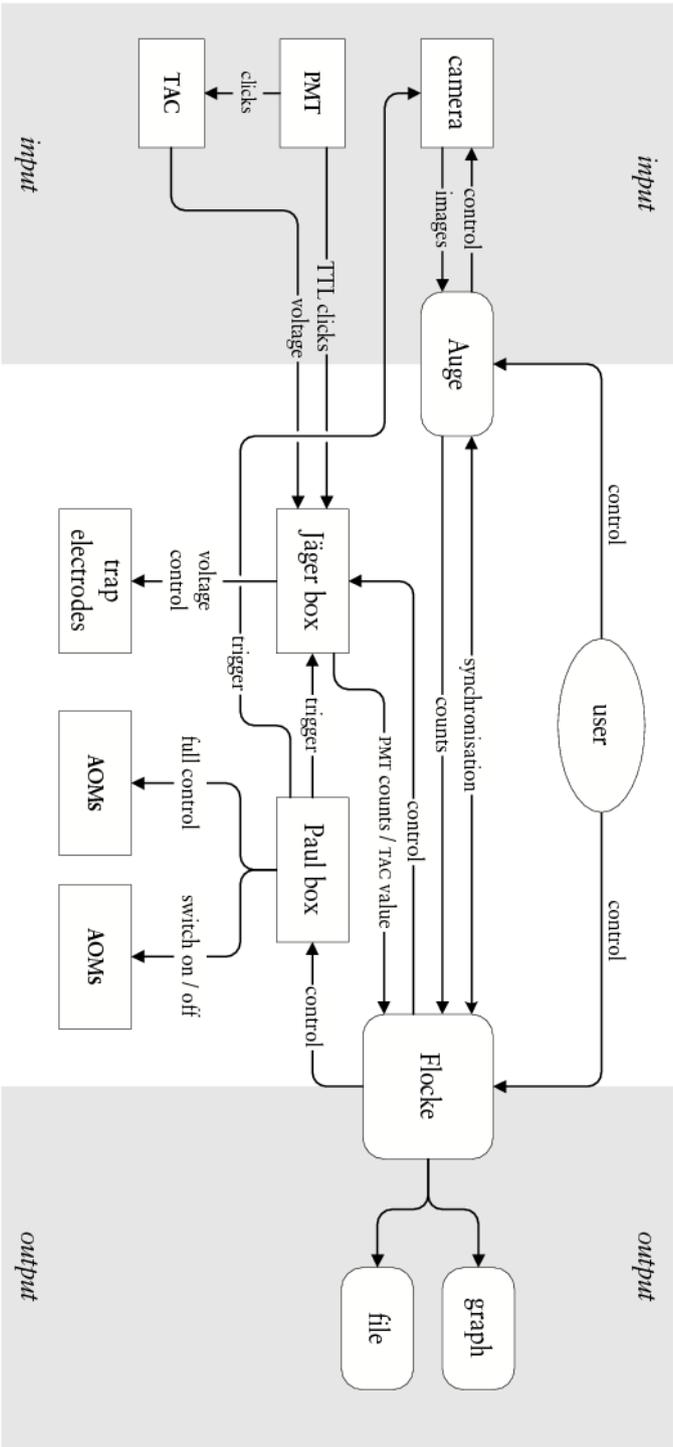}
\caption{Interplay of the components controlling operation of our
experiments.}
\label{fig:interplay}
\end{sidewaysfigure}

\index{TAC@\protect\ac{TAC}|see{time-to-amplitude converter}} The
\index{time-to-amplitude converter} \ac{TAC}---a stand-alone \ac{NIM}
component---is a special input device in the way that it is
exclusively designated for compensation of micromotion (see
\autoref{ch:micromotion}). Its start and stop triggers are supplied by
the \ac{PMT} on the one hand and a rising-edge zero-crossing of the
radial confinement \ac{RF} voltage on the other.

\part{Experimental issues}
\chapter{Simulating the Early Expanding Universe}
\label{ch:universe}

The theory of quantum fields in curves space-times (see
e.\,g.~\cite{birrell}) comprises many fascinating and striking
phenomena---one of them being the amplification of quantum vacuum
fluctuations due to the rapid expansion of space-time. According to
our standard model of cosmology this effect has caused the generation
of the seeds for cosmological structure formation during the
very early moments of the universe. Although these effects are far
removed from every-day experience, they are very important for the
past and the future fate of our universe. Thus, it would be desirable
to render these phenomena accessible to an experimental verification.

Ions confined in a Paul trap are particularly well suited as an
analogue system which can reproduce the Hamiltonian of quantum
fields. The analogue of an initially empty universe is achieved by
cooling the ions to their motional ground state. Subsequent creation
of particles is simulated by phonons which can be sensitively detected.

\section{Theoretical considerations}
\label{sec:ralf}

The considerations presented in this section are part of
\cite{schuetzhold} and were only possible due to the close and
fruitful collaboration with Ralf Schützhold.

\subsection{Original system}

Consider a massless scalar field $\Phi$ in curved space-time coupled
to the Ricci (curvature) scalar $\mathfrak{R}$ via a dimensionless
parameter $\zeta$ \cite{birrell}. Using the Friedmann-Robertson-Walker metric
\be
\text{d}s^2 = \mathfrak{a}^6(t) \text{d}t^2 - \mathfrak{a}^2(t) \text{d}\mathbf{r}^2
\ee
where $\mathfrak{a}(t)$ denotes a time-dependent scale parameter
(corresponding to cosmological expansion\,/\,contraction) we can
derive the wave equation
\be
\left(\frac{\partial^2}{\partial t^2} + \left[\mathfrak{a}^4(t)\mathbf{k}^2 + \zeta\mathfrak{a}^6(t)\mathfrak{R}(t)\right]\right) \Phi_\mathbf{k} = 0
\label{eq:wave_eq_universe}
\ee
where each mode $\mathbf{k}$ represents a harmonic oscillator with a
time-dependent potential $\mathfrak{a}^4(t)\mathbf{k}^2 +
\zeta\mathfrak{a}^6(t)\mathfrak{R}(t)$. For a non-adiabatical
time-dependence the initial vacuum state $\left|0\right>$ containing
no particles will evolve into a \index{squeezed state} squeezed state
\be
\left|\psi(t)\right> = \exp\left\{\sum\limits_\mathbf{k} \xi_\mathbf{k} \hat{a}_\mathbf{k}^2 - \mbox{h.\,c.}\right\} \left|0\right>
\label{eq:squeeze_universe}
\ee
where $\hat{a}_\mathbf{k}$ denotes the annihilation operator for mode
$\mathbf{k}$ and $\xi_\mathbf{k}$ is the respective squeezing
parameter which is determined by the solution of \autoeqref{eq:wave_eq_universe}.

\subsection{Analogue system}

The equations of motion for the axial positions $q_i$ of ions
crystallised and confined to a string along the axis of a linear trap
are given by
\be
\left(\frac{\partial^2}{\partial t^2} + \omega_z^2(t)\right) q_i = \frac{e^2}{4\pi\epsilon_0} \sum_{j \neq i} \frac{\mbox{sign}(i - j)}{(q_i - q_j)^2}
\label{eq:eq_motion_ion}
\ee
where $\omega_z(t)$ is a time-dependent axial oscillator frequency
and the term on the right hand-side encodes the mutual repulsive
Coulomb interaction of the ions, $e$ is the electron charge and
$epsilon_0$ is the vacuum permittivity.

The solution of \autoeqref{eq:eq_motion_ion} is obtained via
the scaling ansatz
\be
q_i(t) = b(t) \cdot q_i^0
\ee
where $q_i^0$ are the initial static equilibrium positions. We thus
obtain an equation for the evolution of the scale parameter $b(t)$:
\be
\left(\frac{\partial^2}{\partial t^2} + \omega_z^2(t)\right) b(t) = \frac{\omega_z^2(t=0)}{b^2(t)} \text{.}
\ee

\noindent Now we introduce \index{quantum fluctuations} quantum
fluctuations by splitting the position operator $\hat{q}_i(t)$ for
each ion into its classical trajectory $b(t) q_i^0$ and a fluctuation
term $\delta\hat{q}_i(t)$:
\be
\hat{q}_i(t) = b(t) q_i^0 + \delta\hat{q}_i(t) \text{.}
\ee
Since the fluctuations $\delta\hat{q}_i(t)$ are very small for
magnesium ions, we may linearise the \hyperref[eq:eq_motion_ion]{full equation of
motion~(\ref*{eq:eq_motion_ion})}. After a normal-mode expansion we get
\be
\left(\frac{\partial^2}{\partial t^2} + \omega_z^2(t) + \frac{\omega_\kappa^2}{b^3(t)}\right) \delta\hat{q}_\kappa = 0
\label{eq:wave_eq_ion}
\ee
for the phonon modes $\kappa$.

Comparing eqs.~\hyperref[eq:wave_eq_universe]{(\ref*{eq:wave_eq_universe})} and~\hyperref[eq:wave_eq_ion]{(\ref*{eq:wave_eq_ion})}
we observe a strong similarity if we identify $\Phi_\mathbf{k}$ with
$\delta\hat{q}_\kappa$. The wavenumber $\mathbf{k}^2$ from
\autoeqref{eq:wave_eq_universe} directly corresponds to the phononic
eigenfrequency $\omega_\kappa^2$ in \autoeqref{eq:wave_eq_ion} and the
scale factors $\mathfrak{a}(t)$ and $b(t)$ can be linked in a similar
way.

In view of the formal equivalence of eqs.~\hyperref[eq:wave_eq_universe]{(\ref*{eq:wave_eq_universe})}
and~\hyperref[eq:wave_eq_ion]{(\ref*{eq:wave_eq_ion})} we expect the system of ions to show the
same effects as the original cosmological system. In particular, we
will be interested in the analogue effect of \autoeqref{eq:squeeze_universe}:
Introducing a non-adiabatic time dependence of the axial oscillator
frequency $\omega_z(t)$ should excite squeezed phonon modes of the
trapped ions \cite{janszky:squeezing,yi:squeezing}. It is important to
note that a squeezed phonon state differs inherently from a classical
oscillation and can only be explained by quantum effects, see
\autoref{fig:squeezed_motional}.

\section{Experimental implementation}

\subsection{General setup}

To perform a first experimental realisation of the proposed experiment we
would confine one single magnesium-25 ion to the axis of our trap (see
\autoref{ch:trap}). The simulation basis will be provided
by the hyperfine ground-state level $3S_{1/2} \left|F=3,m_F=3\right>
\equiv \left|\down\right>$ and the harmonic oscillator levels arising
from the axial confinement of the ion which can be characterised by its resonance
frequency $\omega_\text{axial}$ (see \autoref{ch:mg25}). The constant
magnetic field of \autoref{sec:magnetic_field} will define the
quantisation axis for the laser fields and lift the degeneracy of the
$3S_{1/2}$ hyperfine level manifold.

\subsection{Protocol of the experiment}

At the start of each experiment, the ion will be laser-cooled close to
the ground state of the axial motion (see \autoref{sec:cooling})
and optically pumped into the electronic $\left|\down\right>$
state. We will adiabatically decrease the axial confinement and 
subsequently reset it to its initial value on a non-adiabatic
timescale, similar to a proposal in \cite{meekhof:non-classical}.
Since the width of the ground-state wave function of the ion cannot
adiabatically adapt to the new confinement, it will breathe in the
final trapping potential. This non-classical oscillation can be
described via the squeezed state formalism mentioned in
\autoref{sec:ralf}.

The readout of the final motional state (which is the quantity of
interest for this experiment) occurs via mapping the motional state
onto an internal electronic state $\left|\up\right> \equiv 3S_{1/2}
\left|F=2,m_F=2\right>$ separated from the $\left|\down\right>$ state
by the hyperfine splitting of $\omega_0 = 1789\MHz$ (see
\autoref{fig:mg25}). There are proposed \cite{solano} and already
realised \cite{meekhof:non-classical} methods to provide the mapping
of motional states to internal electronic states. That way, we may
take advantage of the high accuracies available for internal state
detection.

We suggest two-photon stimulated Raman transitions near
$280\nanometer$ for this purpose, see \autoref{sec:tps-raman}.
Their wavevector difference $\Delta\mathbf{k} = \mathbf{k}_2 -
\mathbf{k}_1$ is aligned along the trap axis thus coupling to the
axial motional states. Their frequency difference $\Delta\omega =
\omega_2 - \omega_1$ can be precisely adjusted using acousto-optical
modulators (\ac{AOM}s). Transitions on the carrier
($\left|\down,n\right> \leftrightarrow \left|\up,n\right>$,
$\Delta\omega = \omega_0$), first sidebands ($\left|\down,n\right>
\leftrightarrow \left|\up,n\pm1\right>$, $\Delta\omega = \omega_0 \pm
\omega_\text{axial}$), and second sidebands ($\left|\down,n\right>
\leftrightarrow \left|\up,n\pm2\right>$, $\Delta\omega = \omega_0 \pm
2\omega_\text{axial}$) can be accomplished at high fidelities.

At the end of each simulation experiment, we will have to read out the
motional state population. In our setup, state-sensitive detection is
performed by the \ac{BD} laser beam that resonantly couples the
$\left|\down\right>$ state to the $3P_{3/2}$ level thereby scattering
photons at $10\MHz$ rates, see \autoref{sec:state_detection}. This
allows to distinguish the ``bright'' $\left|\down\right>$ state from
the ``dark'' $\left|\up\right>$ state with high accuracy---even at low
photon detection efficiencies.

For the special case of experiment considered here we suggest a method
for reading out the most important motional state population of the
squeezed state, $n = 2$, via a \index{sideband!second red} second red
sideband transition followed by a carrier transition. The former
transfers the motionally excited population $\left|\down,n=2\right>$
into $\left|\up,n=0\right>$ leaving $\left|\down,n=0\right>$ untouched
as there is no red sideband transition for that population. The latter
transition will exchange the populations of the two electronic
levels\footnote{The attentive reader will have noticed that
co-propagating Raman beams are required for this step, see
\autoref{sec:tps-raman}.}. After this step, the population initially
in the $n=2$ state will have been transferred into $n=0$ whereas the
population initially in the $n=0$ state will reside in the electronic
$\left|\up\right>$ state. State-sensitive detection will then provide
the two-phonon ($n=2$) generation probability (but see below for
complications).

\subsection{Expected results}

Ramping the axial potential from $100\kHz$ up to $2\MHz$ in a non-adiabatical way,
i.\,e. fast compared to the oscillation period of the lower frequency,
requires a rise time of the order of $1\microsecond$. For our trap
geometry, the corresponding voltage range is several $10\volt$. We
expect to observe considerable blind currents during the ramping
process caused by the trap electrodes' low-pass filters (see
\autoref{sec:axial}). We will also have to carefully balance the
applied voltages as they are modified. Otherwise, the ion's axial
motional mode could be excited classically; we might even lose the ion
due to unsufficient confinement. The duration of the experiment~($\sim
3\millisecond$) should be kept short compared to the thermal heating
times (estimated $\sim 0.01$ excited motional quanta per
millisecond)---although the thermal and squeezed motional spectra
should still be distinguishable for higher temperatures.

\bfig
\subfloat[Thermal distribution with $\overline{n} = 0.1$.]{
  \includegraphics{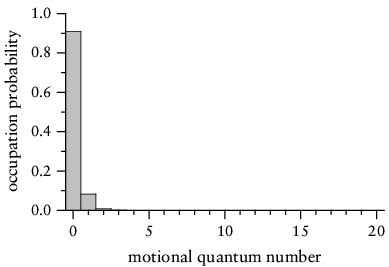}}
\hfill
\subfloat[Squeezed distribution with $\overline{n} = 2$.]{
  \includegraphics{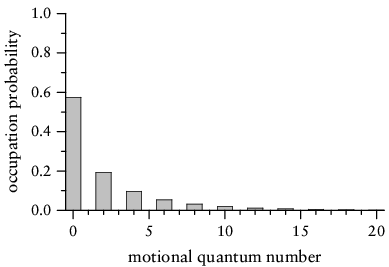}}
\caption{Calculated occupation probabilities of motional states in our
trap at the final axial confinement of $\omega_z = 2\pi \cdot
2\MHz$. The figures show the thermal spectrum after resolved-sideband
cooling and the expected squeezed distribution both characterised by
their mean motional quantum number $\overline{n}$.}
\label{fig:squeezed_motional}
\efig

Our numerical simulations indicate that we should be able to ``squeeze''
more than $20\%$ of the motional state population from the initial
ground state $n=0$ into $n=2$, see \autoref{fig:squeezed_motional}.
Since the state-of-the-art fidelities for the carrier and sideband
transitions as well as the state-sensitive detection exceed $99\%$ we
expect the creation and the parameters of the squeezed state to be
measurable with high accuracy, especially in comparison with a purely
thermal state.

\noindent Relying on driving the second red sideband transition, the suggested
readout scheme will also transfer population from motional states
$n>2$ into their respective lower states. Thus, a heated thermal
population might lead to the same signal as a squeezed one. We will
discriminate these two distributions by an additional signal acquired
using a first red sideband\,--\,carrier sequence (analogous to the
second red sideband\,--\,carrier sequence mentioned above) thereby
detecting the population in $n\ge1$. The ratio of the two signals will
then provide us with the precise occupation probabilites of the
motional state $n=1$.

\subsection{Further applications}

The investigation of non-adiabatic switching of trapping potentials
and the influence on the quantum state of motion might shed light on
possible problems with fast-shuttling schemes for ions in multiplex
trap architectures. These are required for scaling the ion trap
approach towards a universal quantum computer.

\chapter{Compensation of micromotion}
\label{ch:micromotion}
\index{micromotion|main}

As introduced in \autoref{sec:radial_confinement} any radial motion in
our linear Paul trap is associated with micromotion. We shall call
this type of micromotion \textit{natural micromotion} as it is an
inherent part of motion in a Paul trap. In contrast we will speak of
\textit{excess micromotion} when it comes to micromotion caused by
external imperfections such as external electric fields displacing the
ions from the trap axis.

For our quantum simulation experiments it is crucial that excess
micromotion is reduced to a minimum. Uncompensated excess micromotion
will be responsible for the appearance of sidebands. This in turn will
have negative effects on any experiment including
\bitem
\item heating the ions with the \ac{BD} beam (which is actually
intended for cooling them) due to a possible blue-detuning with
respect to red sidebands,
\item reduced Doppler cooling rates as some fraction of the cooling
transition line will be distributed to sidebands and thus be ``lost'',
\item the possibility to drive unwanted transitions due to the
coupling to sidebands,
\item a higher probability for ion loss due to high-energetic
collisions with particles from the residual gas,
\item longer Doppler-cooling durations before the ion(s)
form(s) a proper ion crystal.
\eitem

\section{Theoretical background}

\subsection{Sources of micromotion}

We will consider three sources of excess micromotion:
\bitem
\item External electric \ac{DC} fields in the radial directions. These
displace the ions from their centred positions on the trap axis
defined by the \ac{RF} quadrupole potential. With one \ac{RF}
electrode being closer to the ion than the opposing one, the ions will
be exerted to a force oscillating with the \ac{RF} frequency.
\item Phase difference of the driving \ac{RF} voltage at opposing
electrodes. This will cause the extremum of the quadrupole potential
to shift spatially during an oscillation period of the driving \ac{RF}
voltage.
\item Disadvantageous axial position of the ion. The \ac{DC}
subelectrodes are separated by slits whereas the \ac{RF} electrodes
aren't. An axial off-position from the centre between two opposing
\ac{DC} electrodes will thus induce a net oscillating force on the ion.
\eitem

\subsection{Modulated Bloch equations}
\index{Bloch equations!modulated}

The following derivation of why and how sidebands emerge in the case
of excess micromotion largely follows \cite{berkeland:micromotion}.
Assuming for simplicity that both sources of micromotion only have an
effect in the trap $\hat{x}$ direction, it will be sufficient to
modify the respective \hyperref[eq:radial_motion]{equation~(\ref*{eq:radial_motion})} as follows:
\be
x(t) = \left(x_\text{DC} + x_0 \cos(\omega_x t + \phi_x)\right) \left(1 + \frac{q}{2} \cos \Omega_\text{RF} t\right) - \frac{1}{4} q R \alpha \phi_\text{RF} \sin \Omega_\text{RF} t \text{,}
\label{eq:micromotion_x}
\ee
where $x_\text{DC}$ describes the offset due to external \ac{DC}
fields and the last term takes care of a phase difference
$\phi_\text{RF}$ at the \ac{RF} electrodes. $\alpha$ is a geometrical
factor which quantifies how the opposing \ac{RF} electrodes differ
from an ideal capacitor. Using the software package \ac{SIMION} we
estimate $\alpha \approx 1.4$ for our trap geometry.

An ion oscillating according to \autoeqref{eq:micromotion_x} will---in
its rest frame---perceive a different electric field than an ion at
rest. Assuming that excess micromotion outperforms secular motion as
well as natural micromotion in terms of amplitude, we have
\be
\begin{split}
\mathbf{E}(t) &=  \mathbf{E}_0 \exp\left[i \left(k x(t) - \omega t\right)\right] \\
         &\kern-0.055em\ \dot{=}\ \mathbf{E}_0 \exp\left[i k \left(x_\text{DC} + \tilde{x}(t)\right) - \omega t)\right] \text{,}
\end{split}
\label{eq:micromotion_efield}
\ee
where $k$ is the modulus of the wavevector in the $\hat{x}$ direction
and
\be
k \tilde{x}(t) = \beta \cos(\Omega_\text{RF} t + \delta)
\label{eq:micromotion_cosine}
\ee
``combines'' the contributions of both sources of excess micromotion
and $\beta$ and $\delta$ defined by
\be
\begin{split}
\beta  &= \sqrt{\left(\frac{1}{2} k q x_\text{DC}\right)^2 + \left(\frac{1}{4} k q R \alpha \phi_\text{RF}\right)^2} \text{,} \\
\delta &= \arctan\left(-R \alpha \phi_\text{RF}, 2 x_\text{DC}\right)
\end{split}
\ee
represent the strength and phase of excess micromotion respectively.

The optical Bloch equations for a two-level system interacting with
the \hyperref[eq:micromotion_efield]{modulated electric field~(\ref*{eq:micromotion_efield})}
are similar to the unmodulated case; we will merely have to carry out
the substitution
\be
e^{i \omega t} \rightarrow e^{i \omega t - i \beta \cos(\Omega_\text{RF} t + \delta)}
\ee
resulting in the following set of modulated Bloch equations (don't
confuse the Rabi frequency $\Omega$ and the driving \ac{RF} trap
frequency $\Omega_\text{RF}$!):
\be
\begin{split}
\dot{\rho}_{bb} & = -\frac{i \Omega}{2} \left(\rho_{ab} e^{i \Delta t + i \beta \cos(\Omega_\text{RF} t + \delta)} - \text{c.\,c.}\right) - \Gamma \rho_{bb} \text{,} \\
\dot{\rho}_{ab} & =  \frac{i \Omega}{2} e^{-i \Delta t - i \beta \cos(\Omega_\text{RF} t + \delta)} \left(1 - 2\rho_{bb}\right) + (i\Delta - \frac{\Gamma}{2}) \rho_{ab} \text{,} \\
\dot{\rho}_{ba} & =  \dot{\rho}_{ab}^* \text{,} \\
\dot{\rho}_{aa} & = -\dot{\rho}_{bb} \text{.}
\end{split}
\ee

\bfig
\includegraphics{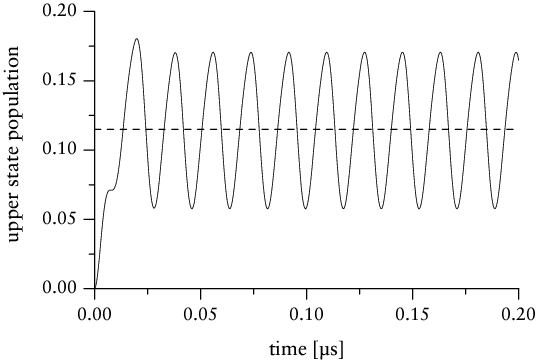}
\caption{Excess micromotion of a ${}^{25}\text{Mg}^+$ ion stored in
our trap at an \protect\ac{RF} voltage of magnitude $1000\volt$. This
graph was calculated for a moderate excess micromotion amplitude
$\beta = 0.7$ resulting from an offset $x_\text{DC} = 0.16\micrometer$
caused by an external electric field $E_\text{ext} =
100\volt/\unitsignonly{\meter}$. Parameters of the detection \protect\ac{BD}
laser were taken from \autoref{tab:bd}.}
\label{fig:micromotion_time}
\efig

\noindent \hyperref[fig:micromotion_time]{Fig.~\ref*{fig:micromotion_time}} shows the time-resolved upper-level
population of a magnesium ion stored in our trap. There is no
steady state any more, rather does the upper-state population---and
thus also the fluorescent scattered intensity---oscillate with the
driving frequency $\Omega_\text{RF}$.

\subsection{Sidebands}
\index{sideband!micromotion}

During normal, non-time-resolved detection we will not be able to
detect oscillations in the population of the upper state but instead
observe the mean population (dashed line in \autoref{fig:micromotion_time}).
We can derive an analytical expression for the mean upper-state
population by utilising the expansion
\be
\exp\left[i \beta \cos(\Omega_\text{RF} t + \delta)\right] = \sum\limits_{n=-\infty}^{\infty} J_n(\beta) \exp\left[in(\Omega_\text{RF}t + \delta + \pi/2)\right]
\ee
where $J_n(\beta)$ signify the Bessel functions of the first
kind. This yields \cite{berkeland:micromotion}
\be
\left<\rho_{bb}\right> = \Omega^2 \sum\limits_{n=-\infty}^{\infty} \frac{J_n^2(\beta)}{(\Delta+n\Omega_\text{RF})^2 + (\Gamma/2)^2} \text{.}
\ee

\noindent \hyperref[fig:micromotion_detuning]{Fig.~\ref*{fig:micromotion_detuning}} illustrates how the associated
mean fluorescent scattering rate $r = \rho_{bb} \Gamma$
varies with the detuning $\Delta$ of the laser with respect to the
actual transition frequency. In addition to the \textit{carrier} peak
at $\Delta = 0$ there are peaks at $\Delta = \pm i \cdot %
\Omega_\text{RF}, i \in \mathbb{N}$. We call these patterns
\textit{sidebands}---in particular, the peaks where $\Delta < 0$ are
referred to as the red sideband while the peaks where $\Delta > 0$ are
called the blue sideband. As the micromotion amplitude $\beta$
increases, more and more contribution is shifted into the
sidebands. At the same time the carrier is decreased (the area
under the curves of \autoref{fig:micromotion_detuning} is the same
for all $\beta$).

\bfig
\subfloat[Resolved-sideband case calculated for parameters achieved in
the \protect\ac{NIST} group: $\Omega_\text{RF} = 2\pi \cdot 150\MHz$,
$\Gamma = 2\pi \cdot 20\MHz$.]{
  \includegraphics{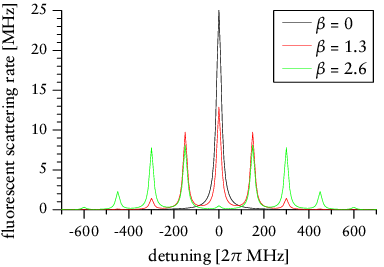}}
\hfill
\subfloat[Nonresolved-sideband case calculated for parameters used in
\cite{schaetz:thesis}: $\Omega_\text{RF} = 2\pi \cdot 6\MHz$,
$\Gamma = 2\pi \cdot 43\MHz$.]{
  \includegraphics{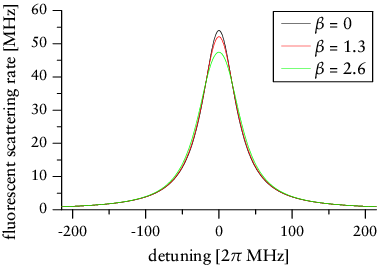}}
\\
\subfloat[Intermediate case for magnesium ions in our trap:
$\Omega_\text{RF} = 2\pi \cdot 56\MHz$, $\Gamma = 2\pi \cdot
43\MHz$.]{
  \includegraphics{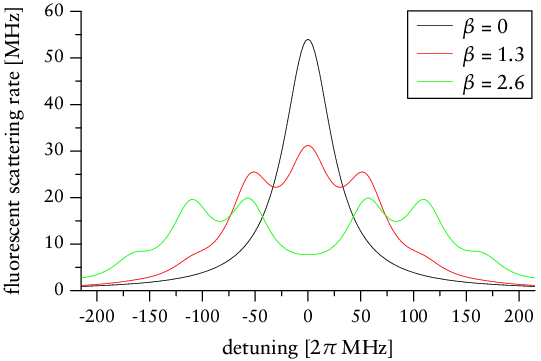}}
\caption{Sideband structure in dependence of the amount $\beta$ of
micromotion for different ratios of the driving \protect\ac{RF}
frequency $\Omega_\text{RF}$ and the linewidth $\Gamma$. All plots
were calculated for a laser beam intensity $I = (2/3) I_\text{sat}$.}
\label{fig:micromotion_detuning}
\efig

\section{Methods for compensation of micromotion}

In principle, it is easy to compensate the sources of
micromotion. A phase difference of the \ac{RF} voltage at opposing
electrodes as the other source of micromotion is compensated by
appropriately adjusting the length of the connecting wires. External
electric \ac{DC} fields in the radial directions can be compensated by
an antiparallel counteracting field effectively reducing the relevant
\ac{DC} electric field at the trap centre to zero. In our trap apparatus (see
\autoref{fig:trap_west}) we can apply compensation fields in the
$\hat{y}$ direction and in the $\hat{x}_\text{ext}$ direction. The
former is achieved by applying differential voltages to the \ac{DC}
sub-electrodes, the latter by applying an appropriate voltage to the
compensation electrode located beneath the trap, see
\autoref{fig:trap_west}.

In reality, things are a bit more complicated due to the interaction
of the radial and axial confinements. As described in
\autoref{sec:trap_confinement_interference} an axial displacement
of the ion will also give rise to micromotion. The reason are the
slits between the \ac{DC} subelectrodes which are not present at the
\ac{RF} electrodes thus introducing an asymmetry of the trap. In order
to suppress this source of micromotion we have to adjust the axial
position of the ion to match the centre of an opposing electrode
pair. That way, we have to take care of three degrees of freedom
(differential voltage, compensation electrode, and axial position),
which is not really surprising for an ion confined in three dimensions.

For a three-dimensional detection and compensation of micromotion we
would need three laser sources shining onto the ions from different
(linearly independent) directions. This is because the ion's motion
will only modulate the electric field if it has a component parallel
to the laser beam wavevector; motion perpendicular to the wavevector
does not alter the fluorescent scattering rate. In fact, two laser
beams will suffice for our applications.

\subsection{Minimising ion displacement}

Assumed that we can rule out an \ac{RF} phase difference as the source
of micromotion, we have to determine is the compensating electric
field. As it compensates external fields that are not related to the
trap operation in any way, it should be the same no matter what
\ac{RF} confinement we choose for the ions.

Based on this consideration, we come up with a method that is very
easy to implement: As long as external fields are not ideally
compensated, ions in the trap will be displaced in the direction of
the residual electric field when their \ac{RF} confinement is reduced. We
adjust the compensation field such that varying the radial confinement
by tuning the \ac{RF} voltage magnitude does not displace the ions any
more.

We use the \ac{CCD} camera to determine the position of an ion inside
the trap. Whereas it is easy to visualise a displacement in the
$\hat{y}_\text{ext}$ and $\hat{z}_\text{ext}$ directions, changes in
the $\hat{x}_\text{ext}$ position (perpendicular to the focal plane of
our objective) can only be detected by different defocusings of the
image. As these changes are relatively small when the objective is
focused, we initially defocus the objective so that the
focal point lies above the ion, $x_\text{ext}(\text{focal point}) >
x_\text{ext}(\text{ion})$. This defocusing setting leads to a circular
structured image, \autoref{fig:defocused2}, whose radius can be associated
with the ion position in the $x_\text{ext}$ direction. The optimum
compensation settings are found if the defocused image remains
unchanged upon changes in the radial confinement.

Although this method provides an easy way to compensate micromotion,
it is not suitable for strong axial confinements. In order to detect
displacement of a stored ion, the radial confinement would have to be
lowered below a value that still guarantees a stable confinement.

\subsection{Maximising the fluorescent scattering rate}

\hyperref[fig:micromotion_detuning]{Fig.~\ref*{fig:micromotion_detuning}} implies that the fluorescent
scattering rate at zero detuning is maximised when there is no
micromotion. This is also true for a detuning of $\Delta = \Gamma/2$,
which we normally use for optimum Doppler cooling, see
\autoref{fig:micromotion_carrier}.

\bfig
\includegraphics{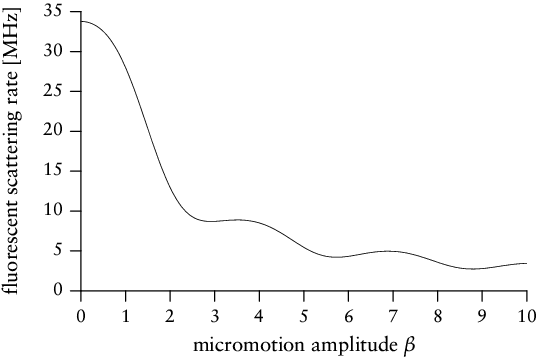}
\caption{Fluorescent scattering rate for different amounts of
micromotion. For this curve a detuning of $\Delta = \Gamma/2$ was
chosen. Other laser parameters were taken from \autoref{tab:bd}.}
\label{fig:micromotion_carrier}
\efig

Using the photomultiplier or the \ac{CCD} camera, we can try to
maximise an ion's brightness by properly adjusting the differential
voltage, compensation electrode, and axial position. As the
fluorescent scattering rate rises almost monotonously with
decreasing $\beta$, this seems to be feasible at first sight. However, during
the adjustment process an ion in the trap will be displaced (basically
in any of the three spatial directions), which means that the detection
device~(\ac{PMT} or \ac{CCD}) also has to be readjusted. And adjusting
three compensation knobs plus a detection device at the same time can
be quite tedious.

One could also try to minimise the fluorescent scattering rate on one
of the sidebands' transition lines, for example on the first line
where $\Delta = \Omega_\text{RF}$. This technique works well for the
resolved-sideband case ($\Omega_\text{RF} \gg \Gamma$). As can be seen
from \autoref{fig:micromotion_first_sb} the aspired minimum in our
intermediate case (sidebands are resolved only partially) is not a
global minimum though. We were thus unable to successfully compensate
micromotion using this technique.

\bfig
\includegraphics{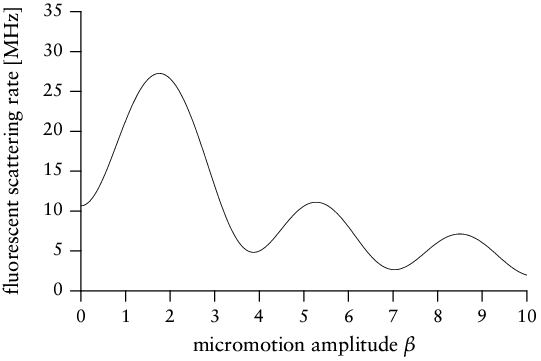}
\caption{Fluorescent scattering rate for different amounts of
micromotion. For this curve a detuning of $\Delta = \Omega_\text{RF}$
(the first sideband line) was chosen. Other laser parameters were
taken from \autoref{tab:bd}. Note that the local minimum at $\beta =
0$ is not the global minimum of the scattering rate.}
\label{fig:micromotion_first_sb}
\efig

\subsection{Minimising the correlation of photons and \texorpdfstring{\protect\ac{RF}}{RF}}
\label{sec:mm_correlation}

\bfig
\subfloat[Uncorrelated photon emission (stray light).]{
  \qquad\includegraphics{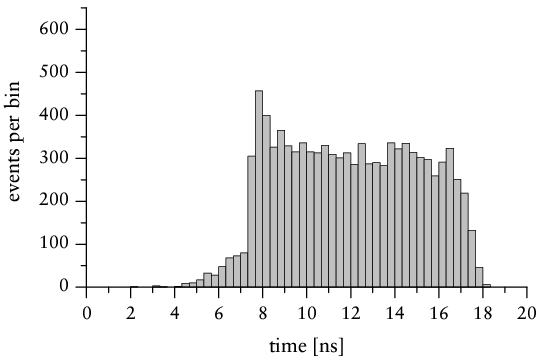}
  \label{fig:micromotion_uncorrelated1}} \\
\subfloat[Correlated photon emission (ion undergoing micromotion).]{
  \includegraphics{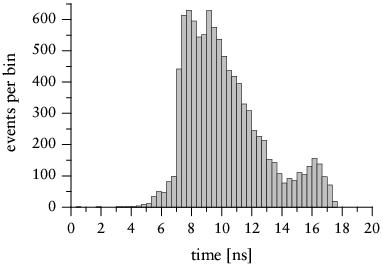}}
\hfill
\subfloat[Uncorrelated photon emission (micromotion has been
compensated).]{
  \includegraphics{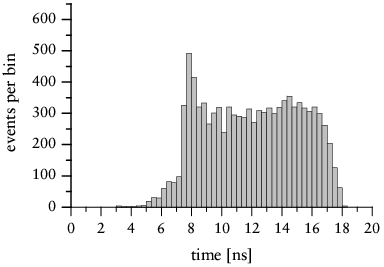}}
\caption{Correlation histograms for different amounts of micromotion.
The \hyperref[fig:micromotion_uncorrelated1]{``teapot curve''~\ref*{fig:micromotion_uncorrelated1}} serves as a
calibration standard. The deviation from a perfectly flat-shaped
histrogram is caused by a technical inadequacy of the \protect\ac{TAC}
module (see main text).}
\label{fig:micromotion_correlation}
\efig

This is the preferred method for compensating micromotion. It however
involves some additional effort in terms of electronic signal
processing. In principle, this method measures the time-resolved
fluorescent scattering signal plotted in \autoref{fig:micromotion_time}:
As long as an ion undergoes micromotion, the fluorescent scattering
rate will be modulated with the driving \ac{RF} frequency
$\Omega_\text{RF}$. Thus, the two signals ``fluorescent scattering
rate'' and ``driving \ac{RF} frequency'' will be correlated. If
micromotion has vanished, the fluorescent scattering rate is constant
over time and will not be correlated to the driving \ac{RF} frequency.

Every time our \ac{PMT} registers a photon, we measure the time it
takes till the next rising edge zero-crossing of the \ac{RF}
voltage. The timespans will be in the range $0 < t
2\pi/\Omega_\text{RF} = 17.9\nanosecond$ for our trap. We use a
time-to-amplitude converter~(\ac{TAC}, Ortec model 437A) to convert
these short spans into a voltage that can be measured using the Jäger
box. After some $10,000$ of these measurements, our control and
analysis software \textit{Flocke} draws a histogram of the acquired
timespan values.

We expect the histogram to be flat for uncorrelated photon emission
and to show a peak somewhere for correlated photon emission (more
photons are emitted at a particular phase of the \ac{RF} voltage than
at other times). Results are shown in
\autoref{fig:micromotion_correlation}. The histogram is not perfectly
flat for uncorrelated photon emission because our \ac{TAC} cannot
properly convert timespan values below $7\nanosecond$. Apparently,
most of the timespan values $t < 7\nanosecond$ are not converted at
all while values $t \approx 7\nanosecond$ are all converted to the
same output voltage (corresponding to a timespan of $t =
7\nanosecond$). Due to its shape we call the histogram a \index{teapot
curve} ``teapot curve''.

\section{Experimental results}

Although our compensation electrode is located just beneath the trap
axis, it only has a small penetration factor of about $...$. For
proper compensation of micromotion we thus have to apply voltages of
the order of $10^2\volt$. The differential voltage applied to opposing
\ac{DC} subelectrodes can be considerably smaller; we would usually
operate with voltages of $10^{-1}\volt$.

If we shine onto the ion using only one laser beam, there are a number
of compensation settings that produce flat teapot curves, see
\autoref{fig:micromotion_unidirectional}. The optimum compensation
setting is the one that also produces a flat teapot curve for a
different laser beam (propagating in a different direction). For
${}^{25}\text{Mg}^+$ this easier said than done as not all laser beams
will be able to efficiently cool the ions at the same time. Actually,
only the $\sigma^+$ polarised \ac{BD} beam propagating parallel to
the magnetic quantisation axis will optimally Doppler-cool the
ions and excite them to emit fluorescent light. Laser beams from
different directions, e.\,g. the \ac{BD} beam entering the chamber via
port~\ac{J} (see \autoref{fig:chamber}), are unable to drive the
cycling \ac{BD} transition and thus unable to cool the ions
efficiently. Eventually, we would lose an ion that is not illuminated
by the $\sigma^+$ polarised \ac{BD} laser.

\noindent We overcome this problem by shining onto the ions using two laser
beams at the same time; we attenuate the $\sigma^+$ polarised \ac{BD}
beam to a minimum that still cools the ions and use the \ac{BD} beam
from chamber port~\ac{J} as a second laser source. That way we can
determine an optimum set of compensation settings. These settings were
not constant over time however. In particular, we always had to
re-adjust the compensation settings after having loaded new ions into
the trap. Variances of up to $100\%$ in both the compensation voltage
and the differential voltage settings are not unusual.

\bfig
\includegraphics{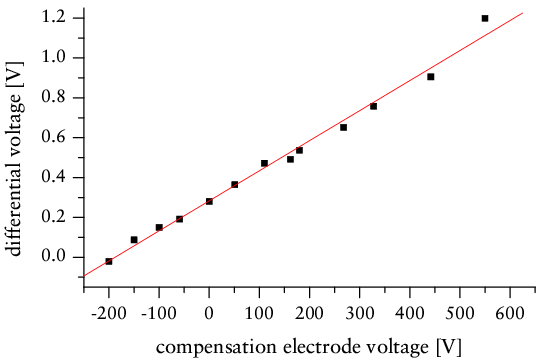}
\caption{Micromotion compensation settings that produce a flat ``tea
pot curve''. These data were measured using a single ion in the trap
illuminated by the $\sigma^+$ polarised \protect\ac{BD} laser. The
optimum compensation settings within the data are those which also
produce a flat tea pot curve for a second laser beam propagating in a
different direction.}
\label{fig:micromotion_unidirectional}
\efig

\chapter{Experiments on coherent transitions}
\label{ch:flopping}

Rabi flopping experiments investigate the Rabi oscillations between
the electronic states $\left|\down\right>$ and $\left|\up\right>$.
All these experiments can be characterised by some
common parameters: the intensity $I_1$, $I_2$ of the two Raman beams,
their mutual frequency difference $\omega_R$, their detuning
$\Delta_R$ with respect to the $3P_{3/2}$ level and the pulse
duration (duration that both Raman beams shine onto the ion)
$\tau$. \index{acquisition!statistics} For each set of parameters, we
repeat $100$\,--\,$1000$ identical experiments to acquire statistics
on superposition states. As described in \cite{kielpinski:thesis}, two
kinds of acquisition are of special interest:
\bitem
\item \index{duration scan} Increasing the pulse duration $\tau$ after
one set of repeated identical experiments while leaving all other
parameters unchanged. With this setup we measure the Rabi flopping in
dependence of time, see \autoref{fig:timescan}.
\item \index{frequency scan} Scanning the frequency difference
$\omega_R$ where all other parameters are kept fixed. If we adjust
$\tau$ to be the $\pi$ pulse duration for distinct carrier or sideband
transitions, we will observe dips in the lower-state population. Their
abscissa (frequency difference $\omega_R$) will provide us with the
frequency of the transitions, see \autoref{fig:frequencyscan}.
\eitem
Depending on the geometric configuration of the laser beams we will be
able to measure different Raman Rabi frequencies corresponding to different
transitions.

\bfig
\subfloat[Rabi flopping in dependence of the pulse duration.]{
  \includegraphics{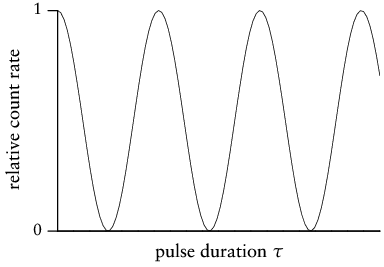}
  \label{fig:timescan}
}
\hfill
\subfloat[Rabi flopping in dependence of the frequency difference
between the two Raman laser beams.]{
  \includegraphics{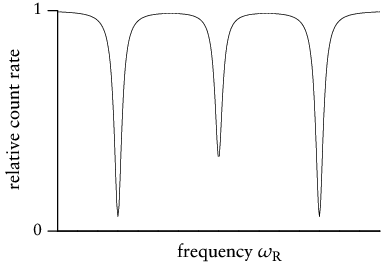}
  \label{fig:frequencyscan}
}
\caption{Sample plots of ideal Rabi flopping curves.}
\label{fig:sample_flopping}
\efig

For all experiments we will have to consider the frequency resolution
of the two-photon stimulated Raman transitions. A quick estimation
(properties of the Fourier transformation) yields
\be
\Delta\omega_R \approx 1/\tau \text{.}
\label{eq:fourier}
\ee
Now assume that we want to drive a Rabi flopping until a particular
oscillation phase $\theta \in [0,\inf)$ has been reached. The duration
$\tau$ it takes to do this can be modified by appropriately adjusting
the Raman Rabi frequency $\Omega_R \rightarrow \theta/\tau$ which in
turn is determined by the intensities of the two Raman beams, see
\autoeqref{eq:coprop_rabi_intensity}. That way, the frequency
resolution can be adapted to the experimental requirements even for a
constant phase~$\theta$.

\section{Common steps of a flopping experiment}

We prepared a single ion in the experimentation region of the trap
\index{confinement strength} confined in a radial potential of
$\omega_\text{radial} \approx 2\pi \cdot 6\MHz$ and an axial potential
of $\omega_\text{axial} \approx 2\pi \cdot 2\MHz$. The magnetic
quantisation field \index{magnetic field!strength} of about
$5.6\gauss$ was aligned to the propagation direction of the \ac{BD} beam
(maximising the fluorescent scattering rate by adjusting the
respective currents flowing through the two spatial compensation
coils). This allows for optimising the $\sigma^+$ polarisation of the
\ac{BD} beam by turning and tilting the $\lambda/4$ waveplate in front
of port~\ac{H} (see \autoref{fig:beamline}). Micromotion was
compensated using a correlation method
(\autoref{sec:mm_correlation}). At the usual operating characteristics
for the \ac{BD} beam~(\autoref{tab:bd}) we observed a \index{count
rate!PMT@\ac{PMT}} \ac{PMT} count rate of around $210\kHz$.

For each set of parameters $(\Delta\omega,\tau)$ we ran a \index{pulse
programme} pulse programme consisting of
\benu
\item Doppler cooling. Both the \ac{BD} and the \ac{BD} detuned beams
were enabled during a period of $2\millisecond$.
\item Two-photon stimulated Raman pulse. The frequency difference
$\Delta\omega$ or the pulse duration $\tau$ of the two Raman beams
could be varied. The detuning $\Delta_R$ of the virtual Raman level
with respect to the $3P_{3/2}$ level was $80\GHz$. \label{it:tps-raman}
\item Detection pulse. With the \ac{BD} detuned beam switched off, the
\ac{BD} beam was engaged for $20\microsecond$. The resulting
fluorescent scattering rate was measured using the \ac{PMT}.
\eenu

\noindent The detection step of the pulse programme will either yield ``bright''
(scattering photons) or ``dark'' (scattering no photons) depending on
whether the qubit state eventually collapses to $\left|\down\right>$ or
$\left|\up\right>$. We will have to conduct a statistical analysis in
order to be able to make a statement on superposition states. Repeating
the pulse programme 100 to 1000 times for each set of parameters
$(\Delta\omega,\tau)$ provides sufficient data for this
purpose.

\section{\texorpdfstring{``Hot''}{{\textquotedblleft}Hot\textquotedblright} flopping using co-propagating Raman beams}

As we saw in \autoref{sec:tps-raman}, the easiest way to observe
Rabi oscillations induced by a two-photon stimulated Raman transition
is to use parallel laser beams. For this geometry the Raman Rabi frequency
does not depend on the motional state number $n$, i.\,e. the transfer
of electronic population between $\left|\down\right>$ and
$\left|\up\right>$ occurs with equal speeds for all motional states. We
thus expect to completely deplete the $\left|\down\right>$ state for
an ion initially in the $\left|\down\right>$ state if we apply an adequate
$\pi$-pulse. Subsequent state detection using the \ac{BD} laser should
not scatter any photons. As the experimental result does not depend on
the particular occupation of the motional states, this type of
experiments is ideally suited to conduct first flopping experiments
after Doppler precooling~(\autoref{sec:doppler_cooling}). We use
the B$1$ and R$1$ beams (see \autoref{fig:beamline} for the beam
nomenclature) for the co-propagating geometry. For the experiments
described below we worked with the following beam parameters:
\index{Raman beam!intensity}
$I_\text{B$1$} = 2.9\cdot10^5\mWpersqcm$, $I_\text{R$1$} =
6.0\cdot10^5\mWpersqcm$, $\Delta_R = 2\pi\cdot80\GHz$.

\subsection{Frequency scan}
\index{frequency scan!parallel geometry}

For a fixed pulse time $\tau_\pi = 5.8\microsecond$ we scanned the
frequency difference $\omega_R$ of the two Raman beams, see
\autoref{fig:coscan}. We can clearly identify the dip in the
fluorescence rate at $\omega_R = 1775.28\MHz$ which represents the
carrier flopping frequency for the transition $\left|\down,n\right>
\rightarrow \left|\up,n\right>$. The width $\Delta\omega_R \approx
0.14\MHz$ fits our expectations from \autoeqref{eq:fourier} within an
uncertainty of $5\%$. For the detection period of $20\microsecond$ the
\index{contrast} contrast (=\,depth of the dip in relation to the
overall signal height) exceeds $93\%$. The fitted curve of
\autoref{fig:coscan} should have a Lorentzian shape arising from the
fact that the pulse time is limited. We however found that a Gaussian
fitted the data better than a Lorentzian. One possible reason might be
modulations of the Raman Rabi frequency $\Omega_R$ during the
experimentation (see also the next section). These cause an actually
Lorentzian-shaped line to broaden to a Gaussian shape.

\bfig
\includegraphics{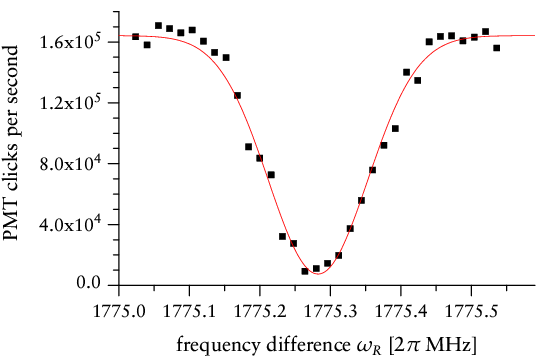}
\caption{Frequency scan for a co-propagating geometry. Note that the
fitted curve is not Lorentzian but Gaussian (see main text for
possible reasons).}
\label{fig:coscan}
\efig

\subsection{Duration scan}
\index{duration scan!parallel geometry}

For the two-photon stimulated Raman pulse~(\hyperref[it:tps-raman]{step~\ref*{it:tps-raman}} of
our pulse programme) we used the co-propagating B$1$ and R$1$
beams (see \autoref{fig:beamline} for the beam nomenclature). The
carrier Rabi flopping data are plotted in \autoref{fig:co_flopping}.
Note that the actual flopping begins at $\tau = 1.2\microsecond$,
which can be attributed to an imperfect synchronisation of the B$1$
and R$1$ beams. Apparently, the two beams do not overlap temporally
for $\tau < 1.2\microsecond$. (This effect will be suppressed in
future experiments by appropriately adjusting the pulse control
output.) Therefore, the actual $\pi$ pulse duration $\tau_\pi$ is not given
by the $\tau$ coordinate where the flopping curve passes its first
minimum. A better method is to measure the duration between two
subsequent flopping curve minima and divide the difference by two. We
thus estimate $\tau_\pi = 2.47\microsecond$.

\begin{sidewaysfigure}\centering
\includegraphics{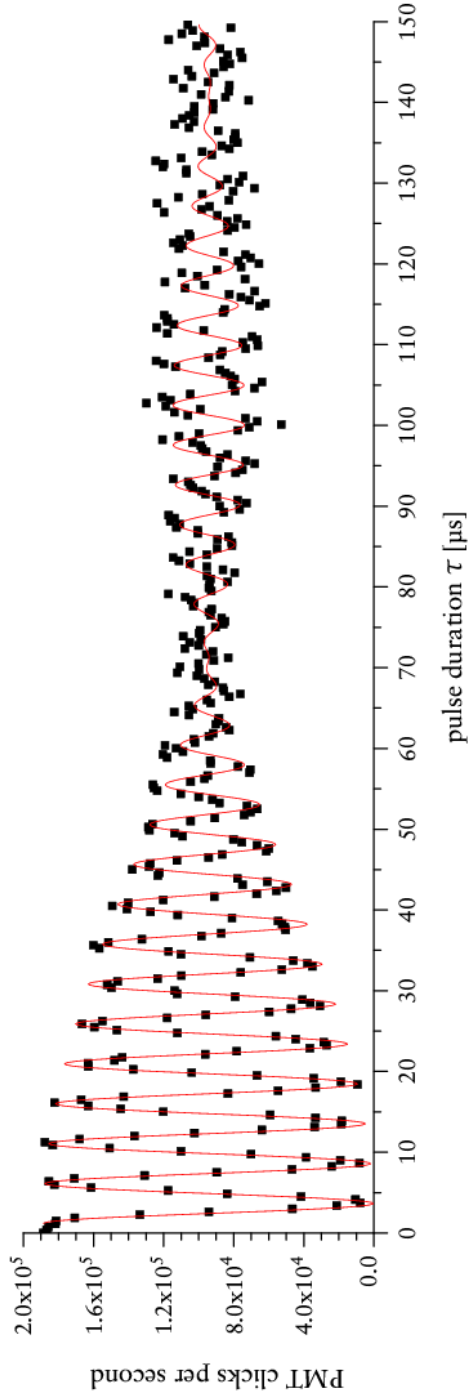}
\caption{Flopping curve for the co-propagating Raman beam
geometry. Due to imperfect temporal overlap of the two Raman beams the
flopping begins at $\tau = 1.2\microsecond$ instead of zero. The
beating structure can be attributed to an unstable Raman Rabi
frequency that shows a maximum deviation of $\pm 2\pi \cdot 7\kHz$
from the central Raman Rabi frequency $\Omega_R = 2\pi \cdot
202\kHz$. See the main text for possible reasons and remedies to this
unstability.}
\label{fig:co_flopping}
\end{sidewaysfigure}

The flopping contrast continuously decreases, but eventually
experiences a revival after $\tau = 70\microsecond$. We explain this
beating signal as follows: Each data point in the plot is the average of
$500$ identical experiments. Assume that the Raman Rabi frequency
$\Omega_R$ is not absolutely constant over time, but fluctuates
slightly from experiment to experiment with $\epsilon$ being the
maximum deviation from $\Omega_R$. Further assume that these
deviations are all equally probable. As the Rabi flopping for the
$\left|\down\right>$ population $\rho_{aa}$ at a constant Raman Rabi
frequency is described by
\be
\rho_{aa,\Omega_R}(t) = \frac{1}{2} \left(1 + \cos\Omega_R t\right)
\ee
we expect the averaged curve fitted to the acquired data to satisfy
\be
\begin{split}
\rho_{aa}(t) &= \frac{1}{2\epsilon} \int\limits_{-\epsilon}^{\epsilon} \rho_{aa,\Omega_R+\xi}(t) \mathrm{d}\xi \\
             &= \frac{1}{2} \left(1 + \frac{\sin\epsilon t}{\epsilon t} \cos{\Omega_R t}\right) \text{.}
\end{split}
\label{eq:revival}
\ee
\hyperref[fig:co_flopping]{Fig.~\ref*{fig:co_flopping}} shows the fit of this function to our data
for a time offset of $\tau = 1.2\microsecond$ (for the same reasons
noted above). According to \autoeqref{eq:revival} we can identify the
envelope frequency with the maximum deviation of the Raman Rabi
frequency. From the fit parameters we deduce $\Omega_R \in \left[2\pi
\cdot 195\kHz \mid 2\pi \cdot 209\kHz\right]$.

The fluctuations of the Raman Rabi frequency can most probably be attributed
to two culprits introducing noise into our system:
\bitem
\item \index{Raman beam!intensity fluctuation} Raman beam
intensity. The complex beam generation process (\autoref{sec:shg})
causes fluctuations of the beam intensity on the microsecond timescale
by as much as $4\%$ \cite{friedenauer}. The Raman Rabi frequency
fluctuations introduced hereby, see \autoeqref{eq:coprop_rabi_frequency},
fit our measurements quite well. In the final experimental setup, we
will use ``noise eaters'' in the Raman beamline that stabilise their
intensity to $\pm1\%$ at the cost of reducing overall beam power.
\item \index{magnetic field!fluctuation} Magnetic quantisation
field. Due to various line-powered electric devices in the vicinity of
our trap the strength of the magnetic field at the trap centre will
probably oscillate with the line frequency~($50\Hz$). Perturbations
might also be caused by external electromagnetic fields at higher
frequencies. The effects are different Zeeman shifts on the one hand
and a tilt of the quantisation axis on the other. The former will
effectively detune the two-photon stimulated Raman process from the
$\left|\down\right> \leftrightarrow \left|\up\right>$ transition
thereby altering the Raman Rabi frequency, see
\autoeqref{eq:effective_rabi}. The latter in contrast causes imperfect
polarisations of the Raman beams (a second order effect). Thus, only a
fraction of their overall intensity will contribute to driving the
two-photon stimulated Raman transition---with the same consequences as
outlined above. We expect to considerably decrease the magnetic field
noise by putting into operation the temporal compensation coil together
with its associated electronics, see \autoref{sec:magnetic_temporal}.
\eitem

\section{\texorpdfstring{``Hot''}{{\textquotedblleft}Hot\textquotedblright} flopping using perpendicular Raman beams}
\label{sec:sb_flopping}

For two Raman beams propagating perpendicularly to each other all
Raman Rabi frequencies depend on the motional state number $n$. This
is even true for the carrier pulse $\left|\down,n\right> \leftrightarrow
\left|\up,n\right>$. Observation of flopping for the population not
residing in the motional ground state will thus not be as clear
as in the co-propagating case. As an additional feature of the
perpendicular geometry however we will be able to drive sideband
transitions $\left|\down,n\right> \leftrightarrow
\left|\up,n\pm1\right>$. This is the most important ingredient for
resolved-sideband cooling, see \autoref{sec:sb_cooling}. In our
setup, the orthogonal Raman beams are B$1$ and R$2$ (see
\autoref{fig:beamline} for the beam nomenclature). For the
experiments described in the following sections, we used the following
laser beam parameters:
$I_\text{B$1$} = 2.9\cdot10^5\mWpersqcm$, $I_\text{R$2$} =
6.0\cdot10^5\mWpersqcm$, $\Delta_R = 2\pi\cdot80\GHz$.

\subsection{Frequency scan}
\label{sec:orthoscan}
\index{frequency scan!orthogonal geometry}

At $\pi$ pulse duration that was $5$ times as long as in the
co-propagating case we could clearly resolve the carrier as well as
the \index{sideband!transition} first red and the first blue sideband
in a frequency scan, see \autoref{fig:orthoscan}. For our Lamb-Dicke
parameter of $\eta \approx 0.3$ \autoeqref{eq:ortho_rabi_frequency}
suggests a $\pi$ pulse duration of only $3.5$ times the co-propagating
$\pi$ pulse duration. This mismatch was probably caused by a reduced
R$2$ laser beam intensity due to a non-optimal overlap with the ion.

\begin{figure}[t]\centering
\includegraphics{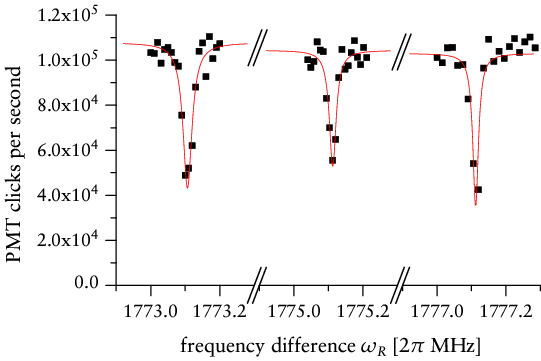}
\caption{Frequency scan for perpendicular Raman beams. The $\pi$ pulse
duration $\tau_\pi$ was adjusted at $5$ times the $\pi$ pulse duration of
the co-propagating case. Note that the abscissa is segmented into
three regions each displaying (from left to right): first red sideband
transition, carrier transition, and first blue sideband
transition. The contrast of the blue sideband dip is increased
compared to the red sideband dip as the red sideband transition cannot
be driven for the motional state $n=0$. The carrier dip appears
suppressed; a different pulse duration is necessary to drive the
associated transition at a higher fidelity.}
\label{fig:orthoscan}
\efig

We can estimate the \index{temperature!ion} temperature of our ion
from the relative contrast of the sideband dips (=\,difference of the
fluorescence rates) in the frequency scan. Due to the fact that we
cannot drive a first red sideband transition of the motional level
$n=0$ (there is no lower motional level), the contrast for the red
sideband dip will be smaller than for the blue sideband dip (where we
can drive transitions for any motional level). The mean motional
quantum number \index{motional quantum number!mean} can be estimated
by comparing the depths of the first red and the blue sideband dips
$A_\text{rsb}$, $A_\text{bsb}$. According to \cite{leibfried:habil} we
have
\be
\overline{n} = \frac{r}{1-r}
\label{eq:mean_quantum_number}
\ee
where $r = A_\text{rsb}/A_\text{bsb}$. Although the fits of
\autoref{fig:orthoscan} could be improved by measuring the dips in
more dense intervals, there is evidence that $\overline{n} \approx
15$. Assumed that the motional spectrum is thermal, we deduce $T
\approx 1.5\millikelvin$.

The spacing between the carrier frequency and the sideband frequencies
is determined by the frequency of the motional quanta coupled to the
electronic qubit states. From \autoref{fig:orthoscan} we evaluate
$\omega_z = 2\pi \cdot 2.00\MHz$. The voltages of the experimentation
region subelectrodes \ac{D}, \ac{E}, and \ac{F} were $10\volt$,
$-30\volt$, and $10\volt$ each. For these values, the axial frequency
estimated using \autoeqref{eq:axial_freq} matches our measurement
within a tolerance of $0.1\%$.

\subsection{Duration scan}
\index{duration scan!orthogonal geometry}

Unless the ion's motion has been cooled to its ground state, flopping
curves for an orthogonal geometry will quickly ``wash out'' due to the
superposition of the different Raman Rabi frequencies involved. We
calculated the expected flopping curves of the carrier and the first
red sideband for an ion in our trap Doppler-cooled to different
temperatures $T_D$, see \autoref{fig:ortho_flopping_sim}.

\bfig
\subfloat[Carrier flopping curve for $T = 0.1\millikelvin$.]{
  \includegraphics{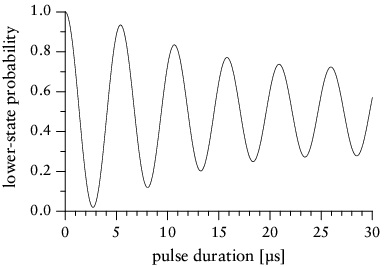}}
\hfill
\subfloat[First red sideband flopping curve for $T = 0.1\millikelvin$.]{
  \includegraphics{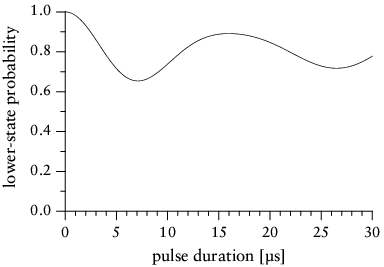}}
\\
\subfloat[Carrier flopping curve for $T = 1\millikelvin$.]{
  \includegraphics{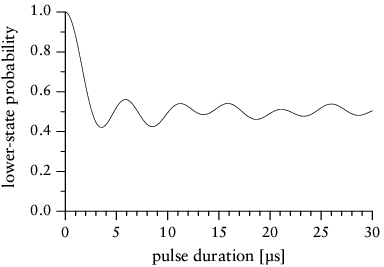}}
\hfill
\subfloat[First red sideband flopping curve for $T = 1\millikelvin$.]{
  \includegraphics{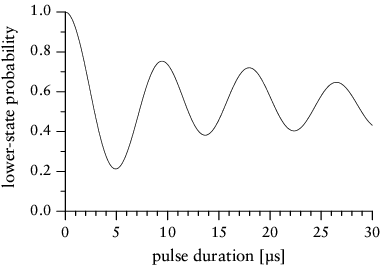}}
\\
\subfloat[Carrier flopping curve for $T = 2\millikelvin$.]{
  \includegraphics{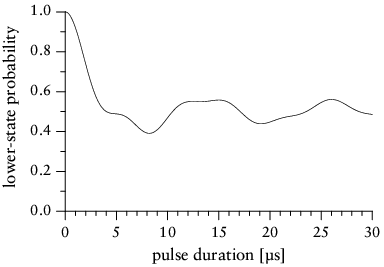}}
\hfill
\subfloat[First red sideband flopping curve for $T = 2\millikelvin$.]{
  \includegraphics{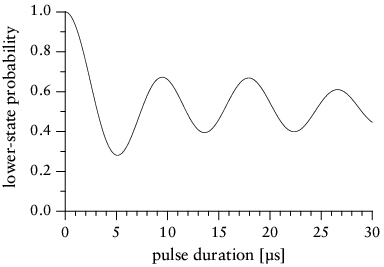}}
\caption{Simulated flopping curves for the electronic population in
the $\left|\down\right>$ state using perpendicular Raman beams for the
two-photon stimulated Raman transitions at an axial confinement of
$\omega_\text{axial} = 2\pi \cdot 2\MHz$ ($\eta = 0.3$), a ``base''
Raman Rabi frequency of $\Omega_R = 2\pi \cdot 202\kHz$, and an ion
temperature of $T_D = 1\millikelvin$.}
\label{fig:ortho_flopping_sim}
\efig

Experimental data of our measurements is plotted in
\autoref{fig:ortho_flopping_exp}. The pulse durations of the
calculated and experimental data differ by a factor of about $4$. This is
possibly caused by a misaligned R$2$ beam, but was still under
investigation at the printing time of this thesis. Even so, we can see
that the characteristics of the curves closely resemble the calculated
curves for $T = 1\millikelvin$.

\bfig
\subfloat[Carrier flopping curve.]{
  \includegraphics{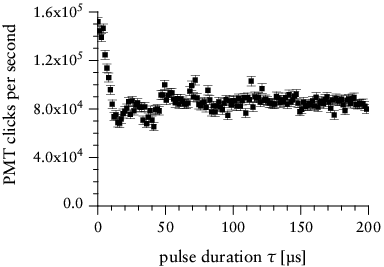}}
\hfill
\subfloat[First red sideband flopping curve.]{
  \includegraphics{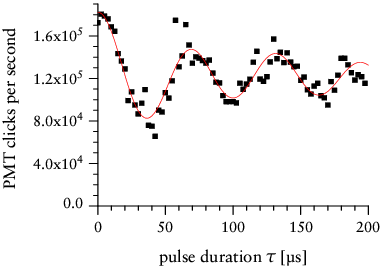}}
\caption{Actually measured flopping curves using perpendicular Raman
beams for the two-photon stimulated Raman transitions at an axial
confinement of $\omega_\text{axial} = 2\pi \cdot 2\MHz$ ($\eta =
0.3$). The ``base'' Raman Rabi frequency had been measured in a prior
co-propagating Rabi flopping timescan to be $\Omega_R = 2\pi \cdot
202\kHz$ (see \autoref{sec:motion_coupling} for further
explanations on the ``base'' frequency). We were unable to provide a
good fit to the carrier flopping data. Still, we can see that the
pulse durations of the calculated (\autoref{fig:ortho_flopping_sim})
and experimental data are rather different. Possibly, the redder Raman
beam R$2$ was not properly aligned thus providing less intensity and
decreasing the Raman Rabi frequency.}
\label{fig:ortho_flopping_exp}
\efig

\section{Shelving}
\label{sec:shelving}
\index{shelving}

As noted in \autoref{sec:state_detection} we might off-resonantly
scatter photons from the $\left|\up\right>$ state when the \ac{BD}
detection beam is engaged. This process eventually pumps
$\left|\up\right>$ into the $\left|\down\right>$ state thereby
unwantedly scattering many photons within the cycling transition.
We suppose to weaken this effect by applying additional shelving
pulses before actual readout occurs. The shelving pulses are
two-photon stimulated Raman transitions which will transfer population
from $\left|\up\right>$ into states of the $3S_{1/2}$ manifold with a
smaller $m_F$ value, see \autoref{fig:shelving}.

\bfig
\includegraphics{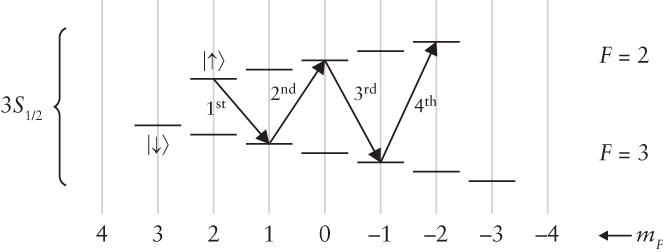}
\caption{Shelving pulses inside the $3S_{1/2}$ manifold. Each shelving
transition is provided via a two-photon stimulated Raman pulse. As the
two Raman beams are polarised $\pi$~(bluer beam) and
$\sigma^+$~(redder beam), the population will be shifted by $\Delta
m_F = -1$ with each pulse.}
\label{fig:shelving}
\efig

We may also use this technique to determine the actual magnetic field
strength \index{magnetic field!strength|main} at the trap centre by
measuring the Zeeman shifts of the involved states. The Zeeman shifts
can be calculated as follows:
\be
\Delta\omega = m_F g_F \mu_B B\,/\,\hbar
\ee
where $g_F$ is the Landé factor ($+1/3$ for $3S_{1/2}
\left|F=3\right>$, $-1/3$ for $3S_{1/2} \left|F=2\right>$), $\mu_B$ is
the Bohr magneton and $B$ is the strength of the magnetic field. Thus,
the $i$th shelving pulse frequency $\omega_i$, i.\,e. the frequency
difference of the two Raman beams providing the $i$th shelving pulse,
is determined by
\be
\omega_i = \omega_0 + \frac{2i-5}{3} \mu_B B\,/\,\hbar
\ee
where $\omega_0$ indicates the hyperfine splitting of the $3S_{1/2}$
manifold. With a current of $I = 6.000\ampere$ flowing through the
field-generating coils of \autoref{sec:magnetic_field} we measured
$\omega_1 = 1781.02\MHz$, $\omega_2 = 1786.24\GHz$, which yields $B =
5.589(15)\gauss$ and $\omega_0 = 1788.850(15)\MHz$. This result
differs from the calculated field strength by about $44\%$ (see
\autoref{sec:magnetic_field}).

\section{Cooling to the motional ground state}

We extended the pulse programme to include a step of resolved-sideband
cooling right after having Doppler-cooled the ion. It consisted of
sideband-cooling and repumping cycles $\left|\down,n\right> \rightarrow
\left|\down,n-1\right>$ for $n = 30 \ldots 1$ where each cooling cycle
was run twice in succession.

\index{frequency scan!ground-state cooled}
At a $\pi$ pulse duration of $\tau_\pi = 40\microsecond$ we observed a
substantial decrease in contrast of the first red sideband, see
\autoref{fig:sb_scan}. The mean motional quantum number
\index{motional quantum number!mean|main} calculated according to
\autoeqref{eq:mean_quantum_number} is $\overline{n} = 0.65$. Assumed
that the occupation of motional states resembles a thermal
distribution, this corresponds to a sideband-cooled temperature of
$T_\text{SB} = 0.1\millikelvin$ which means that more than $60\%$ of
the population are in the motional ground state $n=0$.

Note that this estimation is not valid for the entire population. Part
of the population was ``lost'' in the $3S_{1/2} \left|F=3,
m_F=2\right>$ state (see \autoref{fig:ground_flop}) due to an
unreliable repumper beam during the cooling process (\ac{RD} was
present though!). It is however correct for the portion of the
population that was successfully transferred into the
$\left|\down\right>$ state.

\bfig
\includegraphics{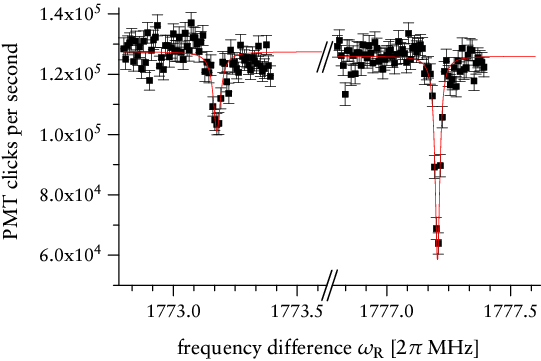}
\caption{Sideband dips of a frequency scan using an orthogonal Raman
beam geometry at a $\pi$ pulse duration of $\tau_\pi =
40\microsecond$. Compared to \autoref{fig:orthoscan} we included an
additional step of resolved-sideband cooling in the pulse
programme. This reduced the contrast of the first red sideband
dip. From the ratio of the dip depths we estimate $\overline{n}=0.65$.}
\label{fig:sb_scan}
\efig

In the case of ground-state cooled ions, we have a predominant
contribution of the motional state $n=0$ to the overall distribution
of motional levels. Thus, flopping curves will not wash out even if
we use an orthogonal Raman beam geometry. See \autoref{fig:ground_flop}
for the graphs of a carrier and a first blue sideband flopping curve.
As flopping of the post-selected population in the
$\left|\down\right>$ is now mainly determined by the $n=0$ state, we
expect from \autoeqref{eq:ortho_rabi_frequency} that the ratio of the
respective $\pi$ pulse durations is given by
\be
\frac{\tau_{\pi,\text{carrier}}}{\tau_{\pi,\text{bsb}}} \approx \eta
\ee
in very good accordance with the ratio determined from the graphs. The
flopping curves have a maximum contrast \index{duration scan!ground-state cooled}
of $40\%$ which differs substantially from the expected near-$100\%$.
As mentioned before, the reason is an unreliably applied repumper
beam. A considerable amount of population will not have been shifted
into the $\left|\down\right>$ state after resolved-sideband cooling
thus causing a loss of contrast. We expect to attain a contrast close
to $100\%$ for future measurements involving a reliable repumper beam.

\bfig
\subfloat[Carrier flopping curve.]{
  \includegraphics{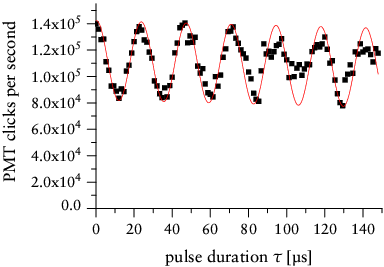}}
\hfill
\subfloat[First blue sideband flopping curve.]{
  \includegraphics{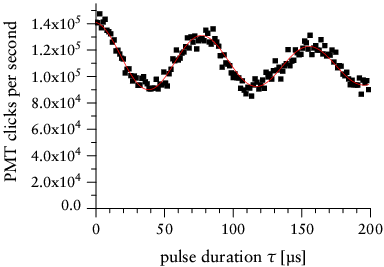}}
\caption{Flopping curves for a ground-state cooled ion using an
orthogonal Raman beam geometry. Although the fit for the carrier
flopping curve might be too optimistic, there is a clear difference to
the curves of \autoref{fig:ortho_flopping_exp}. The ratio of the two
Raman Rabi frequencies (determined from the respective $\pi$ pulse
durations) is $0.3$ in good accordance with the expected ratio of
$\eta$. See the main text for reasons on the low contrast.}
\label{fig:ground_flop}
\efig

\appendix
\settocdepth{chapter}

\ifpdfoutput{
  \cleardoublepage\phantomsection
  \pdfbookmark[-1]{\appendixname}{appendix}
}{}
\part*{\appendixname}
\chapter{Auge visualisation and data acquisition software}
\label{app:auge}
\index{Auge software@\textit{Auge} software}

The \textit{Auge} software has been written in C++ in order to
facilitate inspection and maintenance and at the same time maximise
the programme's execution speed. As we are dealing with large amounts
of data (up to $9\megabyte/\unitsignonly{\second}$) efficient data
handling becomes important. \hyperref[fig:auge_interaction]{Fig.~\ref*{fig:auge_interaction}}
summarises the interaction and data flow between the particular
objects.

We may assign each object to one of the three application
layers. These are: background logic (=\,business logic), user
interface logic, and network logic. The background logic layer deals
with the acquisition, storage, processing, accession, and analysis of
data, the user interface layer presents these data on screen and
provides user feedback, while the network layer provides a
bidirectional communication facility with our control software
\textit{Flocke}.

As a Windows (Win32) application \textit{Auge} continously runs an
event loop thus polling and reacting on Windows or user events. In
parallel, it runs an \index{Auge software@\textit{Auge} software!thread}
infinitely looping thread that unintermittently reads out newly
available camera data. These two threads have to be synchronised on
the following occasions:
\bitem
\item When readout parameters are changed. This ensures that possible
memory allocation\,/\,deallocation processes are carried out properly.
\item When camera acquisition is started\,/\,stopped. This ensures
that an acquisition in progress is not cancelled.
\eitem
\textit{DataController} as the main object for data accession and
processing is used as a synchronising mutex object.

Where function user interface objects call of methods of background
logic objects, these calls occur directly. The other way round,
function calls occur indirectly: User interface objects have to
register themselves as listeners for events originating from
background logic objects. The philosophy behind this approach is to
anchor a scheme within the design of the software that reflects the
user's control.

\begin{sidewaysfigure}\centering
\includegraphics{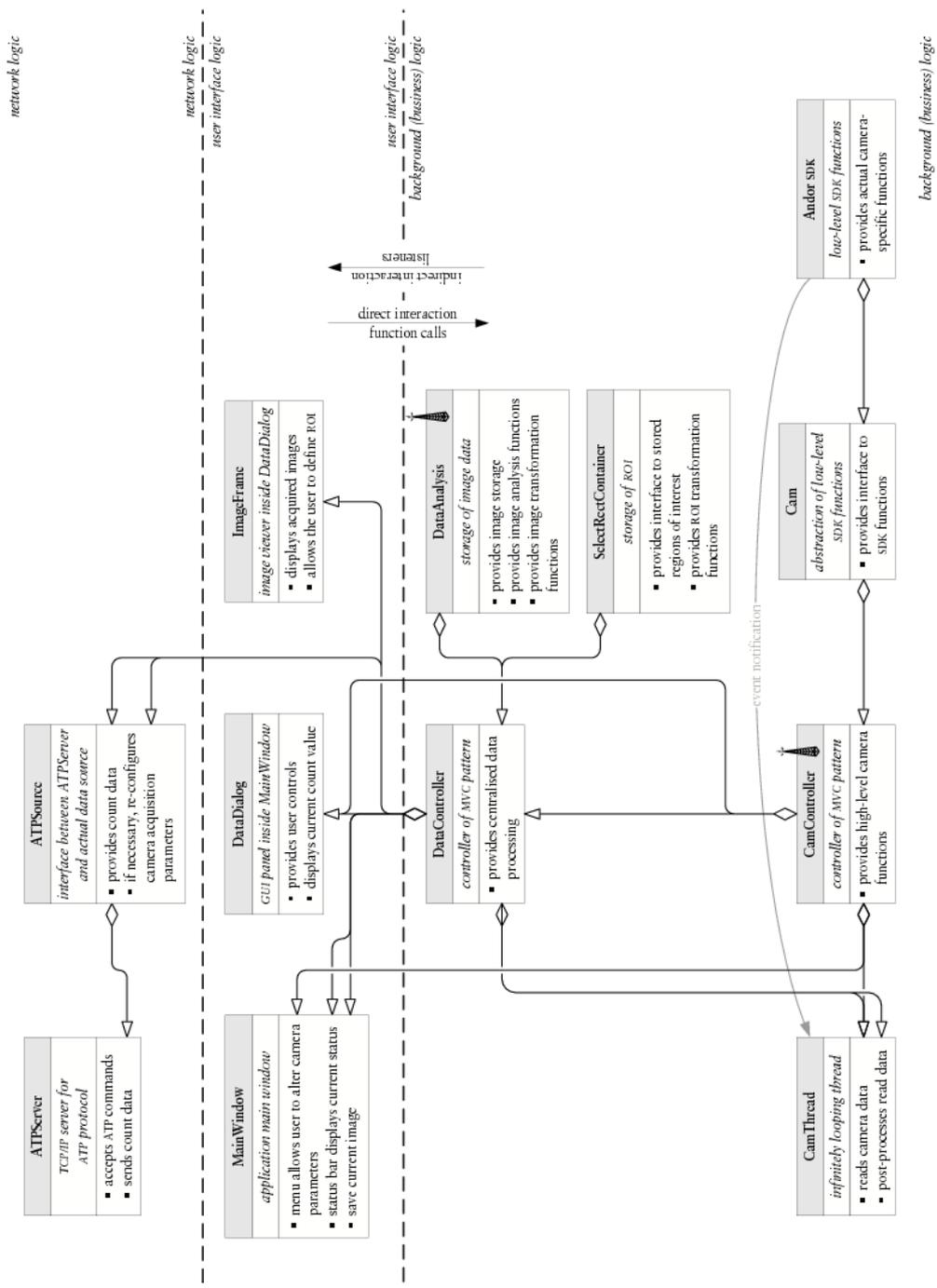}
\caption{Interaction between the particular objects of
\textit{Auge}. Arrows with transparent heads indicate the direction of
data flow. The radio masts indicate event sources that user interface
logic objects can register with.}
\label{fig:auge_interaction}
\end{sidewaysfigure}

Ownership of the involved objects is arranged in a somewhat simpler
way, see \autoref{fig:auge_ownership}. Eventually, all objects are
owned by the \textit{MainWindow}. If the \textit{MainWindow} is
destroyed, i.\,e. when the application quits, all other objects will
thus be disposed of automatically.

\begin{sidewaysfigure}\centering
\includegraphics{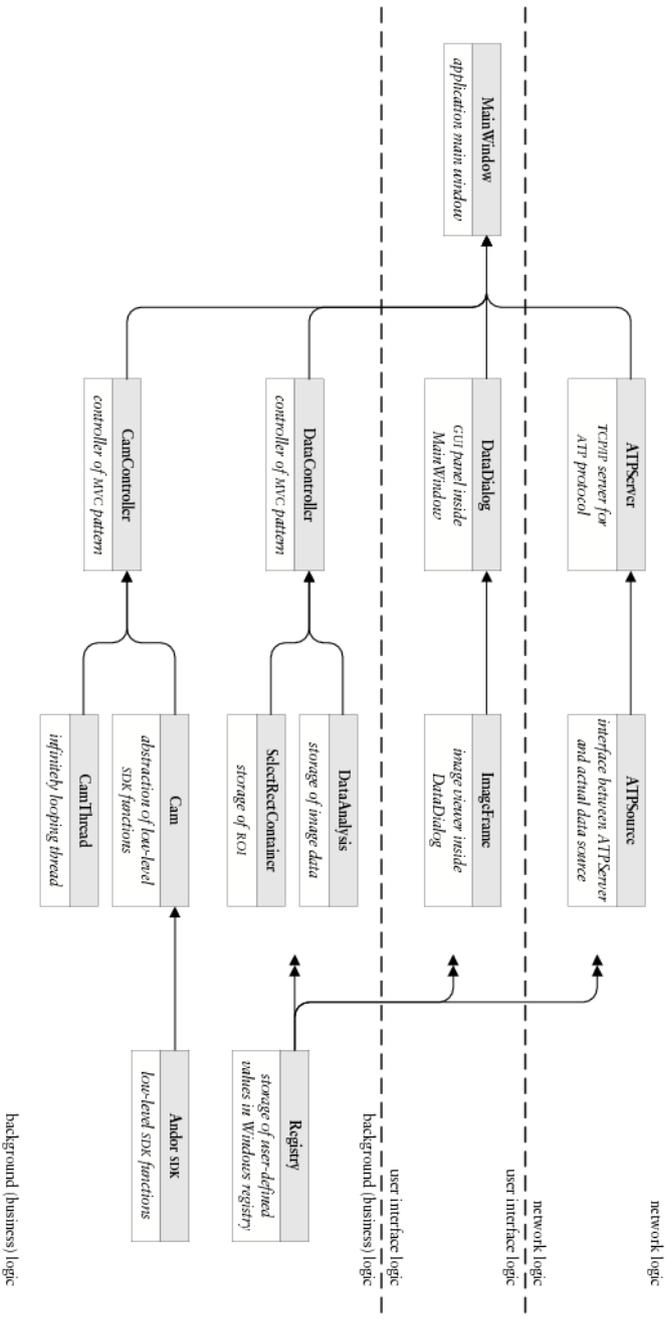}
\caption{Ownership of the particular objects of \textit{Auge}. The
\textit{Registry} object follows the Singleton pattern and can as such
not be assigned to a particular owner.}
\label{fig:auge_ownership}
\end{sidewaysfigure}

\chapter{Auge Transfer Protocol (ATP)}
\label{app:atp}
\index{ATP@\protect\ac{ATP}|see{Auge Transfer Protocol}}
\index{Auge Transfer Protocol}

The Auge Transfer Protocol (\ac{ATP}) defines a communication
interface between our visualisation and data acquisition software
\textit{Auge} and an analysis software (which can be arbitrary as long
as it keeps to the rules defined in this section). In the framework of
this protocol \textit{Auge} provides \index{count rate!Auge@\textit{Auge}}
count rates (confined to previously defined elliptically shaped
regions of interest, \ac{ROI}) which can in turn be read by the
analysis software.

\textit{Auge} provides the data as a \ac{TCP/IP} server. Normally, the
server will run on port 1899. The analysis software connects to
\textit{Auge} as required and directs commands to \textit{Auge}. These
commands will instantaneously be processed and replied to where the
replies follow an \ac{HTTP}-like scheme\footnote{\href{http://www.faqs.org/rfcs/rfc2616.html}{http://www.faqs.org/rfcs/rfc2616.html}}.

\section{Typical course of an ATP connection}

\begin{center}
\btab{rrcl}\toprule
                  & Client                 &                   & Server \\ \midrule
0.                &                        & $\longleftarrow$  & Status \\
\multirow{2}*{1.} & command \texttt{CT}    & $\longrightarrow$ & \\
                  &                        & $\longleftarrow$  & reply \texttt{CT} \\
\multirow{2}*{2.} & command \texttt{START} & $\longrightarrow$ & \\
                  &                        & $\longleftarrow$  & reply \texttt{START} \\
\multirow{2}*{3.} & command \texttt{STOP}  & $\longrightarrow$ & \\
                  &                        & $\longleftarrow$  & reply \texttt{STOP} \\
\multirow{2}*{4.} & command \texttt{QUIT}  & $\longrightarrow$ & \\
                  &                        & $\longleftarrow$  & reply \texttt{QUIT} \\ \bottomrule
\etab
\end{center}

\section{Reply and status codes}

A reply consists of a header and a data block where the header is
compulsory. The data block may be omitted depending on the type of
reply. Header and data block are separated from each other by
\ac{CRLF}. The syntax of the header is
\begin{quote}
\texttt{nnn{\space}xxxxx}
\end{quote}
where \texttt{nnn} is the answer code expressed as a three-digit
decimal integer number and \texttt{xxxx} contains a textual
representation of the reply. Each reply code can be assigned to a
family, see \autoref{tab:status_codes}. The reply data block may
contain arbitrary data and does not follow a predefined formatting
scheme.

\bt
\btab{rl}\toprule
reply code & description \\
\cmidrule(r){1--1}
\cmidrule(r){2--2}
$100-199$ & status notifications \\
$200-299$ & replies containing numerical values \\
$300-399$ & replies containing binary data in their data block \\
$400-499$ & error messages \\ \bottomrule
\etab
\caption{Reply code families.}
\label{tab:status_codes}
\et

\section{Commands and their replies}

\subsection{Error replies}

Error replies can be sent any time during an \ac{ATP}
connection. Two error codes are defined:

\btab{rp{11cm}}
\texttt{400} & \texttt{syntax error} \\
             & The command directed at the server contains a syntax error. \\
\texttt{401} & \texttt{unknown error} \\
             & An unknown error has occurred.
\etab

\subsection{Start of an \texorpdfstring{\protect\ac{ATP}}{ATP} connection}

Upon the start of an \ac{ATP} connection the \ac{ATP} server will
provide the client with a status code that signifies the current
status of the server. Possible status codes are:

\btab{rp{11cm}}
\texttt{100} & \texttt{ready} \\
             & \textit{Auge} is ready to accept commands. \\
\texttt{101} & \texttt{busy} \\
             & \textit{Auge} cannot process and commands. The \ac{ATP}
connection is closed immediately after this message. \\
\texttt{102} & \texttt{no data available} \\
             & \textit{Auge} cannot provide any data. The \ac{ATP}
connection is closed immediately after this message.
\etab

\subsection{The \texttt{CT} command (Cycle Time)}

With the \texttt{CT} command the client tells \textit{Auge} what
maximum timespan it expects between subsequent acquisitions of the
\ac{CCD} camera. This \textit{must} be the first command of any
\ac{ATP} connection. The syntax is
\begin{quote}
\texttt{CT f}
\end{quote}
where \texttt{f} is the duration between subsequent acquisitions in
seconds. It must be formatted as a decimal number with the period
(\texttt{.}) as a decimal separator.\\
Possible reply codes:

\btab{rp{11cm}}
\texttt{200} & \texttt{n} \\
             & \texttt{n} is the number of acquisitions per readout
process. The client must ensure that the total number of acquisitions
equals an integer multiple of \texttt{n}. \\
\texttt{402} & \texttt{desired cycle time cannot be achieved} \\
             & The camera cannot deliver data with the desired speed.
\etab

\subsection{The \texttt{ET} command (Exposure Time)}

With the \texttt{ET} command the client tells \textit{Auge} to
determine and return the current exposure time of the \ac{CCD}
camera. This command is \textit{always} valid. The syntax is
\begin{quote}
\texttt{ET}
\end{quote}

\pagebreak\noindent Possible reply codes:

\btab{rp{11cm}}
\texttt{201} & \texttt{f} \\
             & \texttt{f} is the exposure time of the \ac{CCD} camera
in seconds. It is formatted as a decimal number with the period
(\texttt{.}) as a decimal separator.
\etab

\subsection{The \texttt{TM} command (Trigger Mode)}

With the \texttt{TM} command the client tells \textit{Auge} to open or
close the camera shutter. This command cannot be executed prior to a
successfully executed \texttt{CT} command. The syntax is
\begin{quote}
\texttt{TM n}
\end{quote}
where \texttt{n} is \texttt{1} if the shutter should be opened and
\texttt{0} otherwise. \\
Possible reply codes:

\btab{rp{11cm}}
\texttt{100} & \texttt{okay} \\
             & The shutter state has been set successfully.
\etab

\subsection{The \texttt{START} command}

With the \texttt{START} command the client tells \textit{Auge} to
transmit measurement data. This command cannot be executed prior to a
successfully executed \texttt{CT} command. The syntax is
\begin{quote}
\texttt{START}
\end{quote}
Possible reply codes:

\btab{rp{11cm}}
\texttt{300} & \texttt{here we go} \\
             & The data block of this reply contains the count rates
within the regions of interest defined by the experimentator. See
below for the formatting of the data.
\etab

\subsection{The \texttt{STOP} command}

With the \texttt{STOP} command the client tells \textit{Auge} to
cancel the transmission of measurement data. This command cannot be
executed prior to a \texttt{START} command. The syntax is
\begin{quote}
\texttt{STOP}
\end{quote}
Possible reply codes:

\btab{rp{11cm}}
\texttt{100} & \texttt{ready} \\
             & The transmission has been cancelled.
\etab

\subsection{The \texttt{QUIT} command}

With the \texttt{QUIT} command the client tells \textit{Auge} to close
the \ac{ATP} connection. This command is \textit{always} valid. The
syntax is
\begin{quote}
\texttt{QUIT}
\end{quote}
Possible reply codes:

\btab{rp{11cm}}
\texttt{199} & \texttt{bye} \\
             & The \ac{ATP} connection is being closed.
\etab

\subsection{Formatting of the transferred count rate data}

The count rate values for the defined regions of interest~(\ac{ROI})
are transmitted sequentially. The data for each \ac{ROI} encompasses
five bytes $b_1 \cdots b_5$. The first byte $b_1$ contains the
\ac{ROI} index. If the \ac{ROI} is focused during readout, the highest
(most significant) bit of $b_1$ will be set. The following bytes $b_2
\cdots b_5$ contain the respective count rate as a 32-bit long integer
in network byte order (Big Endian).

\chapter{Paulbox C compiler}
\label{app:lxx}

The Paulbox compiler translates a pulse
programme written in a human-readable language into binary code which
can subsequently be sent to the Paul box. A pulse programme consists of
lines separated by a \ac{CRLF} character sequence. Each line contains
tokens which may be separated by whitespace characters (spaces and
tabs).

\section{General notes}

\subsection{Arithmetics}

Integer numbers can be specified using either an octal (prefix
\texttt{0}), decimal, or hexadecimal (prefix \texttt{0x})
notation. Rational numbers must be specified using the decimal
notation with the period (\texttt{.}) as a decimal separator. An
integer part of zero may be omitted.

The compiler supports simple arithmetic operations:
summation~(\texttt{+}), subtraction~(\texttt{-}),
multiplication~(\texttt{*}), division~(\texttt{/}), and association
using parentheses. \texttt{pi} (the circle number) is recognised as a
special constant.

\subsection{Comments}

Comments begin with the number sign~(\texttt{\#}) and reach to the end
of line.

\section{Syntax}

\subsection{\texorpdfstring{\protect\ac{TTL}}{TTL} control}

\begin{quote}
\texttt{ttlX = Y}
\end{quote}
where \texttt{X} is the index of the \ac{TTL} output expressed as a
decimal number between $0$ and $15$ without leading zeroes and
\texttt{Y} is the desired \ac{TTL} value---\texttt{0} encodes \ac{LO}
and \texttt{1} encodes \ac{HI}.

\subsection{\texorpdfstring{\protect\ac{DAC}}{DAC} control}

\begin{quote}
\texttt{dacX = Y}
\end{quote}
where \texttt{X} is the index of the \ac{DAC} expressed as a decimal
number between $0$ and $3$ and \texttt{Y} is a rational number between
$0$ and $1$.

\subsection{\texorpdfstring{\protect\ac{DDS}}{DDS} control}

The four \ac{DDS} boards each use four internal profiles which can be
independently written. Each profile contains a frequency register and
a phase register. Note that a call to \texttt{ddsX.update()} is needed
to force the \ac{DDS} boards to output the currently set register values.

\subsubsection{Writing frequency registers}

\begin{quote}
\texttt{ddsX.profileY.frequency = Z}
\end{quote}
where \texttt{X} is the index of the \ac{DDS} board expressed as a
decimal number between $0$ and $3$, \texttt{Y} is the profile index
expressed as a decimal number between $0$ and $3$, and \texttt{Z} is
the desired frequency in megahertz expressed as a rational number.

It is also possible to use the shortcut
\begin{quote}
\texttt{ddsX.frequency = Z}
\end{quote}
as an equivalent to \texttt{ddsX.profile0.frequency = Z}.

\subsubsection{Writing phase registers}

\begin{quote}
\texttt{ddsX.profileY.phase = Z}
\end{quote}
where \texttt{X} is the index of the \ac{DDS} board expressed as a
decimal number between $0$ and $3$, \texttt{Y} is the profile index
expressed as a decimal number between $0$ and $3$, and \texttt{Z} is
the desired phase in rad expressed as a rational number.

It is also possible to use the shortcut
\begin{quote}
\texttt{ddsX.phase = Z}
\end{quote}
as an equivalent to \texttt{ddsX.profile0.phase = Z}.

\subsubsection{Setting the active profile}

\begin{quote}
\texttt{ddsX.profile = Y}
\end{quote}
where \texttt{X} is the index of the \ac{DDS} board expressed as a
decimal number between $0$ and $3$ and \texttt{Y} is the index of the
profile to be activated expressed as an integer number between $0$ and
$3$.

\subsubsection{Outputting the current register values}

\begin{quote}
\texttt{ddsX.update()}
\end{quote}
where \texttt{X} is the index of the \ac{DDS} board expressed as a
decimal number between $0$ and $3$ or ``X'', in which case all four
\ac{DDS} boards are updated at the same time. \textit{Note:} The
parentheses \textit{must} be supplied.

\subsection{Pausing}

\begin{quote}
\texttt{wait X}
\end{quote}
where \texttt{X} is the duration that the Paul box should pause. It is
either specified as the integer number of processor cycles (one cycle
is $10\nanosecond$) or as a value in seconds. The latter specification
consists of a number (integer or rational) followed by a time unit
(\texttt{s}, \texttt{ms}, \texttt{us}, or \texttt{ns}). The unit
specification may be separated from the value by one (1) space.

\subsection{Triggering}

\begin{quote}
\texttt{wait triggerX}
\end{quote}
where \texttt{X} is the index of the \ac{TTL} input that---if set to
\ac{HI}---will cause the program to continue. It must be a decimal
number between $0$ and $7$ or ``X'', in which case any set input \ac{TTL}
will cause the program to continue.

\subsection{Looping}

\subsubsection{Infinite loops}

\begin{quote}
\texttt{loop \{} $\cdots$ \texttt{\}}
\end{quote}
The commands to be repeated are enclosed in curly brackets.

\subsubsection{Finite loops}
\begin{quote}
\texttt{loop (X) \{} $\cdots$ \texttt{\}}
\end{quote}
where \texttt{X} is the number of iterations expressed as an integer
number.

\subsection{Subroutines}

Subroutines are code blocks labelled with a unique identifier. They
can be called from within the main programme using their identifier.

\subsubsection{Definition of subroutines}

\begin{quote}
\texttt{sub X \{} $\cdots$ \texttt{\}}
\end{quote}
where \texttt{X} is the case-sensitive identifier of the
subroutine. It must start with a letter (A\,--\,Z, a\,--\,z); the use
of reserved words (\texttt{dds\_update}) is forbidden.

\subsubsection{Calling subroutines}

\begin{quote}
\texttt{X()}
\end{quote}
where \texttt{X} is the case-sensitive identifier of the subroutine to
be called. The parentheses \textit{must} be supplied. Note that a
subroutine must be defined prior to calling it.

% undo \settocdepth{chapter} for list of figures and list of tables
\settocdepth{section}

\backmatter

\bibliographystyle{amsunsrt}
\begin{flushleft}
\ifpdfoutput{
  \cleardoublepage\phantomsection
  \pdfbookmark[-1]{\bibname}{bibliography}
}{}
\bibliography{thesis}
\end{flushleft}

\ifpdfoutput{
  \cleardoublepage\phantomsection
  \pdfbookmark[-1]{\indexname}{index}
}{}
\printindex

\end{document}